\documentclass[5p,times,authoryear]{elsarticle}
\usepackage{amsmath,pifont,graphicx,amssymb,color,setspace,lineno,comment,
	booktabs,dcolumn,tabularx,makecell,pdflscape,caption,subcaption,
	longtable,adjustbox,dcolumn,subcaption,graphicx,lscape,caption}
\usepackage[absolute,overlay]{textpos}
\usepackage{afterpage,lipsum}
\usepackage{multirow}
\usepackage{changepage}
\usepackage[table,xcdraw,dvipsnames]{xcolor}
\usepackage{xspace}

\begin{document}
	\begin{frontmatter}
		\title{Saturn's ancient regular satellites}
		
		\author[ELSI]{E. W. Wong\corref{cor}}
		\ead{emily79@elsi.jp}
		
		\author[CSFK]{R. Brasser}
		
		\address[ELSI]{Earth Life Science Institute, Tokyo Institute of Technology; Meguro-ku, Tokyo, 152-8550, Japan}
		\address[CSFK]{Origins Research Institute, Research Centre for Astronomy and Earth Sciences; 15-17 Konkoly Thege Miklos St., H-1121 Budapest, Hungary}
		\cortext[cor]{Corresponding authors}
		
		\author[UiO]{S. C. Werner}
		\address[UiO]{Centre for Planetary Habitability, Department of Geosciences, University of Oslo; N-0315 Oslo, Norway}
		
		\author[SW]{M. R. Kirchoff}
		\address[SW]{Southwest Research Institute, 1050 Walnut St, Suite 300, Boulder, CO, USA}
		
		\begin{abstract}
			\begin{linenumbers}
				Are Saturn's regular satellites young or old? And how old are Enceladus' cratered plains? To answer these questions we computed model surface ages of the most heavily cratered terrains on Saturn's regular icy satellites using new high-resolution outer Solar System evolution simulations, and coupled with improved estimates of the trans-Neptunian objects populations. 
				The output of the simulations allowed us to construct a model impact chronology onto Saturn which automatically applies to its regular satellites. We used crater densities and our impact chronology onto Saturn to construct model impact-crater isochrons, i.e., the scaling of the satellite crater production function with time. 
				The surface ages derived for the cratered plains on Mimas, Enceladus, Tethys, Dione and Rhea range from 4.1 Ga to 4.4 Ga, with the surfaces of Mimas and Enceladus roughly 200 Myr younger than those of the outer three satellites. Uncertainties in these ages are less than 300~Myr. The calculated model surface ages of these satellites are consistent over as much as two orders of magnitude in the observed crater diameter. The similarity of the crater production function amongst all satellites suggests that they were bombarded by a single impactor source. This work supports the idea that Saturn's regular satellites are ancient, and has implications for their formation and their tidal evolution.
			\end{linenumbers}
		\end{abstract}
		
	\end{frontmatter}
	\section{Introduction}
	What are the surface ages of the icy satellites of the giant planets? In the absence of samples suitable for radiometric dating answering this question can best be achieved through crater chronology of said satellites. The cratering rates in the outer Solar System have been studied by \citet{Zahnle1998} for the Galilean satellites and by \citet{Zahnle2003} for all of the regular satellites of the outer planets from heliocentric impactors. Both crater studies relied on the dynamical simulations of \citet{DuncanLevison1997}, which had a limited impact resolution. \citet{Schenk07} mapped the craters on Triton and argued for a mixed source of impactors consisting of Neptune-centric and heliocentric planetesimals. These studies focused on the apparent steady-state cratering flux rather than the earliest potential decline related to planet formation. The works by \citet{Zahnle1998,Zahnle2003} assumed that the total number of planetesimals crossing the orbits of the giant planets declined with time as $N(>t) \propto t^{-1}$ \citep{HolmanWisdom1993}. With this assumption the heavily cratered regions of most of the regular icy satellites were computed to be older than 4 Ga, and often as old as the Solar System itself; alternatively, the surfaces were computed to be rather young when a much higher flux of planetesimals from the trans-Neptunian region we assumed. 
	Yet, as \citet{Wong2020} showed, the assumed decline in impact flux is not applicable during the early stages of giant planet migration.\\ 
	
	In \citet{Wong2020} we simulated the outer Solar System forward in time from the onset of giant planet migration, which we set at 4.5~Ga \citep{Mojzsis2019}. 
	They updated the impact probabilities from those computed by \citet{Zahnle2003} as well as the expected crater densities using Monte Carlo impact experiments assuming two different size-frequency distributions for the trans-Neptunian objects encountering the giant planets. They combined the decline of the planetesimals from their giant planet migration simulations and their Monte Carlo simulation output to compute surface ages of the most heavily cratered surfaces of the icy regular satellites of Jupiter, Saturn and Uranus, and generally found that these ages ranged from 3.8 Ga to 4.4 Ga. \\
	
	Previous simulations of \citet{Wong2020} had poor resolution for impacts onto the giant planets in the last 3.5 Gyr of their outer Solar System simulations due to the limited number of planetesimals in their simulations. This prevented them from obtaining accurate ages of surfaces with low crater densities, such as Enceladus' ridged plains \citep{Kirchoff2009}. Furthermore, their obtained ages showed some variation due to the order of magnitude difference in the number of small planetesimals for the two different size-frequency distributions used. Their ages were also tied to craters with diameter $D_{\rm cr}>20$~km rather than based on the crater production function of these satellites.\\ 
	
	In this work, we obtained a more accurate crater chronology onto the icy regular satellites that can be applied to the whole history of the Solar System. We computed model surface ages for said satellites for various crater diameters from age isochrons similar to those that exist for the Moon and Mars in the inner Solar System. We applied this new chronology to the inner Saturnian satellites to compute more accurate model surface ages of their most densely cratered surfaces. Application to the Jovian and Uranian satellites is reserved for future work. We stress that our work here relies {\it only} on heliocentric impactors. We do not account for potentially planetocentric impactors because circumplanetary bodies should be eliminated in less than 1~Myr \citep{Cuk2016}. Effects of sesquinary craters still rely on primary impacts for supply, whose delivery is tied to the flux of heliocentric impactors. \\
	
	In this paper, we uses ``{\it crater chronology}" and ``{\it cratering}" to refer to the bombardment of the icy regular satellites, and ``{\it impact chronology}" and ``{\it impacting}" for those of the giant planets. Because craters form on the solid crust of the satellites whereas no long-term bombardment remnant can be preserved on the atmospheric surfaces of the giant planets.
	``{\it Size-frequency distribution}" refers to a model of a statistical relationship between the diameters of craters or projectiles (interchangeable with impactors) and their frequency of occurrence, which is express as spacial density (km$^{-2}$) for craters and number of object for projectiles.
	An expression for a diameter ranges is: 
	$n_{\rm >D} \propto (D_0/D)^\alpha$, 
	where $D_0$ is a scaling diameter in km, and $\alpha$ is the cumulative slope for that size ranges. 
	Its vertical axis can be adjusted for different conditions on satellites or terrains. E.g., an older surface would have an up-shifted distribution, assuming the same source and rate of bombardment. ``{\it Size-frequency measurement}" has the same physical nature as the {\it size-frequency distribution}, plots the diameter of observed craters surveyed from icy satellites' images. 
	In short, {\it size-frequency distribution} are models that can be fitted to or derived from the observed data -- {\it size-frequency measurement}.
	The figures in this paper show {\it size-frequency distribution} and {\it measurement} in cumulative frequency, where the number of craters larger than a specific diameter is plotted against diameter, and the population decreases with size. 
	Regarding the unit for time,  ``Myr" represents million years and ``Gyr" represents billion years,   denoting durations or specific times within the simulation, e.g., Neptune began to migrate outward 10 Myr into the simulation or at 10 Myr; ``Ma" and ``Ga" denote the ages, counting the time before present, e.g., the surface ages of Enceladus is $\sim$4.1 Ga, i.e. 4.1 billion years old.

	\section{Methods, inputs and parameter choices}
	\label{sec:method}
	
	To determine the surface age on an icy satellite of the giant planet, we evaluated which crater isochron best represents the observed crater size-frequency measurement. The crater isochron --- crater size-frequency distribution as a function of surface age, derived using our improved crater chronological model; this model was constructed from the results of multiple sets of 4.5 Gyr dynamical simulations of the outer Solar System evolution with a tran-Neptunian scattered disc and the giant planets migration. \\
	
	Here we briefly outline the steps for surface age estimation. \\
	
	First, we ran N-body simulations of giant planet migration and their effect on the planetesimals in the trans-Neptunian disc to establish the rate of decline in the planetesimals.
	Second, we used the simulation results to construct the crater chronologies for the icy satellites by fitting the impact flux of planetesimals onto their corresponding host giant planet, and scaled the planets' impact chronologies with each satellite's impact probability and crater scaling law. Third, we computed crater isochrons by combining the decline in the cratering rate from our modelled crater chronologies, the planetesimal size-frequency distribution and crater production functions. Last, we evaluated the observed crater size-frequency measurement of each icy satellite's cratered plain, as reported by \citet{Kirchoff2009,Kirchoff2010}, with the isochrons to search for the best-fitted surface ages.\\
	
	In this work, we substantially enhance the modelled chronologies through increased impact resolution, thereby improving the precision of the estimated surface ages. We achieved this by cloning planetesimals after 1 Gyr of giant planet migration and running them in another set of simulations to sustain an adequate amount of impacts over the last 3.5 Gyr, especially near present. 
	
	\subsection{Giant planet migration simulations of the outer Solar System}
	\label{sub:mig}
	Our purpose in conducting these simulations is to establish the rate and extent of planetesimals depletion through ejection and collisions with the giant planets. This information is necessary for constructing the profile of the satellite crater chronologies, which are essential tools for estimating the surface ages of the icy satellites.\\ 
	
	We ran 320 N-body simulations of the outer Solar System. Each simulation starts with the four giant planets on a compact configuration partially taken from \citet{Nesvorny2012}, and a primordial trans-Neptunian disc consisting of 16,000 planetesimals that extends from 24 AU to beyond 30 AU (the disc's outer edge will be defined later), with a surface density scaling as $\Sigma \propto r^{-1}$. 
	The initial semi-major axes from Jupiter to Neptune are: 5.2 AU, 9.58 AU, 16.4 AU, and 22.8 AU. Their eccentricities range from 0.004 to 0.006 and their inclinations are 0.006$^\circ$ for Jupiter, 0.016$^\circ$ for Saturn, and 0.001$^\circ$ for the ice giants. 
	We utilised the initial orbital elements from \citet{Brasser2015} and \citet{Wong2020}, which were selected based on multiple test simulations to achieve a post-migrated configuration that aligns with the current state of outer Solar System.\\
	
	The dynamical evolution of the giant planet migration simulations consists of two parts. 
	First, we simulated the migration of Uranus and Neptune using fictitious forces over the course of the first 1 Gyr, which corresponds to 4.5 Ga to 3.5 Ga. To conduct this simulation, we employed a modified version of the SWIFT RMVS4 integrator \citep{LevisonDuncan1994} as described in \citet{Levison2008}. This integrator not only migrates the planets, but also damps their eccentricities, and stabilises their orbital evolution to increase the chance of replicating the current configuration of the outer Solar System.
	For the second part, the last 3.5 Gyr, we utilised the unmodified RMVS4 integrator, and giant planets remained on their current orbits.
	We refer to \citet{Levison2008,Nesvorny2012,Brasser2013,Brasser2015} for the planets' evolutionary pathway which reproduces the current condition of the outer Solar System, including the final orbital elements of the giant planets and leftover planetesimals (i.e., Scattered Disc objects). A more in-depth discussion of our methodology and its limitations is discussed in Section~\ref{sec:gpm}.\\
	
	\begin{table}[ht]
		\centering
		\begin{tabular}{lccccc} 
			Type of disc & \multicolumn{2}{c}{ Compact disc }  & \multicolumn{3}{c}{ Extended disc } \\
			\cmidrule(lr){1-1}\cmidrule(lr){2-3}\cmidrule(lr){4-6}
			Batch  & ~1$^{st}$ & 2$^{nd}$~ & ~3$^{rd}$ & 4$^{th}$ & 5$^{th}$~ \\
			Outer edge [AU]~~ & ~31 & 33~ & ~35 & 37 & 39~ \\
			No. simulations & \multicolumn{2}{c}{128}  & \multicolumn{3}{c}{192} \\
		\end{tabular}
		\caption{\label{tab:sdsim} Overview of the two types of giant planet migration simulations with their corresponding batches, distance of the disc's outer edge to the Sun, and number of simulations.} 
	\end{table}
	
	Regarding the outcomes of the simulations, since the depletion rate and percentage of the planetesimal depend on the location of the planetesimal disc's outer edge, we account for that uncertainty by separating our simulation into five batches, with 64 simulations each. The inner edge of the disc was fixed at 24 AU from the Sun, while the outer edge distance was varied: 31 AU, 33 AU, 35 AU, 37 AU, and 39 AU from the Sun. The first two values were chosen based on the assumption that Neptune stopped migrating near the outer edge of the disc, which is estimated to be around 30 AU \citep{Gomes2004}. However, we wanted to explore how the location of the disc's outer edge affects the simulation result and especially the removal of the planetesimals. The first two batches, comprising 128 simulations, involved close-in and more compact discs with the initial disc extending from 24 AU to 31 AU and 33 AU. The remaining three batches, comprising 192 simulations, involved extended discs with the outer edge at 35 AU, 37 AU, and 39 AU.\\
	
	Each simulation varies only in the initial vectors of the planetesimals. We generated 16,000 planetesimals randomly with initial eccentricities of 0.05$\pm$0.02, inclinations of 1.5$^\circ\pm$0.6$^\circ$, and perihelion distances greater than 23 AU. The longitude of the ascending node, the argument of perihelion, and the mean anomaly were distributed between 0$^\circ$ to 360$^\circ$. The time step was set to 0.4 year, and planetesimals were removed upon collision with a massive body or if they ventured beyond 3000 AU from the Sun \citep{Wong2020}.
	
	\subsection{Restriction on using giant planet migration simulations alone}
	\label{sub:restrict}
	Planetesimals are rapidly removed due to encounters with the giant planets. Previous results showed only 0.39-0.78\% (50 -- 125 per simulation) remaining after 1 Gyr of evolution \citep{Wong2020}, which is insufficient to accurately model the impact rate on the giant planets for the next 3.5 Gyr. This leads to a truncated impact chronology and high statistical uncertainty. \\
	
	The truncated chronology can be exemplified by our previous simulations, in Figure 6 of \citet{Wong2020}, the chronological curves for the Saturnian and Uranian systems terminated at about 1.2 Gyr. Analysing the impact data from all 320 migration simulations written in Section~\ref{sub:mig} of this work, we found that the last impacts onto Jupiter, Saturn, Uranus, and Neptune occurred at 3.82, 4.05, 3.45, and 1.39 Gyr, respectively, and were preceded by $\sim$1 -- 2 Gyr without impact (except Neptune). Since our model crater chronology is based on the impact flux onto the giant planets, the sparse impacts at late simulation time restricts the length of the applicability and accuracy of the chronology curve. Therefore, using the migration simulations alone, we can only accurately construct the impact chronology up to 1.5--2 Gyr, at best.\\
	
	Regarding the accuracy of the impact chronology, we note that each rare stochastic impact on the giant planets occurring after 1~Gyr of simulation time would significantly offset the tail of the impact chronology profile, making the impact rate at later time inaccurate, if not unrepresentative. E.g., the second last impact on Saturn occurred at $\sim$1.3 Gyr: the impact frequency forms a straight line from 1.3 Gyr to the last collision at 4.05 Gyr (see Appendix  figure.~\ref{fig:fit_step}). Due to the sparseness and randomness of impacts, an extrapolation between the last two or three recorded impacts would be arbitrary, and it becomes neigh impossible to derive any reliable impact rate for the recent 3 billion years. 
	Therefore, accurate surface age estimations can only be reliably performed for old and heavily cratered plains using the output from migration simulations, as demonstrated in \citet{Wong2019,Wong2020}. \\
	
	The insufficient number of initial particles in the migration simulation cannot withstand the rapidly declining impact flux, which it restricts the surface ages ranges that the modelled chronology could determine, especially for young surfaces such as that of Enceladus. As such, to extend the chronology reliably to 4.5 Gyr more planetesimals or recorded impacts are needed.\\
	
	\subsection{Additional simulations with cloned planetesimals}
	\label{sub:clone}
	We thus ran two batches of 32 additional simulations to better model the planetesimal evolution after 1 Gyr, when the giant planets were in their final orbital configuration. These extra 64 simulations will be dubbed as the ``{\it clone simulations}", contrasting with the ``{\it giant planet migration simulations}" in Section~\ref{sub:mig}. \\
	
\begin{figure*}
		\centering
		\resizebox{\hsize}{!}{\includegraphics{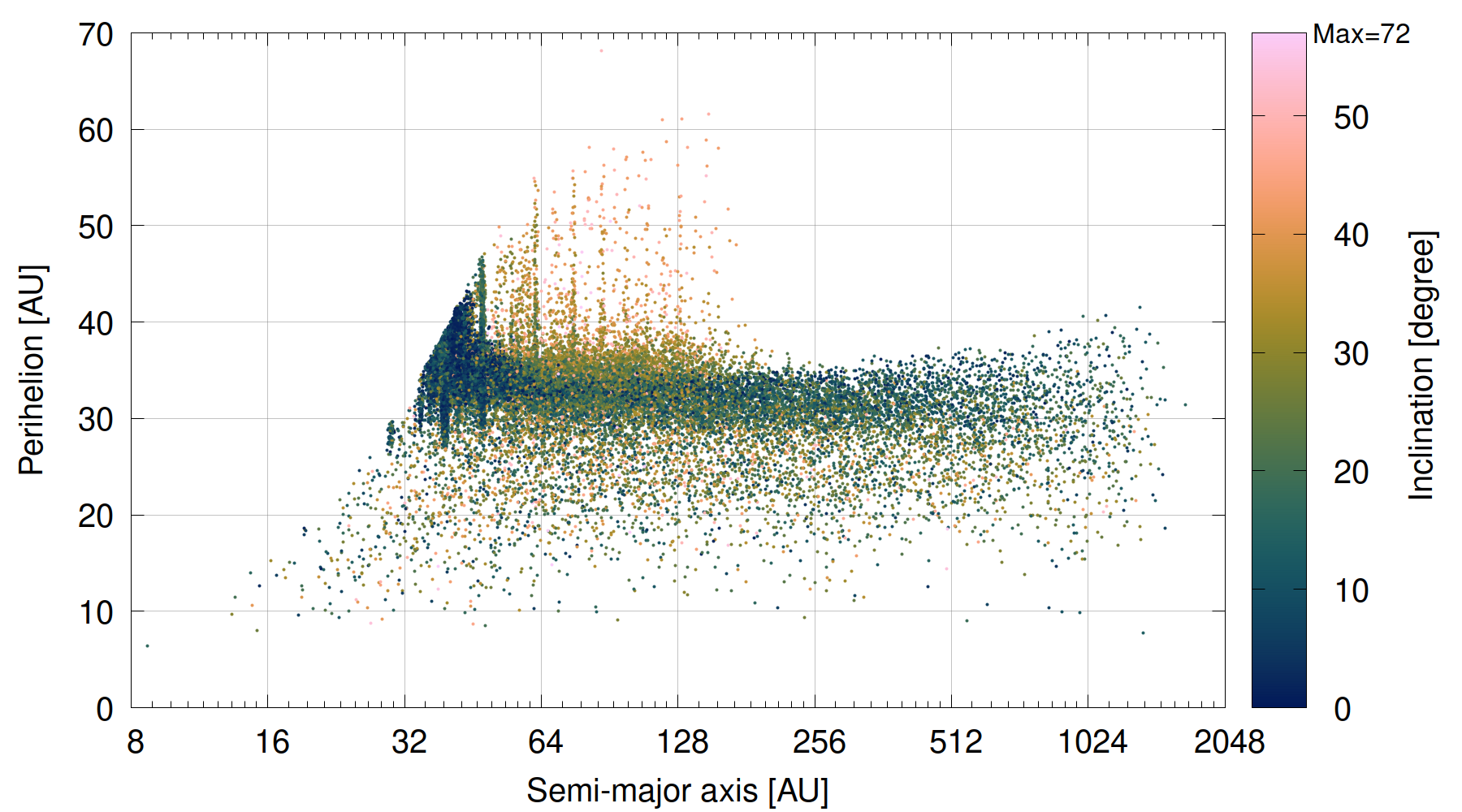}}
		\caption{\label{fig:tp1g}Perihelion (q) versus semi major axis (a) distribution of surviving planetesimals after 1 Gyr from 320 giant planet migration simulations. It combines survivors from both the compact disc and extended disc models with dot colours indicating inclination ($i$) according to the scale of the colour-bar. Most planetesimals are on Neptune-crossing orbits with q$\sim$30 AU. Some of them resemble the current plutinos: $\sim$39 AU and in 2:3 resonance with Neptune, and Cubewanos: 41.8 AU $< a <$ 48 AU and in several orbital resonance with Neptune.  Beyond 48 AU, particles generally have higher eccentricity and inclination (e.g., e $>$ 0.2 \& i $>$ 40$^\circ$), then become more highly elliptical (e.g., e $>$ 0.6) yet losing a few highly inclined objects beyond 130 AU. 
		}
	\end{figure*}

	Each clone simulation started with the four giant planets in their final configuration and 4096 planetesimals (clones), which are randomly replicated from the orbital elements of the surviving planetesimals (survivors) at 1 Gyr of the regular giant planet migration simulations. There are 39 to 241 survivors per simulation, depending on whether the disc was compact or extended. The semi-major axis, perihelion distance and inclination of the survivors are depicted in Fig.~\ref{fig:tp1g}, and the clones' mean anomalies were randomized while keeping the other orbital elements fixed. \\
	
	The first batch of clone simulations utilised survivors from the compact giant planet migration simulations, where the disc's outer edge was at 31 or 33 AU, while the second batch used survivors from the extended disc simulations, where the outer edge was at 35, 37 or 39 AU. The difference between the two batches of clones simulation lies in the source groups of survivors from which they were replicated.\\
	
	A high initial depletion rate of the clones is expected because many of them will be on unstable orbits due to their altered initial mean anomalies. Most planetesimals that survived for 1 Gyr in the migration simulation did so on meta-stable orbits, most likely in a temporary resonant configuration with Neptune \citep{DuncanLevison1997}. Since we changed their mean anomalies in the clone simulations, many clones will not be in a resonance with Neptune and are quickly scattered by the ice giant. \\
	
	The majority of clones resided in the scattered disc (see Fig. \ref{fig:tp1g}), while some were on Neptune-crossing orbits, so that many of them were removed in the first $\sim$100 Myr. We consider this initially enhanced depletion and collision rates to be artefacts of the cloning procedure. Therefore, when constructing the impact chronology from collision counts, we exclude the first 100 Myr of clone particle evolution. \\
	
	In Section~\ref{sub:fit}, we will explain how we incorporated the collision counts from the clone simulations with those from the original giant planet migration simulations; this step enhanced the impact statistics, and smoothed and extended the crater chronology to the present. A description of how the clone simulations improved our overall results are given in Appendix B.
	
	\subsection{The initial number of trans-Neptunian objects and their size-frequency distribution}
	\label{sub:nsd}
	The expected number of craters formed on the icy satellite's surfaces depends on the initial number of planetesimals in the trans-Neptunian source region at 4.5 Ga ($N_{\rm SD}$). Based on Outer Solar System Origins Survey observations and the Centaur distribution, the current number of scattered disc objects with diameter $D_{\rm i}>10$~km is $(2.0 \pm 0.8) \times 10^7$ \citep{Nesvorny2019}. Our giant planet migration simulations showed the nominal average remaining percentage of the scattered disc objects (SDOs) after 4.5 Gyr of evolution is 0.31\%, consistent with independent estimates 
	by \citet{Nesvorny2018}. This efficiency implies there were nominally $6.4 \times 10^9$ SDOs with $D_{\rm i}>10$~km in the beginning, before giant planet migration.
	To scale the above number to a population with different minimal sizes, a cumulative size-frequency distribution is used: $N(>D) \propto (D_0/D)^{ \alpha}$, where $\alpha$ is the cumulative slope (negative, as the cumulative number of object decrease with size), and $D_0$ is a scaling diameter.\\ 
	
	Two observations help us to constrain the size-frequency distribution: the current small bodies' size-frequency distribution (e.g. Jupiter-family comets, Jupiter Trojans), and the crater size-frequency distribution on Pluto and Charon. \\
	
	Based on the former observations, several studies suggested the size-frequency distribution of SDOs has a slope $\alpha \sim 4.5$ for impactor/planetesimal diameter ($D_{\rm i}$)$>100$~km \citep[e.g.,][]{Sheppard00,Trujillo01,Gladman2001,Zahnle2003,Shankman2013,Fraser2014,Lawler2018}. 
	However, the slope might not truly reflect the historical size-frequency distribution of the impactors in the distant past, as the initial size distribution in the early Solar System could have been altered by collisional evolution \citep{KenyonBromley2004}. 
	Moreover, the diameter ranges of the observed objects in the current SDOs or Centaur region are larger than the impactor sizes considered in this study, which are mainly objects with $D_{\rm i}\lesssim 1.5$ km, for hallowing out a 20~km diameter craters on icy satellites (please see Table~\ref{tab:crater}). Compared with the studies of large objects, the size-frequency distribution for outer Solar System planetesimals with $D_{\rm i} <1$~km or even $D_{\rm i}<10$~km are poorly constrained \citep[e.g.][]{Lawler2018,Nesvorny2018}.\\
	
	Various crater studies on the outer Solar System have provided insights into the size-frequency distribution of planetesimals.
	For example, Enceladus exhibits a cumulative crater diameter slope of $\alpha_{\rm cr}=-0.9$ for $D_{\rm i}<1$~km \citep{Kirchoff2009}, while the $\alpha_{\rm cr}$ becomes steeper for larger diameters.
	Ganymede's craters in the Gilgamesh region have $\alpha_{\rm cr}=-1$ to $-1.2$ for $D_{\rm i}<2$~km \citep{Zahnle2003}. 
	On Triton, \citet{Schenk07} reported $\alpha_{\rm cr}=-2.25$ for crater diameter ($D_{\rm cr}$) ranging from 5 -- 25~km, approximately corresponding to $D_{\rm i}$ of 0.4 -- 1.4~km. Additionally, it is possible that Triton's craters between 3 -- 5~km in diameter, which roughly correspond to $D_{\rm i}$ of 0.2 -- 0.4~km, have a steeper slope of $\alpha=-3$. Furthermore, craters on the Pluto-Charon binary have revealed a size-frequency distribution with one kink\footnote{Depends on the crater diameter ($D_{\rm cr}$), the kink at $D_{\rm cr}>10$~km as knee, and $D_{\rm cr}<1$~km as elbow}, where $\alpha_{\rm cr} = -0.7$ for $D_{\rm cr}=$ 1 -- 13~km, which corresponds to $D_{\rm i}$ of 0.1~km to less than 1~km, and a steeper slope of $\alpha = -2$ for $D_{\rm i}$ from 1~km to 10~km \citep{Singer2019}.\\
	
	It is important to note that the slope derived from the observed crater remnants may not be entirely accurate. $\alpha_{\rm cr}$ may be shallower due to the preferential erasure of small craters by material burial inside the crater floor, geological activities, and removal or overlap by a large impact that came at a later time. On the other hand, $\alpha_{\rm cr}$ may be steeper due to contamination by generally smaller secondary and sesquinary craters --- both produced by minor impact ejecta. However, data from Cassini suggests a trend of shallow $\alpha_{\rm cr}$ for small diameters \citep{Kirchoff2018}, indicating that contamination by secondary and sequinary craters may not be a significant factor.\\
	
	We scaled the number of impactors from the observed number of SDOs with $D_{\rm i}$. We have, 
	\begin{equation}
		\label{eq:n_sd_d}
		N_{\rm SD,~>D_{\rm i}} =  N_{\rm scale} \left(\frac{D_{\rm i}}{10~\rm km}\right)^{\alpha}.
	\end{equation}
	
	For this work we shall adopt $\alpha=-2$ from \citet{Singer2019} for an impactor size-frequency slope with $D_{\rm i}>10$~km, because the slope of Triton's craters could be affected by Neptune-centric debris \citep{Schenk07,MahBrasser2019}, while those of the Pluto-Charon pair does not indicate strong contamination from planetocentric secondary craters \citep{Singer2019}.
	
	Furthermore, this slope seems to comport with several observational studies of trans-Neptunian objects \citep{Shankman2013,Fraser2014,Lawler2018}. 
	At the smaller end, for impactors with diameter $D_{\rm i}<1$~km, we shall occasionally adopt the slope of \citet{Singer2019}, i.e. $\alpha=-0.7$, while at other times we keep $\alpha=-2$. For planetesimals with $D_{\rm i} \ge 100$~km we adopt $\alpha=-4.5$, which is consistent with the high-end slope for the dynamically hot SDOs \citep{Shankman2013,Fraser2014,Lawler2018}. Jupiter's Trojans appear to yield a similar faint-end slope \citep{WongBrown2015,YoshidaTerai2017}.\\
	
	According to observations, at present $N_{\rm scale}=2.0 \times 10^7$ \citep{Nesvorny2019}, and with our adopted values of $\alpha$ the current number of SDOs with diameter $D_{\rm i}>1$~km is approximately $2.0\times10^9$, which is consistent with the values obtained by previous works \citep[e.g.][]{Brasser2013,VolkMalhotra2008,Shankman2013,Lawler2018}. 
	Assuming a nominal depletion of 99.69\% (which is the derived from our migration simulation result; see Section~\ref{sub:result_nsd}), the initial number of objects in the disc with $D_{\rm i}>10$~km at 4.5 Ga is estimated to be $N_{\rm scale,0}=6.4\times 10^9$, which is in agreement with independent estimates obtained from Jupiter's Trojans \citep{Nesvorny2013,Nesvorny2018}.
	Adopting $\alpha=-2$, the initial number of SDOs in the disc with $D_{\rm i}>1$~km is approximately $N_{\rm SD,>1}=6.4\times 10^{11}$.  This comports with estimates obtained with Monte Carlo simulations using the adopted $\alpha$ for objects larger than 1~km and a disc of 18~$M_\oplus$ \citep{Wong2019}, which yielded $N_{\rm SD,>1}=7.23\times 10^{11}$ and is within the ranges of $2\times 10^{11}$ and $2\times 10^{12}$ listed in \citet{Wong2020}. \\
	
\begin{figure}
		\centering
		\resizebox{\hsize}{!}{\includegraphics{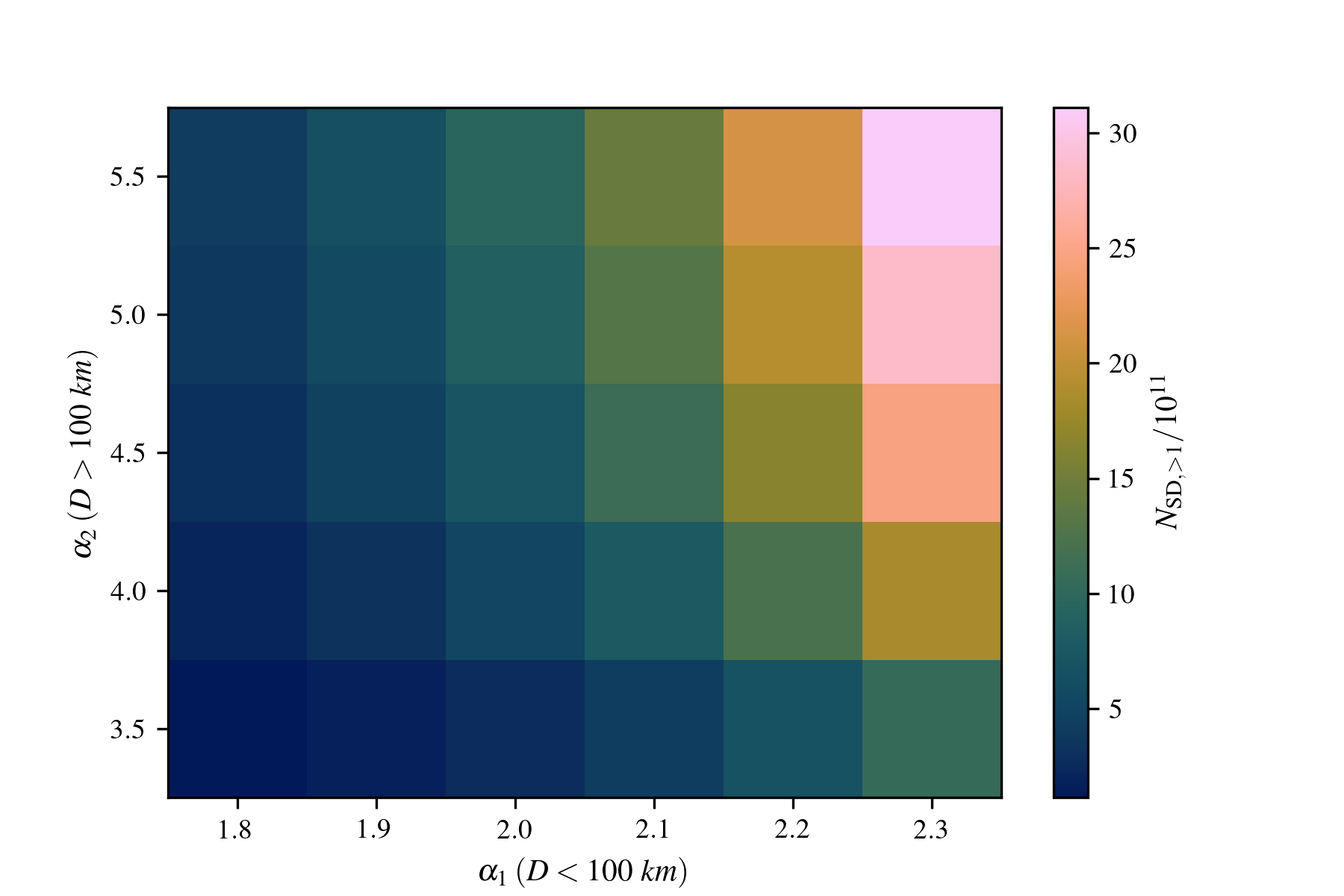}}
		\caption{\label{fig:nsd}Number of scattered disc objects with diameter larger than 1~km ($N_{\rm SD,~>1}$) computed from Monte Carlo simulations adopting different combinations of cumulative slope ($\alpha_{1}$ and $\alpha_{2}$) in the size-frequency distribution. The colour-bar shows $N_{\rm SD,~>1}$ expressed in colour scale. The matrix is plotted with $\alpha_{2}$, the cumulative slope for a diameter larger than 100~km, versus $\alpha_{1}$, the cumulative slope for a diameter smaller than 100~km.}
	\end{figure}

	The actual number of primordial objects with $D_{\rm i}>1$~km can vary by an order of magnitude depending on the chosen slopes at the low and high end \citep{Wong2020}. This is demonstrated in Fig.~\ref{fig:nsd}, which displays the number of primordial objects with $D_{\rm i}>1$~km for an 18~$M_\oplus$ disc as a function of the low- and high-end slopes. However, the estimates presented in the previous paragraphs suggest that our choice of slopes comports with independent estimates of the number of objects and dynamical simulations.
	
	\subsection{The cratering chronology model}
	\label{sub:fit}
	Ideally, such a model would represent all impacts onto the solid crust and extend forward in time. But due to the lack of calibration samples from the icy crust of the giant planets' satellites, such a crater chronology cannot be created. To overcome this challenge, we modelled the crater chronology from the impacts onto the giant planets in our simulations (migration simulation and clone simulation combined). We plotted the chronology as a declining curve that tracks the diminishing number of impacts onto a body with time. \\
	
	The crater chronologies of the icy satellites are calibrated based on the impact chronology onto the host planet since the bombardment onto the satellites originates from the same population of planetesimals that intruded within the planets' Hill radius. The planet's gravity controls those close encountering planetesimals, which in turn governs the bombardment onto the icy satellites. Therefore, the decay curvature of the crater chronological model for the icy satellites shares the same profile as the impact chronological model of host planet. \\
	
	We utilised a stretched exponential function, also known as the complementary cumulative Weibull distribution, to fit the complementary cumulative impact count of a planet, which is denoted as  $F(t)~=~1-N(<t)/N(<4.5)$. Here, $t$ is the time in Gyr after the simulation began, $N(<t)$ is the cumulative number of impacts before time $t$, and $N(<4.5)$ represents the total number of impact throughout the entire simulated time on a planet. A stretched exponential decline is expressed as,
	\begin{equation}
		\ln F(t) = - \left( t/\tau \right) ^{\beta}. 
		\label{eq:weibull}
	\end{equation}
	It gives the best fit in describing the declining impact frequencies of the planets and the removal rate of the planetesimals from the outer Solar System \citep{Dobrovolskis2007}. 
	This function contains the e-folding time ($\tau$) and stretching parameter ($\beta$).
	In our SWIFT RMVS4 simulations, we recorded every impact on the giant planets, allowing us to track the impact count over time. As the migrating giant planets stabilize and the number of planetesimals decreases, the impact count also decreases. 
	By fitting the declining cumulative impact counts with the above function, we acquire $\tau$ and $\beta$ and used them to construct the modelled impact chronology for each planet. \\
	
	As discussed earlier, the scarcity of the impacts after 1 Gyr introduced uncertainty in data fitting and constrained the length of the modelled impact chronology. Only around 108 planetesimals (0.67\%) remained after 1 Gyr per migration simulation.
	Our solution is to incorporate the impact data from the clone simulations with those from the migration simulations to construct a joint impact profile.
	The combined impact frequency extended longer over time and traced the declining curves better, thereby increasing our confidence in the later estimates of the satellites' surface ages. Incorporating the results from the clone simulations is achieved in three steps. \\
	
	First, we computed the cumulative impact distributions from both the migration and clone simulations. 
	Second, we selected only a specific range of the impact chronology from the two simulations, incorporating impacts from the clone simulation only after 1.1 Gyr and terminating the chronological curve of the giant planet migration simulations at 1.1 Gyr. In other words, we use the first 1.1 Gyr of impact data from the giant planet migration simulation output and the last 3.4 Gyr from the clone simulation output.
	Next, we scaled the impact decay profile of the clone simulation vertically to match the value of the migration simulations at 1.1 Gyr and link up the head of the clone simulation's curve to the tail of the migration simulation's curve. 
	The two impact frequencies are aligned at 1.1 Gyr to produce a single, smooth profile. 
	(Fig.~\ref{fig:fit_step} illustrates the steps involved in scaling the decay profile of the clone simulation and combining it with those from the migration simulations.) 
	As a response to the difficulties mentioned in Section~\ref{sub:restrict}, the impact chronology modelled from the combined impact frequencies extended further in time, benefiting from the resupply of planetesimals from the clone simulation; and has a much lower fitting uncertainty as there are more impact counts to fit, especially after 1 Gyr. \\
	
	Fitting the impact frequency poses a challenge in both the late and early stage of the simulation. \\
	
	The early stage of simulation is characterized by an extremely high number of impacts and rapid decline in the rate of impacts: in the giant planet migration simulations, 97.5\% of the impacts onto Saturn happened within the first 100 Myr, whereas only 0.5\% impacts occurred in the later 100 Myr, between 200 Myr and 300 Myr. Such an extreme number of impacts in the beginning introduce biases in the fitting towards the early stage.  Fitting every single impact from the start closely traces the rapidly declining curve of the early stage, but deviates from the slower declining curve at much later time. \\
	
	To address this second issue, we revised our fitting method. A good fit for the impact frequency should trace the declining curve over the entire 4.5 Gyr, regardless of the uneven distribution of impact counts. The interval between the impacts in the early time are shorter, thus the data points are more concentrated near the beginning of the impact frequency curve.
	To achieve an even and fair fitting, we did not fit all the impact counts. Instead, we sampled the data every 30 Myr for each planet and fit through approximately evenly distributed data points in time.
	We extracted one impact count per 30 Myr time interval, starting with the first impact at 30 Myr. The 30 Myr interval itself was obtained by conducting a Kolmogorov-Smirnov goodness-of-fit test between the sampled impacts and all impacts combined for different values of the data spacing. The revised fitting method avoids biases toward the beginning of the simulation and results in a better fit to the impact frequency curve without significant deviations at either end. 
	Additionally, the impact frequency curve approaches a stretched exponential decline after 30 Myr, enabling more accurate fitting using the Weibull distribution. It is worth noting that assigning different weights to impacts occurring at different times can also address the fitting biases, but such statistical manipulation may introduce additional artifacts to the results of the dynamical simulations.\\
	
	A linear least-squares algorithm was used to fit the cumulative impact frequency for each giant planets through $\log | \ln F(t) | = \beta \log t - \beta \log \tau$ \citep{Dobrovolskis2007}, an alternate presentation of Eq.~\ref{eq:weibull}. Using the fitted $\tau$ and $\beta$, we constructed modelled impact chronologies for the four giant planets. The impact chronology of Saturn was then scaled to match the expected crater densities formed on the icy satellites' crusts over the course of 4.5 Gyr, to model the crater chronologies of the Saturnian satellites.\\
	
	Our crater chronology models serves as an age-estimation tool calibrated for each icy satellite in the outer Solar System, showing the expected crater densities as a function of age. By inserting the observed crater densities surveyed from the satellite image into the corresponding chronology, we can calculate a model surface age. 
	Given that these individual chronologies derive from the same overall set of simulations, we have produced an outer Solar System wide crater chronology model and being capable to remotely date all icy regular satellites' surface ages with crater counting.
	
	\subsection{Expected crater density of the icy satellites}
	\label{sub:exp_crater}
	Before we elaborate on how to derive the crater density, we would like to clarify the two sets of crater densities that are used to calculate the ages of the crust. In this subsection, ``{\it expected} crater density" is the average global crater density of a 4.5 Ga surface on a icy satellite, calculated by dividing the total number of craters that were {\it expected} to be formed throughout the 4.5 Gyr of bombardment by the surface area of the satellite ($A_{\rm s}$). Meanwhile, ``{\it observed} crater density" is the number density of craters measured from a satellite's image. Both crater densities are size-dependent.\\
	
	The expected number of craters formed during the episode of giant planet migration can be deduced from the initial number of SDOs and the impact probabilities onto a icy satellite, $P_{\rm s}$. For craters that can be created by an impactor smaller than $D_{\rm i}$, their expected crater density ($n_{>D_{\rm i}}$) can be expressed as:
	\begin{equation}
		\label{eq:n_cr}
		n_{>D_{\rm i}} = \frac{N_{\rm SD, >D_{\rm i}} P_{\rm s}}{A_{\rm s}}.
	\end{equation}
	Here $P_{\rm s}$ is derived from the collision counts onto the giant planets from the dynamical simulations (Sec.~\ref{sub:mig}) and planet-satellites system flyby simulations from \citet{Wong2020}. 
	The impact probability ratios of the Jovian, Saturnian, and Uranian icy satellites to their host planets are obtained from Table 3 of \citet{Wong2020}.
	In this study, we updated the impact probabilities of giant planets by the heliocentric planetesimals ($P_{\rm P}$) using the collision count from the giant planet migration simulations performed here. $P_{\rm P}$ is either an unitless quantity or in percentage. 
	Following \citet{Wong2020}, impact probability with a satellite is computed as 
	\begin{equation}
		P_{\rm s} = \frac{N_{\rm i}}{N_{\rm tot}}\ P_{\rm enc}, 
		\label{equ:psat}
	\end{equation}
	where $P_{\rm enc}$ is the probability to encounter a giant planet within 40 planetary radii, $N_{\rm i}$ is the recorded number of impacts on the satellite, and $N_{\rm tot}$ is the total number of planetesimals generated for the flyby simulations performed by \citep{Wong2019}. \\
	
	The intense gravitational focusing of planetesimals by the giant planets results in the cumulative peri-planet distribution increasing linearly approximately up to the Hill radius \citep{Vokrouhlicky2008}. Hence the impact probability with the satellite is scaled to that of the planet via
	\begin{equation}
		\frac{P_{\rm s}}{P_{\rm P}} \sim \frac{R_{\rm s}^2}{a_{\rm s} R_{\rm P}},
	\end{equation}
	where $a_{\rm s}$ is the semi-major axis of the satellite, $R_{\rm s}$ and $R_{\rm P}$ is the physical radii of satellite and planet, respectively.
	This relation stems from studies on the impact flux or breakup rate of small bodies orbiting Saturn \citep{Morfill1983, Colwell1994}. It has been extended and simplified to estimate the impact probability of icy satellites by \citet{Zahnle1998, Zahnle2003}, and subsequently utilised in \citet{Wong2020}. 
	
	\subsection{Crater scaling laws}
	\label{sub:scaling}
	The scaling law by used by \citet{Zahnle2003} is based on the $\pi$-scaling law \citep{Housen1983}, which predicts the final crater diameter as a function of parameters, such as the impact velocity ($v$), impactor's diameter ($D_{\rm i}$), impactor density ($\rho_{\rm i}$), target density ($\rho_{\rm t}$), and target gravity ($g$). We inverted the scaling law to estimate the required $D_{\rm i}$ that can excavate a crater with final diameter of $D_{\rm cr}$, which is the diameter we use to calibrate our model and compute surface ages. This relationship is given by below, where the densities of both impactor and target only enter as a ratio, and we have the simple-to-complex crater transition as $D_{\rm sc}$.
	\begin{eqnarray}
		D_{\rm i}
		&=&0.143\ \Bigl(\frac{D_{\rm cr}}{1\,{\rm km}}\Bigr)^{1.09} 
		\Bigl(\frac{g}{1\,{\rm m}\,{\rm s}^{-2}}\Bigr)^{0.28} 
		\Bigl(\frac{\rho_{\rm i}}{\rho_{\rm t}}\Bigr)^{-0.43} \nonumber \\
		&\times&\Bigl(\frac{v}{1\,{\rm km}\,{\rm s}^{-1}}\Bigr)^{-0.56} 
		\Bigl(\frac{D_{\rm SC}}{1\,{\rm km}}\Bigr)^{0.19}
		\, {\rm km}.
		\label{eq:scaling}
	\end{eqnarray} 
	We assume the bulk density of the impactor and icy satellites as described in \citet{Wong2020}: where $\rho_{\rm i} \sim 400\rm\ kg\ m^{-3}$ for the impactors, and generally $\rho_{\rm t}=1000\rm\ kg\ m^{-3}$ for the satellites' icy crust. For simplicity, we substituted $v$ with the mean impactor velocity obtained from the fly-by simulations from \citet{Wong2019}.The impact velocity takes into account the encounter velocity of the planetesimal at Saturn's Hill radius and the gravitational focusing of Saturn. The incidence angle of the impact is assumed to always be 45$^{\circ}$, which is accounted for in the coefficient of the expression above. For most icy satellites the simple-to-complex crater diameter depends on the surface gravity as $D_{\rm SC} = 4\ (1~{\rm m~s}^{-2}/g)$ \citep{Schenk2003}, while for Mimas, Enceladus and Miranda it is 15~km \citep{Schenk1991}.
	
	\subsection{Constructing crater isochrons}
	\label{sub:isochrons}
	In our previous works, we computed surface ages of the most highly cratered surfaces based solely on the expected density of craters larger than 20~km in diameter \citep{Wong2019, Wong2020}. In order to enhance the reliability of our age estimates, we plan to incorporate crater densities from a wider range of diameters, spanning at least one order of magnitude. To achieve this, we constructed crater isochrons, which is a modelled chronology of cratering on the satellites projected as the crater production function at different ages. We then compared the isochrons with the crater size-frequency measurements in cratered plains of the icy satellites, which have been published in previous studies \citep[e.g.,][]{Kirchoff2009, Kirchoff2010}.\\
	
	Here we will explain how we construct the isochrons for Dione's cratered plain as an example.
	
	\subsubsection*{Step 1: Identify the crater size excavated by a 1~km impactor.}
	Using our adopted crater scaling law (Eq.~\ref{eq:scaling}), we estimated that an impactor with $D_{\rm i} >$1~km can blast a crater with $D_{\rm cr}~>16.3$ km on the surface of Dione. We chose the cumulative number of craters created by a 1~km diameter projectile as a reference quantity because the size-frequency distribution of SDOs below 1~km is still poorly understood and could have a change in slope \citep{Singer2019}. 
	Furthermore, 
	the largest observed crater size on Enceladus is $\sim$30~km, on Mimas it is $\sim$50~km, and on Dione, Tethys, and Rhea they are around 90 -- 100~km \citep{Kirchoff2009}. 
	A 2~km planetesimal would be able to produce a crater with a diameter of over 52~km on Mimas and 40~km on Enceladus, both exceeding the expected size range of those that were recorded. Using crater sizes that exceed the available ranges of observed crater measurements would lead to inaccurate calibration in the later stage. 
	In Table~\ref{tab:crater}, for all five satellites, we list the diameters of craters excavated by a 1~km impactor ($D_{\rm cr,1~km}$), which is the reference diameters; as well as the required sizes of impactors that can excavate craters with diameters of 1~km, 2~km, 5~km, 10~km, and 20~km. Later, we scale the cumulative number of craters from the larger reference diameter to a smaller diameter to compare with the observed crater size ranges.\\
	
	\subsubsection*{Step 2: Calculate the expected crater densities} 
	We used Eq.~\ref{eq:n_sd_d} to estimate the initial number of SDOs with diameters larger than 1~km that have the potential to form a 16.3~km or larger diameter crater on Dione. With $\alpha=-2$, we find that just after Solar System formation $N_{\rm SD,>1}=6.4 \times 10^{11}$. 
	The probability of collision with Dione is $P_{\rm s}=7.58\times 10^{-8}$ and its surface area $A_{\rm s}=3.96$ million~km$^{2}$. Applying Eq.~\ref{eq:n_cr}, we find that only $1.24\times 10^5$ impactors are expected to collide with Dione over the 4.5 Gyr of evolution, resulting in an expected crater density of $n_{\rm cr,>16.3}=0.012$ crater per km$^{2}$ on a 4.5 Gyr old terrain.\\
	
	\subsubsection*{Step 3: Obtain the crater production function (CPF).} 
	The previous step only focused on calculating the crater density for $D_{\rm cr}>16.3$~km. We further utilised two sets of CPFs -- the profile of the size-frequency distribution of craters on a surface -- to determine the number of craters larger than any specific diameter. We adopted the CPFs from \citet{Singer2019} and \citet{Kirchoff2009,Kirchoff2010}. The CPF from \citet{Singer2019} is based on the craters on Charon and the derived impactor size-frequency distribution is the most up to date. \\
	
	The results of \citet{Singer2019} were previously introduced in Section~\ref{sub:nsd} as the impactors' size frequency distributions in this study. This impactor size-frequency was extrapolated from the cumulative crater size-frequency distribution obtained from the craters on the Pluto-Charon system and primarily associated with heliocentric projectiles. They concluded that for larger craters the slope of the crater size-frequency distribution is $\alpha_{\rm cr}$ = -2.0.  
	As the Saturnian system is believed to have shared the same source of projectiles, we apply $\alpha_{\rm cr}$ = -2 to all Saturnian satellites. \\ 
	
	\citet{Kirchoff2009,Kirchoff2010} studied the crater densities on various Saturnian satellites and reported the crater size-frequency slope $\alpha_{\rm cr}$ for the surveyed terrain and for certain range of crater diameters. In contrast to \citet{Singer2019}, each satellite 
	has a unique CPF with varying $\alpha_{\rm cr}$ for different ranges of crater diameter, and there are a different number of kinks in the distribution. As an example, apart from a scaling constant, the continuous, non-smooth CPF for Dione is as follows:
	\begin{equation}
		n_{\rm cr,~>D_{\rm cr}} = 
		\setlength{\arraycolsep}{0pt}
		\renewcommand{\arraystretch}{1.2}
		\left\{\begin{array}{l @{\quad} l r l}
			\ (1.0/D_{\rm cr})^{1.64},		    
			& \mbox{for } D_{\rm cr}<10~{\rm km},\\
			\ (1.94/D_{\rm cr})^{2.32},
			& \mbox{for } 10~{\rm km}<D_{\rm cr}<20~{\rm km},~~~~~~~~~\\
			\ (2.74/D_{\rm cr})^{2.70},
			& \mbox{for } D_{\rm cr}>20~{\rm km}.
		\end{array}\right.
	\end{equation}
	Assuming that the size-frequency distribution of the planetesimals in the scattered disc and the CPF remain constant over time, we used the previously obtained $n_{\rm cr,~>16.3}$ as an anchor point to model the entire size-frequency distribution as if the surface had recorded all the bombardment since 4.5 Ga.
	
	\subsubsection*{Step 4: Adjust the CPF for different ages.} 
	In this study, we updated the crater chronology function of the three giant planet satellite systems fitted with a stretched exponential profile characterised by $\tau$ and $\beta$. 
	The theoretical CPF obtained for 4.5 Ga is then vertically scaled downward to match the crater densities on younger surfaces, where $n_{{\rm cr},~t}=n_{{\rm cr},4.5} \times F(t)$, with $F(t)$ being the stretched exponential fit to the cumulative impacts onto Saturn. By plotting a set of isochrons as the CPFs expected for terrains with different surface ages, readers can easily identify the narrow age ranges where the isochron closely resemble the crater size-frequency measurements.\\
	
	We utilise a non-linear least squares fitting method to statistically determine the surface ages and uncertainties by identifying the best-fitted reliable ranges of $D_{\rm cr}$ among the craters measurements on each of Dione's terrains. 
	The reliable range represents the diameter ranges of the size-frequency measurement that can be accurately fitted by a power law without rolling-off. Table~\ref{tab:crater} provides the minimum reliable values of $D_{\rm cr}$, excluding the smaller end of the size distribution where the cumulative slope gradually rolls off due to limitations in image resolution.
	To compare our results with previous studies' estimated surface ages (e.g., \citet{Kirchoff2009, Zahnle2003, Wong2019, Wong2020}), we also provided the estimated surface ages derived from the observed crater densities of craters larger than 1~km, 2~km, 5~km, 10~km, and 20~km, depending on the availability of the observed craters on the satellite. \\
	
	In summary, the cratering onto the icy satellites depends not only on their individual physical properties, such as surface gravity and crustal densities, but also on the size-frequency distribution of the projectiles, and the declining impact frequency of the outer Solar System. 
	
	\section{Results}
	\label{sec:result}
	
	\begin{table*}[ht]
		\begin{center}
			\begin{tabular}{lccccccc} 
				\hline\\
				\multirow{3}{*}{Property} &
				\multicolumn{6}{c}{Estimate from crater scaling law}&
				\multirow{2}{*}{$D_{\rm cr,min}$}\\
				\cmidrule(lr){2-7}
				& \multicolumn{5}{c}{$D_{\rm i}$ required to excavate crater of $D_{\rm cr}$} & \multirow{2}{*}{$D_{\rm cr, 1~km}$}\\
				& 1 km & 2 km & 5 km & 10 km & 20 km & & \small{km} \\
				\hline\\
				Mimas & 0.03 & 0.06 & 0.15 & 0.33 & 0.70 & 27.9 &
				5\\
				Enceladus & 0.03 & 0.07 & 0.19 & 0.42 & 0.88 & 22.6 &
				2\\
				Tethys & 0.04 & 0.09 & 0.25 & 0.53 & 1.12 & 18.0 &
				5\\
				Dione & 0.05 & 0.10 & 0.28 & 0.59 & 1.25 & 16.3 &
				6\\
				Rhea & 0.05 & 0.11 & 0.31 & 0.65 & 1.39 & 14.8 &
				8
			\end{tabular}
			\caption{\label{tab:crater}Several parameters and results regarding the craters on the icy satellites. Starting from the left, we adopted the $\pi$-scaling law from \citet{Zahnle2003} to estimate the diamaters of the impactor ($D_{\rm i}$) required to excavate craters of 1~km, 2~km, 5~km, 10~km and 20~km in diameters. We list the crater diameters ($D_{\rm cr,min}$) from which we started fitting the crater size-frequency measurement from \citet{Kirchoff2009,Kirchoff2010} (KS9) into the isochron.}
		\end{center}
	\end{table*}
	
	\subsection{Interaction of the scattered disc objects with the giant planets}
	\label{sub:result_nsd}
\begin{figure*}
		\centering
		\resizebox{\hsize}{!}{\includegraphics{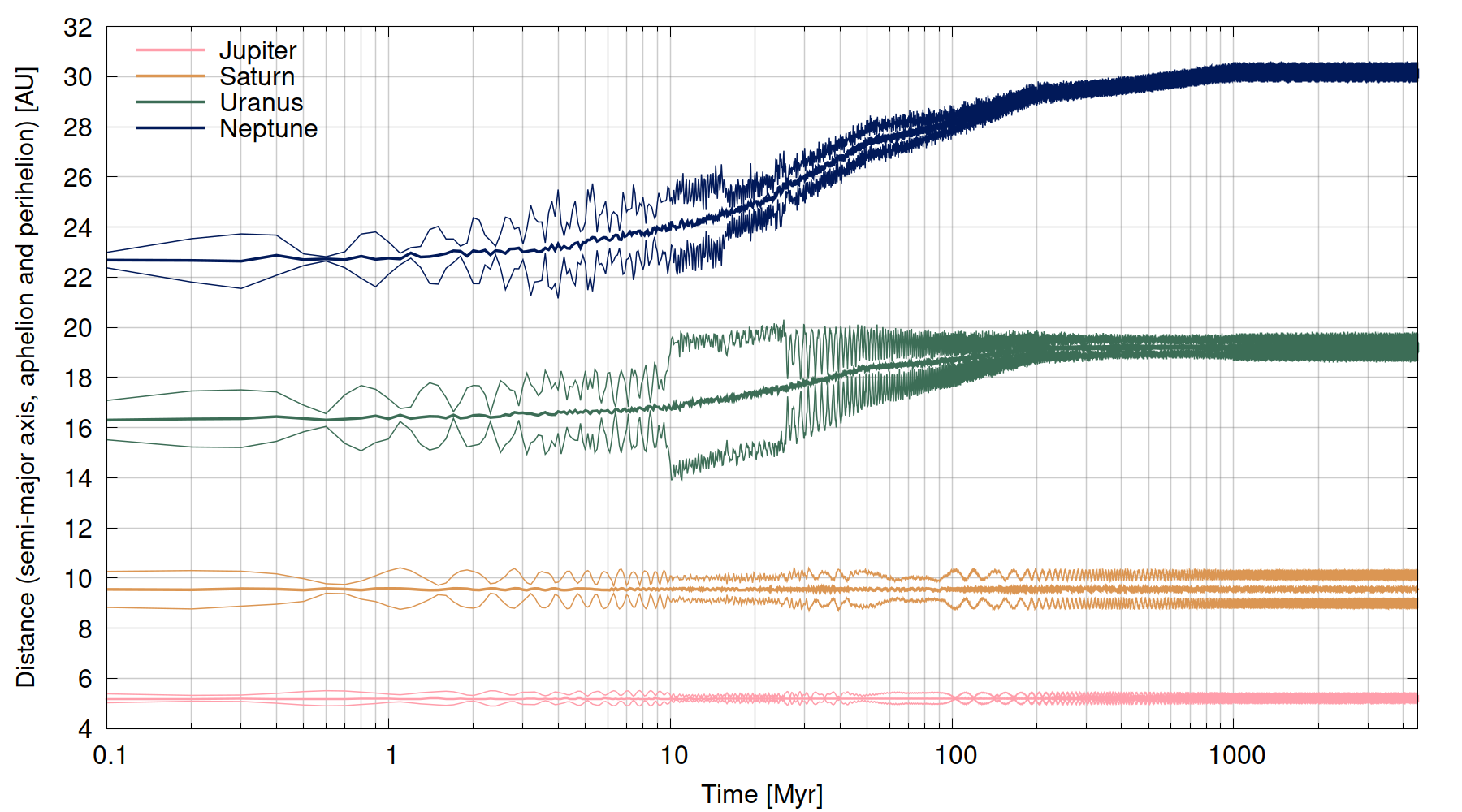}}
		\caption{\label{fig:evo}From bottom to top, the evolution of the semi-major axis, perihelion and aphelion of Jupiter, Saturn, Uranus and Neptune in the migration simulations of the giant planet migration. The presented evolution extends much beyond the giant planet migration simulation demonstrated in Fig.~2 of \citet{Wong2020}, which ceases at 1 Gyr.}
	\end{figure*}

	We illustrate the orbital evolution of the giant planets in Fig. \ref{fig:evo}. Following \citet{Levison2008,Wong2020}, Jupiter and Saturn were fixed while Uranus and Neptune began to migrate immediately, ultimately settling into their current positions at 19.2 AU and 30.1 AU, respectively. Both ice giants experienced kinks at $\sim$50 Myr and $\sim$200 Myr, where we 
	guided their migration to match the roughly logarithmic increase in their semi-major axes observed in N-body simulations. In the giant planet migration simulations, $\sim$99\% of planetesimals were removed within the first 500 Myr. The migration of the ice giants played a significant role in ejecting planetesimals from the Solar System, with $\sim$46\% of planetesimals removed by 10 Myr and $\sim$89\% by 100 Myr as they pass planetesimals onto Saturn and Jupiter, who then efficiently eliminate them from the Solar System \citep{LevisonDuncan1997}. Once the migration of the ice giants ceased at approximately 1 Gyr (Fig. \ref{fig:evo}), only 108 $\pm$ 49 (1-$\sigma$) scattered disc objects (SDOs) survived per simulation, indicating a survival probability of 0.67\% at that time. This is an average amongst both the compact and extended disc simulations. For the compact disc the survival is 0.39\% while it is 0.96\% for the extended disc simulations; after 1 Gyr, the rate of removal decreased rapidly. 
	For the resultant quantities report in this paper, we adopted the average, which is the average value between the mean calculated from 128 compact disc simulations and the mean calculated from 192 extended disc simulations, rather than a simple average across all 320 simulations.\\
	
	At the end of the 4.5 Gyr evolution, a mere 50 $\pm$ 35 SDOs remained per simulation, 
	indicating that approximately 99.69\% of planetesimals that originated from the scattered discs had been removed. In Section~\ref{sub:nsd}, we have applied this percentage of remaining planetesimals to estimate the initial number of SDOs in the Solar System.\\
	
	A stretched exponential function provides a good fit to the cumulative decline rate of SDOs after a few million years. Notably, the e-folding time for ejection, $\tau$ = 12.5 Myr, is longer compared with that for planetary collisions. This indicates that the removal SDOs via ejection is generally slower compared with removal due to impacts with the giant planets.  
	Our simulation shows a roughly simple exponential decline in the fraction of particles remaining over the past 3 billion years, suggesting that close to the present there is an approximately constant flux of objects leaking from the scattered disc. Previous studies by \citet{Holman1993} and \citet{Malyshkin1999} suggested a decline of the remaining fraction following a power law with a $t^{-1}$ for the last 3.5 billion years. Overall the rates of decline from various studies are consistent with each other, though the average rate is generally higher when a longer time space is considered, suggesting that the flux is not quite constant, but nearly so.\\
	
	\begin{table*}[ht]
		\centering
		\begin{tabular}{lccccc} 
			\hline\\
			Planet & $P_{\rm P}$ & $P_{\rm P,GPU}$ & $P_{\rm Z03}$ & $\beta$ & $\tau$\\ 
			\footnotesize{Unit} & \footnotesize{\%} & \footnotesize{\%} & \footnotesize{\%} 
			&& \footnotesize{Myr} \\
			\cmidrule(lr){1-1}\cmidrule(lr){2-4}\cmidrule(lr){5-6}
			\\
			Jupiter & 1.22 & 0.78 & 0.88 & 0.297 & 1.159\\
			Saturn  & 0.45 & 0.28 & 0.37 & 0.317 & 1.596\\
			Uranus  & 0.19 & 0.56 & 0.22 & 0.302 & 1.618\\
			Neptune & 0.46 & 3.85 & 0.24 & 0.296 & 1.224\\
			Ejected & 97.0 & /    & /    & 0.398 & 12.47\\
		\end{tabular}
		\caption{\label{tab:sim}Impact probabilities in percentage onto each planet by the scattered disc objects and the ejection probability of scattered disc objects from the Solar System. Data in the 2$^{\rm nd}$ column is calculated from the impact counts in the scattered disc simulations, corresponding to the average percentage of the compact and extended disc models. In 3$^{\rm rd}$ and 4$^{\rm th}$ columns we list the impact probabilities reported in the previous studies: from GPU simulations of \citet{Wong2019}, which used to compute the impact probabilities with the satellites, and from \citet{Zahnle2003}. In the last two columns, we listed the stretching parameters ($\beta$) and e-folding time ($\tau$) acquired by fitting the impact frequency curve of each planet and the decaying curve of the ejected planetesimals with a Weibull distribution. The typical uncertainty for the planet-collision's $\beta$ is $\sim$1\% and $\tau$ is $\sim$7.5\%.}
	\end{table*}
	
	We computed the impact probability of the SDOs with the giant planets ($P_{\rm P}$) by counting the impact in the simulations. The difference between 
	the compact and extended disc contributes to the slight difference in the $P_{\rm P}$, as listed in Table~\ref{tab:sim}.  \\
	
	
	\subsection{Calibrating our simulations: ecliptic comets}
	\label{sub:nec}
	\begin{table*}[ht]
		\centering
		\begin{tabular}{lccccccc} 
			\hline\\
			Model & $\vert r_{\rm SD}\vert$  & $f_{\rm EC}$  & $\tau_{\rm EC}$ & Depletion &
			$N_{\rm SD,>1}$ & $N_{\rm EC,>1}$ & $F_{\rm EC,>1}$ \\
			\footnotesize{Unit} &\footnotesize{Gyr$^{-1}$} & \footnotesize{$\%$} &
			\footnotesize{Myr} & \footnotesize{$\%$} & \footnotesize{$\times10^{11}$} & \footnotesize{$\times10^{7}$} & \footnotesize{yr$^{-1}$} \\
			\cmidrule(lr){1-1}\cmidrule(lr){2-5}\cmidrule(lr){6-8}\\
			Compact  & 0.128 & 81.6 & 285 & 99.90 & 19.3 & 5.9 & 0.21\\
			Extended & 0.055 & 55.2 & 250 & 99.48 & 3.8 & 1.5 & 0.06 \\
			Average & 0.092 & 68.4 & 267 & 99.69  & 6.4 & 3.7 & 0.13 \\
		\end{tabular}
		\caption{\label{tab:disc}Quantities derived from the results of the scattered disc simulations with compact and extended discs.
			Columns 2 to 4 show the declining rate of scattered disc objects ($r_{\rm SD}$), the fraction of ecliptic comets ($f_{\rm EC}$), the dynamical lifetimes of the ecliptic comet ($\tau_{\rm EC}$), and the percentage of total depletion of scattered disc object. 
			The last three columns provide estimates for the {\it initial} number of scattered disc objects ($N_{\rm SD}$) in the disc, the number of ecliptic comets ($N_{\rm EC}$), the current injection rate of ecliptic comets ($F_{\rm EC}$). We adopted and discuss with the average values, in the last row, in the main text.} 
	\end{table*}
	
	Our objective is to evaluate the injection rate of SDOs into ecliptic comets, and compare it with previous studies to verify the accuracy of our simulations and to compute the current cratering rates onto the icy satellites in future works. \\
	
	Ecliptic comets are SDOs that have left their source region, ventured into the realm of the giant planets. Their motion is dominated by encounters with these planets. A subset of these are the short-period Jupiter-family comets, which are concentrated along the ecliptic plane and with Tisserand's parameter with respect to Jupiter $T>2$ \citep{LevisonDuncan1997}. We qualified the particles in our simulations that met all the following criteria as an ecliptic comet (EC). Our definition focuses on the EC's injection rate, and not so much their actual fate. Our criteria are:
	\begin{enumerate}
		\item the planetesimal survived the first 1 Gyr into the simulation;
		\item its perihelion reached $q<30$ AU in the last 3.5 Gyr of the simulation, and  
		\item it is lost eventually due to collision with the planets, ejection from the Solar System, or survived till the end of simulation at 4.5 Gyr.
	\end{enumerate}
	As such, only those SDOs that leaked in from beyond Neptune and were subsequently removed are counted as ecliptic comets. \\
	
	The number of ecliptic comets is computed as \citep{Duncan1995}
	\begin{equation}
		N_{\rm EC} = N_{\rm SD}~\vert r_{\rm SD}\vert~f_{\rm EC}~\tau_{\rm EC},
		\label{eq:nec}
	\end{equation}
	where $N_{\rm SD}$ is the current number of objects in the Kuiper Belt and scattered disc (not the initial number at 4.5 billion years ago), $r_{\rm SD}$ is the rate at which the scattered disc population declines, $f_{\rm EC}$ is the fraction of SDOs that become ecliptic comets 
	and $\tau_{\rm EC}$ is the mean lifetime of the ecliptic comets. 
	\subsubsection{The rate of decline of scattered disc objects}
	\cite{Duncan1995} defined the rate of decline ($r_{\rm SD}$) as the number of object that left the system ($\Delta N$) divided by the final number of the particles at the end ($N_{\rm fin}$) and the time interval ($\Delta t$). In other words
	\begin{equation}
		r_{\rm SD} = \frac{\Delta N}{N_{\rm fin} \Delta t}. 
	\end{equation}
	It is worth mentioning that $r_{\rm SD}$ depends on the time range over which it is calculated because the decline is not constant. Here, we calculated the $\Delta N$ and $N_{\rm fin}$ for each simulation over the last 0.5 Gyr in the simulations to determine the current rate. 
	Our the nominal rate of decline is $r_{\rm SD}=-0.092$ Gyr$^{-1}$, which is the average value and lower than the value reported by \citet{Levison2006}: $-0.27$~Gyr$^{-1}$, using the simulations of \citet{DuncanLevison1997}, but consistent with more recent estimates from newer simulations \citep{Brasser2013,DiSisto2007,Fernandez2004,VolkMalhotra2008}. 
	For more details and comparison regarding the time-dependence of $r_{\rm SD}$ and the survival of the SDO, please refer to the discussion section. 
	
	\subsubsection{Fraction of ecliptic comets}
	\label{sub2:fec}
	The next quantity to be determined is the fraction of SDOs that eventually became ecliptic comets after the giant planets' migration phrase. This fraction is the ratio of the number of qualified ecliptic comets in the post-migration simulation to the number of surviving SDOs. We do not compare with the total number of the SDOs we put in the disc, because it is the remaining fraction that is used to predict the current rate of trans-Neptunian objects gradually leaking inward to become ecliptic comets, where the dynamical conditions more closely resemble those after 1 Gyr into the simulation when the migration has settled down. By defining ecliptic comets with the three criteria mentioned, we computed a average of 68.4\% of SDOs became ecliptic comets. 
	
	\subsubsection{The dynamical lifetimes of the ecliptic comets}
	\label{sub2:tec}
	There is no clear consensus on the dynamical lifetimes of the ecliptic comets ($\tau_{\rm EC}$) obtained by previous studies. For example, two simulations using the same initial condition gave 47 Myr and 200 Myr as the median and mean lifetime when averaging over 1~Gyr and 4~Gyr, respectively \citep{LevisonDuncan1997,Levison2000a}. 
	\citet{DiSisto2007} found a mean Centaur lifetime of 72 Myr, while \citet{Tiscareno2003} calculated 9 Myr from a smaller simulation data set. \\
	
	Specifically \cite{Levison2000a} state that 
	\begin{equation}
		\tau_{\rm EC} \approx \int_{0}^t \frac{N_{(s)}}{N_{\rm init}}ds \approx \frac{1}{N_{\rm init}} \sum_{i=1}{N_{\rm init}} t_i,
	\end{equation}
	where $N_{\rm init}$ is the initial number of test particles and $N_{(s)}$ is the number of test particles remaining at the time $s<t$. The last step is approximately valid because of the slow decline of planetesimals. The simulation data is written at regular time intervals of 0.1~Myr. 
	We counted how many times a particular particle had $q<30$~AU and computed $\tau_{\rm EC}$ for each particle by multiplying the total count by the data output interval. 
	This is the total time that a particular SDO spends time as an ecliptic comet, even if it periodically switches between these two reservoirs, i.e. if it switches between having $q<30$~AU and $q>30$~au. We then averaged these individual times to compute $\tau_{\rm EC}$. With the above approach, we found an average of $\tau_{\rm EC} = 267 \pm 101$ Myr (1$\sigma$), which is comparable to the value of 200~Myr found by \citet{Levison2000a}, but longer than found by \citet{DiSisto2007} and \citet{Tiscareno2003}.
	
	\subsubsection{Current number and injection rate of ecliptic comet}
	\label{sub:result_nec}
	We now have the ingredients to compute $N_{\rm EC}$ with Eq.~\ref{eq:nec}. Taking the current number of SDOs with diameter $D_{\rm i}>1$~km as $N_{\rm SD,>1} = 2\times10^{9}$, along with the average result for the rate of decline $r_{\rm SD} = -0.092$ Gyr$^{-1}$, adopting $f_{\rm EC}$ = 68.4\%, and $\tau_{\rm EC}$ = 0.267 Gyr, we estimated the total number of ecliptic comets, i.e., planetesimals with $q<30$ AU, to be $N_{\rm EC,>1} = 3.7\times10^{7}$. This value is about an order of magnitude higher than that expected from the analysis by \citet{Nesvorny2019}, but this difference is probably the result of them only considering objects with perihelion $q>7.5$~AU and semi-major axis $a<a_{\rm N}$ whereas our sole criterion is that $q<30$~au.\\
	
	The current injection rate of new comets with $D_{\rm i}> 1$~km from beyond Neptune can be calculated as $F_{\rm EC} = N_{\rm SD,>1}\vert r_{\rm SD}\vert f_{\rm EC} = 0.13$~yr$^{-1}$. All three parameters have been derived to reflect the current non-migrating outer Solar System conditions from the results of the simulations. Therefore, this rate describes the current values. The output from the clone simulations yield virtually identical results for $f_{\rm EC}$ and $\tau_{\rm EC}$.\\

	Multiplying $F_{\rm EC}$ by the impact probability of Jupiter (1.22\%) gives a rough estimation of 1.57 $\times$ 10$^{-3}$ impacts per year onto Jupiter for objects with $D_{\rm i}>1$~km. Our estimated Jovian impact rate falls towards the lower end, but remains consistent with the reported values from previous studies. \citet{Zahnle2003} reported an impact rate of approximately 4 $\times$ 10$^{-3}$ per year based on six encounters (smallest $\sim$ 1 km in diameter) within four Jovian radii over $\sim$350 years. \cite{Lamy2004} observed nine comets crossing Callisto's orbit over approximately 50 years, suggesting an impact rate of about 1.75 $\times$ 10$^{-3}$ per year for $D_{\rm i}>1$~km. Other estimations, using record collisions and close encounters, measurement of excess carbon monoxide content, crater counting on Ganymede, and inference from Near Earth Objects, range from 0.4 to 44 $\times$ 10$^{-3}$ impacts per year. The range of these impact rates for $D_{\rm i}>1$~km spans two orders of magnitude, and the uncertainties in these estimates also vary.

	\subsection{Calibrating the simulations: impact chronologies of the giant planets}
	Assuming that all projectiles originate from heliocentric SDOs, we constructed impact chronologies for impactors larger than 1~km in diameter.\\
	\begin{figure*}[ht]
		\centering
		\resizebox{\hsize}{!}{\includegraphics{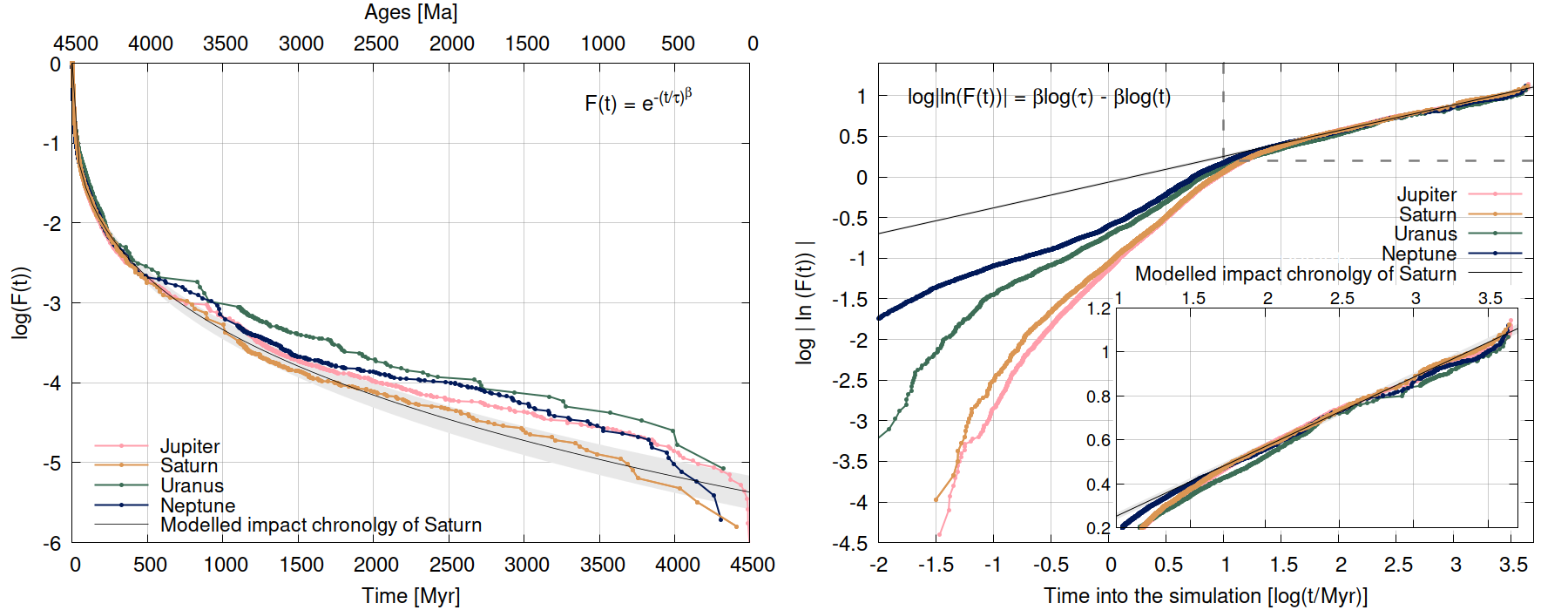}}
		\caption{\label{fig:chron}Impact chronologies onto the giant planets according to the combined results of the migration and clone simulation. {\it Left}: Normalized complementary cumulative distribution of impacts in log-linear scale were fitted by the stretched exponential function (Eq.~\ref{eq:weibull}) in the legend.
			{\it Right}: The Weibull plot, obtained by taking the log and natural logarithm of the function, expresses in the legend as the alternative representation of the stretched exponential function. 
			The best fit for Saturn's impact is shown as a black line with fitting parameters of stretching parameter $\beta$ = 0.317 $\pm$ 0.003 and initial e-folding time $\tau$ = 1.596 $\pm$ 0.091 Myr. 
			The gray band represents the range of uncertainty in fitting the curve of Saturn.
			The inset figures enlarged area bounded by the gray dash lines after 10~Myr of the evolution, when the impact chronologies of all giants gradually become identical and linear in the Weibull plot. 
			For the Y-axis, $F(t)$ represents the cumulative number of impacts onto a giant planet that occurred before a certain time, normalized to the total number of impacts of the corresponding planet recorded over the entire 4.5 Gyr of simulation. Specifically, $F(t=0)$ is set to 1 as the starting point. Within an individual impact decay curve of a planet, $F(t)$ can be interpreted as the relative cumulative number of impacts occurring before a specific time. }
	\end{figure*}
	
	Fig.~\ref{fig:chron} displays the results of our fitting and modelling procedures for the impact chronologies of the four giant planets. Each dot represents an impact on a planet at a specific time. The declining impact chronological curves closely resemble Figure 6 from \citet{Wong2020}, but the length of the chronological curves for all four planets has been significantly extended, e.g., the new chronological curve for Saturn goes 3 Gyr further and the rate is supported by much more reliable simulation outputs.\\
	
	The normalized complementary cumulative distributions of bombardment onto the surface of the four giant planets are displayed in Fig. \ref{fig:chron}. The left panel presents a log-linear scale. 
	We surveyed the timing of impacts on each planet and plotted them as cumulative number of impacts that occurred after specific time (in Myr) into the simulation, and normalised them to 1. 
	This chronology was modelled to cover the entire duration of evolution in the simulation from 4.5 Ga, which referred to the instability bought by the giant planet migration that began before 4.48 Ga \citep{Mojzsis2019}. Even though the migration might occurred earlier, it would only add a maximum 80 Myr to the surface ages of the regular satellites, which is smaller or comparable to the uncertainties from other sources.\\
	
	The same data is displayed as a Weibull plot in the right panel and its inset. A Weibull plot depicts $\log \vert \ln{F(>t)} \vert$ against $\log(t/\rm Myr)$. Notably, after $\sim$30 Myr (or $\log(t/\rm Myr)\sim 1.5$) the profiles for all giant planets transition into a Weibull distribution with similar parameters, manifesting as straight lines in the right panel. These linear profiles persist until the last impact records: at 4.50 Gyr, 4.41 Gyr, 4.32 Gyr, and 4.31 Gyr, correspondingly for Jupiter through Neptune. Still, the lines become more jagged towards the end as the remaining particles are insufficient to sustain a smooth impact frequency curve, and edge effects become noticeable. The initial divergence among the four impact chronologies before 1 Gyr is attributed to each giant's unique migration and evolution path, while the latter convergence is because the impact flux after migration is mainly controlled by the injection of SDOs into the Centaur region by Neptune.\\
	
	We fitted the data of impacts onto the four giant planets with a Weibull distribution (Eq.~\ref{eq:weibull}) using the method as described in Section~\ref{sub:fit}, and present the best-fitted stretching parameter ($\beta$) and e-folding time ($\tau$) for each planet in the last two columns of Table~\ref{tab:sim}.
	All the impact chronologies demonstrated an initial rapid decay which slows down considerably after a few e-folding times. 
	The typical value for $\beta\sim$ 0.3. Jupiter and Neptune display a shorter $\tau\sim$1.2 Myr, attributed to Jupiter's dominant mass and Neptune's intrusion into the disc. Saturn and Uranus have slightly longer $\tau\sim$1.6 Myr, indicating a slower decline in impact frequency. With the supplement impacts from the clone simulations, the typical uncertainties for $\beta$ is $\sim$1\% and for $\tau$ is $\sim$7.5\%, considerably smaller compared with the uncertainties reported in \citet{Wong2020} where the uncertainties for $\beta$ and $\tau$ are 4.9\% and 26\%, respectively. 
	
	\subsection{Absolute surface ages and crater chronologies}
	\label{sub:age}
	In this section, we focus on the five regular icy satellites of Saturn: Mimas, Enceladus, Tethys, Dione and Rhea; hence we adopt the impact data, Weibull distribution fitting and planet's parameters of their host planet, Saturn.\\
	
	We report the absolute surface ages on the cratered plain on all five Saturnian satellites deduced using two approaches: 
	(1) the best-fitted age for the observed cumulative crater size-frequency measurement among the isochrones, and 
	(2) the ages calculated using the stretched exponential function, Equation 11 of \citet{Wong2020}. With the latter approach, the estimated ages vary depending on the minimum crater size used to calculate the ages. Therefore, for each terrain, we report up to five estimated ages for craters larger than 1~km, 2~km, 5~km, 10~km, and 20~km in diameter. These absolute surface ages can be compared with those listed in Table 3 of \citet{Kirchoff2009}. In some regions or satellites, we refrained from estimating surface ages using craters larger than 1~km and/or 2~km due to the lack of observations at those diameters. \\
	
	\begin{table*}[ht]
		\begin{center}
			\begin{tabular}{lcccccccc} 
				\hline\\
				\multirow{3}{*}{Satellites} &
				\multicolumn{2}{c}{$n_{\rm exp, D_{\rm i} >1\ km}$} &~&
				\multicolumn{2}{c}{Previously reported ages}&\multicolumn{2}{c}{This work's ages}\\
				\cmidrule(lr){2-3}\cmidrule(lr){5-6}\cmidrule(lr){7-8}
				&S19 &  MC &~& KS9 & W21 & S19's CPF & KS9's CPF\\
				&\footnotesize{10$^{\rm -3}$ km$^{-2}$} & \footnotesize{10$^{\rm -3}$ km$^{-2}$} &~& \small{Ga} & \small{Ga} & \small{Ga} & \small{Ga}\\
				\hline\\
				Mimas & 26 & 34 &~& 4.0 & 4.01 & 4.21 & 4.16\\
				Enceladus & 18 & 23 &~&  4.6 & 3.80 & 4.29 & 4.10\\
				Tethys & 12 & 16 &~& 4.5 & 4.12 & 4.37 & 4.37\\
				Dione & 11 & 14 &~& 4.6 & 4.18 & 4.39 & 4.38\\
				Rhea & 8 & 10 &~& 4.5 & 4.29 & 4.41 & 4.42
			\end{tabular}
			\caption{\label{tab:age} We calculated the expected crater densities for craters generated by impactors with diameter $D_{\rm i}>1$ km in diameter ($n_{\rm exp, D_{\rm i} > 1}$) as predicted using the size-frequency distribution from \citet{Singer2019} (S19). We calculate that N$_{{\rm SD},D_{\rm i}>1}=6.5\times 10^{11}$ with $\alpha=-2$, and then multiply this with the satellite impact probability and divide by surface area. We compare this with our Monte Carlo impact simulations. In the last four column, we list the absolute surface ages for the cratered plains reported by \citet{Kirchoff2009} (KS9), \citet{Wong2020} (W21) and two ages from this work as the best-fitted ages among the isochrones constructed with the CPF derived by \citet{Singer2019} and \citet{Kirchoff2009}.}
		\end{center}
	\end{table*}
	
	Regardless of the approach used, the surface ages of the satellites generally follow a pattern of increasing age with increasing orbital distance, except Mimas and Enceladus: these two inner and smaller satellites are reversed and their surfaces are around 200 million years younger than those of Tethys, Dione, and Rhea. In this work, we report the absolute surface ages as the best-fitted ages calculated by the first approach, which are listed in the last column of Table~\ref{tab:age} and written as legends in Fig.~\ref{fig:mimS19KS9} to Fig.~\ref{fig:rheS19KS9}. 
	In the figures, we also present the uncertainty in the estimated ages by altering the initial number of SDOs according to a compact or extended disc. The surface ages estimated for the cratered plains on Mimas and Enceladus are 4.16 Ga and 4.10 Ga, respectively, while those for Tethys, Dione, and Rhea are around 4.4 Ga.
	
	\subsubsection{Best-fitted ages and the isochrones}
	\label{sub:}
	In Fig.~5, we show how we fit the surface ages with isochrones for the cratered plains of each satellites. The panels are arranged in order of increasing orbital distance from Saturn, from left to right and top to bottom. The isochrones are plotting as polychromatic functions, while the cumulative size-frequency measurement are shown as black filled dots, which appear as a continuous line at intermediate crater size ranges due to the abundance of craters around those diameters. The colour variation, ranging from dark-to-light (or blue-to-red in coloured versions), displays the predicted crater size-frequency distribution for a surface with age ranging from 3.7 to 4.5 Ga. These figures allow us to intuitively determine the ages of the cratered plains by identifying the coloured lines that best fit the overall measurement (black dots). This approach helps us take full advantage of crater counts by using the entire range of crater dimensions, from several kilometers to over 100~km, as shown on the $x$-axis.\\
	
	In Table~\ref{tab:crater}, we provide the range of crater diameters used for the isochron fitting process. The minimum diameters ($D_{\rm cr,min}$) were chosen to avoid fitting from the sub-kilometer crater size or near the small-end of the cumulative size-frequency measurement where the curve roll over due to image resolution limitations in recognizing and measuring craters.
	The curve for Mimas cratered plain and Enceladus' mid-latitude cratered plain roll over at smaller crater sizes, because their images were at higher resolution at the time this data was compiled.
	Despite the similar crater-scaling conversion unit for Enceladus and Mimas, the overall observed crater size-range of Enceladus' cratered plain is only about one-third that of Mimas.\\
	
	Here, we would like to unpack the structure of the isochron in three layers.\\
	
	The {\it first} layer comprises the profile of the lines in isochrones, which represents the crater production function (CPF) of a cratered plain on the respective Saturnian satellites. 
	The profile of S19, as determined by \citet{Singer2019}, exhibits a constant cumulative slope $\alpha_{\rm cr} = -2$, which remains the same for all satellites. We caution the reader that now we are talking about the slope of the {\it crater} size frequency distribution, and {\it not} that of the impactors.
	
	As for KS9, we derived the crater size-frequency profile of each satellite from Table 2 of \citet{Kirchoff2009}. 
	Each satellite's cratered plain appears to have a unique production function with various cumulative slopes, ranging from $-1.4$ to $-3.0$, and the slope varies with the crater diameter. 
	Ideally, the CPFs should be consistent across all satellites within the same system, as the projectiles entering the Saturnian system and colliding with the satellites are expected to originate from the same source. 
	However, it is not evident what causes the different production functions among the satellites. The differences could be due to actual distinct cratering records and/or individual scaling of the projectile to craters for the specific satellites.\\ 
	
	Comparing the best-fitted surface ages from the two sets of CPFs (S19 and KS9), we find that the ages from S19 are generally the same or older than those from KS9 (Table~\ref{tab:age}).
	These differences in age arise from variations in the cumulative slope ($\alpha_{\rm cr}$) between the two CPF sets for each Saturnian satellite.
	Notably, Mimas and Enceladus show a steeper $\alpha_{\rm cr}\sim-3$ for larger craters and $\alpha_{\rm cr}\leq-1.55$ for smaller craters. Their deviation from the constant $\alpha_{\rm cr} = -2$ at the high and low ends of the CPF results in a large difference in best-fitted surface ages between the two sets. Overall, the obtained values of $\alpha_{\rm cr}$ in KS9 are steeper than $-$2, leading to higher predicted crater densities, especially for intermediate-size craters ($D_{\rm cr} \sim$ 10 km). 
	Rhea stands out as an exception in KS9's CPF, with crater diameters between 10 to 80 km following $\alpha_{\rm cr} =-2.02$, which closely matches S19's CPF. However, it flattens to -1.44 for $D_{\rm cr}<$10 km. As a result, the predicted crater density with KS9 for Rhea is lower than S19 at the smaller crater range, leading to slightly older best-fitted ages. \\ 
	
	By eyes, the fitting with a single power law provided less confident surface ages, as the cumulative crater measurement near the two ends deviated from the function and cut across multiple isochrones. This discrepancy is particularly pronounced for Mimas and Enceladus. Ancient and likely more pristine cratered plains found on Tethys, Dione, and Rhea exhibit an overall slope that is roughly parallel to the single power law function. Consequently, the best-fitted absolute ages for S19 and KS9 for the outer three satellites are similar.\\
	
	The {\it second} layer is the baseline of the isochron, which corresponds to the uppermost pale pink line at 4.5 Ga, and represents the expected crater size-frequency distribution on a pristine and most ancient surface of the icy satellite. The baseline is established by fixing the CPF to the expected cumulative density of craters with a diameter equal to or larger than that created by an impactor $D_{\rm i}\geq$1~km on the surface of the respective satellites. In Table~\ref{tab:crater}, we determine the anchoring size (denoted as $D_{\rm cr,1\,km}$) using the crater-scaling law, while the expected cumulative crater density (labelled to as n$_{\rm exp, D_{i}=1~km}$) highly dependents on the number of SDOs with $D_{\rm i}\geq$1~km in the primordial disc. Overestimating the initial number of objects would cause the entire isochron to shift upward, leading to a younger estimated surface age when fitting the crater measurement, and vice versa.\\
	
	To verify the accuracy of our analytical estimates for the crater densities, we conducted 64 Monte Carlo (MC) impact experiments for the Saturnian system with six satellites, including Titan, to determine the number of impacts onto each satellite with $D_{\rm i} > 1$~km. 
	Using the same particle size-density relationship and coding structure as \citet{Wong2019, Wong2020}, we updated the size-frequency distribution and impact probabilities to assess our analytical calculations' validity. The average global crater densities of craters evacuated by a 1~km projectile on the five satellites, calculated analytically using Eq. \ref{eq:n_cr} and derived from MC simulation are presented in Table~\ref{tab:age} and denoted as MC under $n_{\rm exp,D_{\rm i}>1\ km}$. The difference in the crater density calculated between the MC simulations and analytical calculation can be attributed to the variation in the initial $N_{\rm SD}$: 7.23 $\times 10^{11}$ for MC simulations versus 6.4 $\times 10^{11}$ for the analytical calculation. This slight discrepancy can also be interpreted as the different removal percentages of the planetesimals: 99.72\% for MC simulations and 99.69\% for the analytical calculation. 
	Further information on the setup and execution of the MC simulations can be found in Section 2.4 of \citet{Wong2020}.\\
	
	The {\it final} layer is the separation distance between the polychromatic lines within the isochrones. We assume that the CPF remains constant throughout the Solar System evolution. As the overall crater density on a surface increases with age and the function shifts upwards. 
	Conversely, when constructing the isochrones, by taking 4.5 Ga as the baseline, the other lines representing younger ages are scaled down according to the stretched exponential function.\\
	
	The transition among the age lines is more pronounced for older ages since the impact frequency decays faster in the beginning, i.e., the gap between 4.4 Ga and 4.5 Ga is wider than 4.3 Ga and 4.4 Ga, despite both being separated by 0.1 Ga. Since all icy satellites are hosted by Saturn, the relative amount of transition along the $y$-axis with respect to age is the same for all five satellites and is determined by the modelled crater chronology of the Saturnian satellites system. If the e-folding time ($\tau$) decreases and/or the stretching parameter ($\beta$) increases, the gaps within the isochrones will widen, which can be interpreted as the impacts having been more intense in the beginning and subsided faster, and/or there were more substantial changes in the impact rate over time.
	In this scenario, the absolute surface ages obtained from the fit will be older because the isochron, except for the 4.5 Ga baseline, will be pushed downwards relative to the observed crater measurements.\\
	
	To explore the effects of another predominate uncertainty in the estimated ages, we present two set of modified isochrones with the corresponding best-fitted surface ages in Appendix A. 
	In Fig.~A1, we employed the size-frequency distribution of craters on Triton from \citet{Schenk07} with $\alpha=-$2.25 to estimate the initial number of trans-Neptunian objects, while maintaining the CPFs as reported by KS9. The sole distinction between Fig.~A1 and Fig.~6 lies in the assumed size-frequency distribution of the SDOs. The increase in the cumulative slope leads to a higher total number of impactors with $D_{\rm i}>$1~km, causing an up-shift of the overall isochron and consequently yielding younger best-fit surface ages.
	Additionally, in Fig.~A2, we further adopted $\alpha_{\rm cr}=-2.25$ \citep{Schenk07} as both the size-frequency distribution of the SDOs and a consistent CPF for all Saturnian satellites. By comparing Fig.~A1, which uses KS9's CPFs, with Fig.~A2, where the same single power law function was applied to scale the craters' size, the resulting best-fitted ages reported in Fig.~A1 would be older if the cumulative slope of the respective satellite is shallower than $-2.25$, and vice-versa. 
	
	
	\subsubsection{Ages derived from a specific cumulative crater density}
	\label{sub:2nd}
	The second approach has been adopted in \citet{Wong2020}, which derived the surface ages by comparing the observed cumulative crater density of 20~km or larger craters and the expected cumulative crater density of the same dimension under the framework of the stretched exponential function. 
	In this paper, we expanded on the approach to estimate the surface ages with the cumulative crater densities of all observed crater diameters (see Fig.~\ref{fig:age}), and specifically listed the age estimations with craters larger than 1~km, 2~km, 5~km, 10~km, and 20~km in diameters (see Table~\ref{tab:ks9}). \\
	
	We estimated the expected cumulative crater density and absolute surface ages and reported these in columns 4 and 5 of Table~\ref{tab:ks9}. The cratered plains on Enceladus are the youngest (all within 4.08 to 4.11 Ga), follows by Mimas (mostly close to $\sim$4.15 Ga), then Tethys, Dione and Rhea being older than 4.3 or even 4.4 Ga. 
	The ages deduced from different cumulative crater density's crater diameters are in good agreement with each other, at most $\pm$ 30 Myr of difference, except the surface ages of Mimas deduced with 2~km, which are over 150 Myr younger than the ages deduced with larger diameters. By referring to Fig.~5, the much younger ages can be explained by the rolling-off at the small end of the cumulative crater size-frequency distribution; the rolling-off started at $\sim$4~km for Mimas. Depending on the satellites' cratered plains, the rolling-off part causes age variations of 100 to 200 Myr. At the large end --- $D_{\rm cr}\gtrsim10$~km for Enceladus, and $D_{\rm cr}\gtrsim50$~km for other satellites --- the fluctuation in the cumulative crater density values changes the estimated ages to a smaller extent. Overall, the individual surface ages estimated with the second approach are consistent with the best-fitted ages.\\
	
	Comparing the former best-fitting with the isochron approach and the latter individual crater diameter with stretched exponential function approach, the former is preferable, as it fits the entire cumulative crater density distribution and deduces an age, working around the ages difference caused by the rolling-off of the crater size-frequency distribution at the small end and small-statistic oscillations at the large end. 
	
	\begin{table*}[]
		\begin{center}
			\begin{tabular}{lrccccccc}
				\hline
				\multirow{2}{*}{\small{Source}} & \multicolumn{2}{c}{\small{Crater count}} &~& 
				\multicolumn{4}{c}{\small{Estimates from this work}} \\
				& \multicolumn{2}{c}{\small{from KS9}} &~& 
				\multicolumn{2}{c}{\small{S19's CPF}} &
				\multicolumn{2}{c}{\small{KS9's CPF}} \\
				\cmidrule(lr){2-3}\cmidrule(lr){5-6}\cmidrule(lr){7-8}
				\small{Property} & $D_{\rm cr}$ & $n_{\rm obs}$ &~&
				$n_{\rm exp}$ & Ages & $n_{\rm exp}$ & Ages \\
				\footnotesize{Unit} &\footnotesize{km} &
				\footnotesize{10$^{\rm -6}$ km$^{\rm -2}$} &~&
				\footnotesize{10$^{\rm -3}$ km$^{\rm -2}$} & \footnotesize{Ga} &
				\footnotesize{10$^{\rm -3}$ km$^{\rm -2}$} & \footnotesize{Ga} \\
				\hline\\
				\multirow{4}{*}{Mimas}     
				& 2  & 9,320  &~& 5,618   & 3.94  & 4,255 & 4.02 \\
				& 5  & 4,497  &~& 926	    & 4.19  & 1,030 & 4.17 \\
				& 10 & 1,382  &~& 231	    & 4.22  & 352    & 4.15 \\
				& 20 & 285    &~& 57	    & 4.19  & 81     & 4.12 \\
				\\
				\multirow{5}{*}{Enceladus} 
				& 1  & 53,717 &~& 10,115  & 4.20  & 16,882 & 4.10 \\
				& 2  & 20,745 &~& 2,529   & 4.27  & 6,371  & 4.11 \\
				& 5  & 4,500  &~& 405	    & 4.32  & 1,436  & 4.10 \\
				& 10 & 652    &~& 100	    & 4.24  & 227    & 4.08 \\
				& 20 & 78     &~& 24	    & 4.10  & 29     & 4.08 \\
				\\
				\multirow{4}{*}{Tethys}    
				& 2  & 13,850 &~& 780	    & 4.36  & 818   & 4.37    \\
				& 5  & 3,102  &~& 178	    & 4.37  & 168   & 4.37 \\
				& 10 & 845    &~& 44	    & 4.38  & 51    & 4.36 \\
				& 20 & 154    &~& 11	    & 4.34  & 11    & 4.35 \\
				\\
				\multirow{4}{*}{Dione}     
				& 2  & 4,039  &~& 779	    & 4.30  & 525    & 4.34 \\
				& 5  & 2,187  &~& 125	    & 4.37  & 117    & 4.38 \\
				& 10 & 683    &~& 31	    & 4.39  & 37     & 4.37 \\
				& 20 & 128    &~& 8	    & 4.36  & 8      & 4.37 \\
				\\
				\multirow{2}{*}{Rhea}      
				& 10 & 594    &~& 20  & 4.42  & 20    & 4.41 \\
				& 20 & 149    &~& 5	& 4.42  & 5     & 4.42 \\
			\end{tabular}
			\caption{\label{tab:ks9}
				In the "Estimates from this work" column, we provide the expected crater density ($n_{\rm exp}$) for craters larger than a given diameter ($D_{\rm cr}$) on the cratered plains of each icy satellite. 
				The results are provided for several crater diameters, along with the corresponding estimated surface ages in Ga. Two sets of results ($n_{\rm exp}$ and Ages) are presented: one based on a single power law crater production function with a cumulative slope of $-$2 from \citet{Singer2019}, and the another using varying crater production functions specific to each icy satellite from \citet{Kirchoff2009}. 
				To estimate the ages, we compare the $n_{\rm exp}$ to the observed crater density ($n_{\rm obs}$) within the framework of our modelled crater chronologies. The $n_{\rm obs}$ for various crater diameters were obtained from the list provided by \citet{Kirchoff2009,Kirchoff2010}}.
	\end{center}
\end{table*}

\begin{figure*}
	\begin{subfigure}{.45\textwidth}
		\caption{\label{fig:mimS19S19}Mimas cratered plain}
		\centering
		\includegraphics[width=\linewidth]{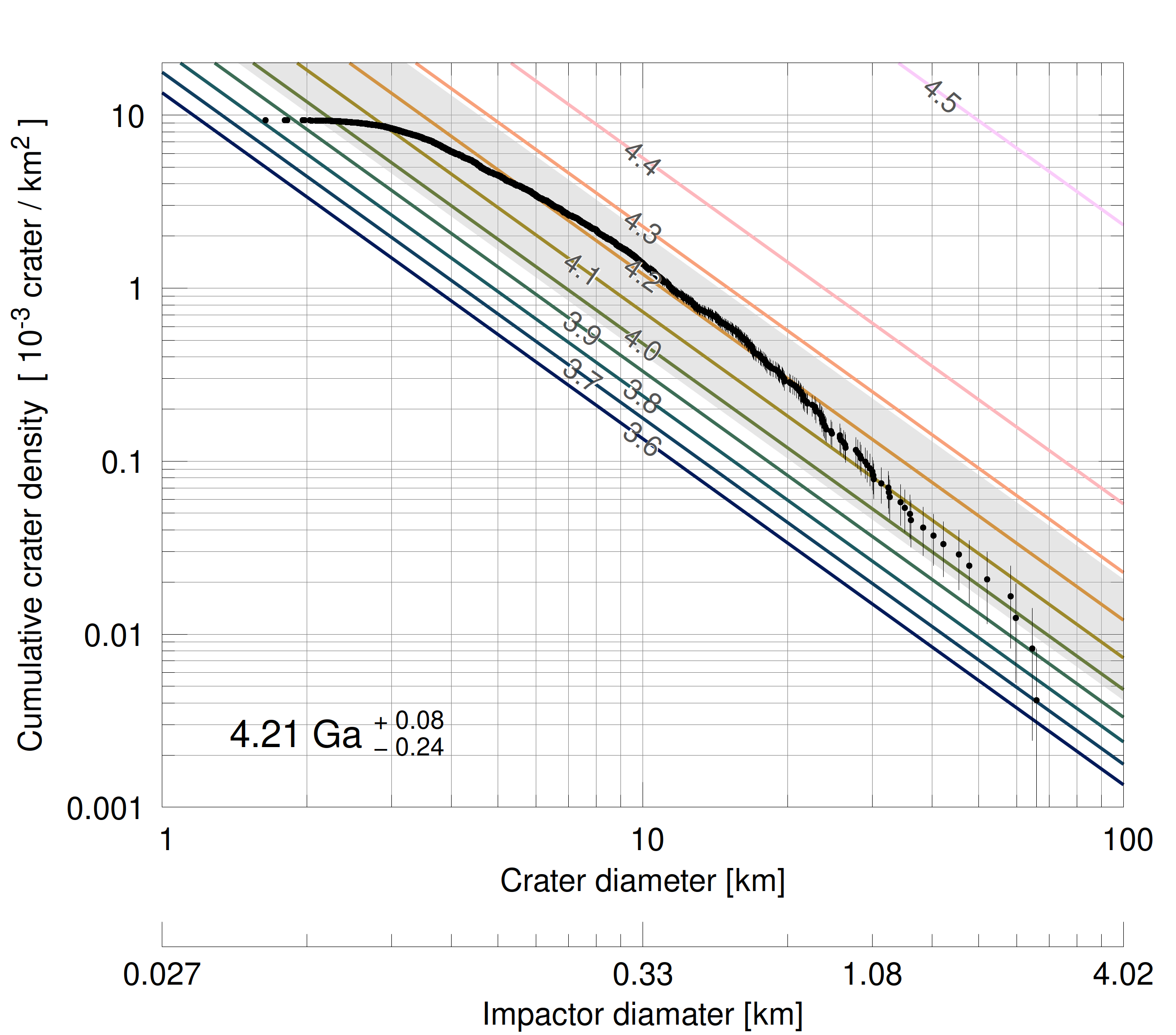}  
	\end{subfigure}
	\begin{subfigure}{.45\textwidth}
		\caption{\label{fig:encS19S19}Enceladus mid-latitude cratered plain}
		\centering
		\includegraphics[width=\linewidth]{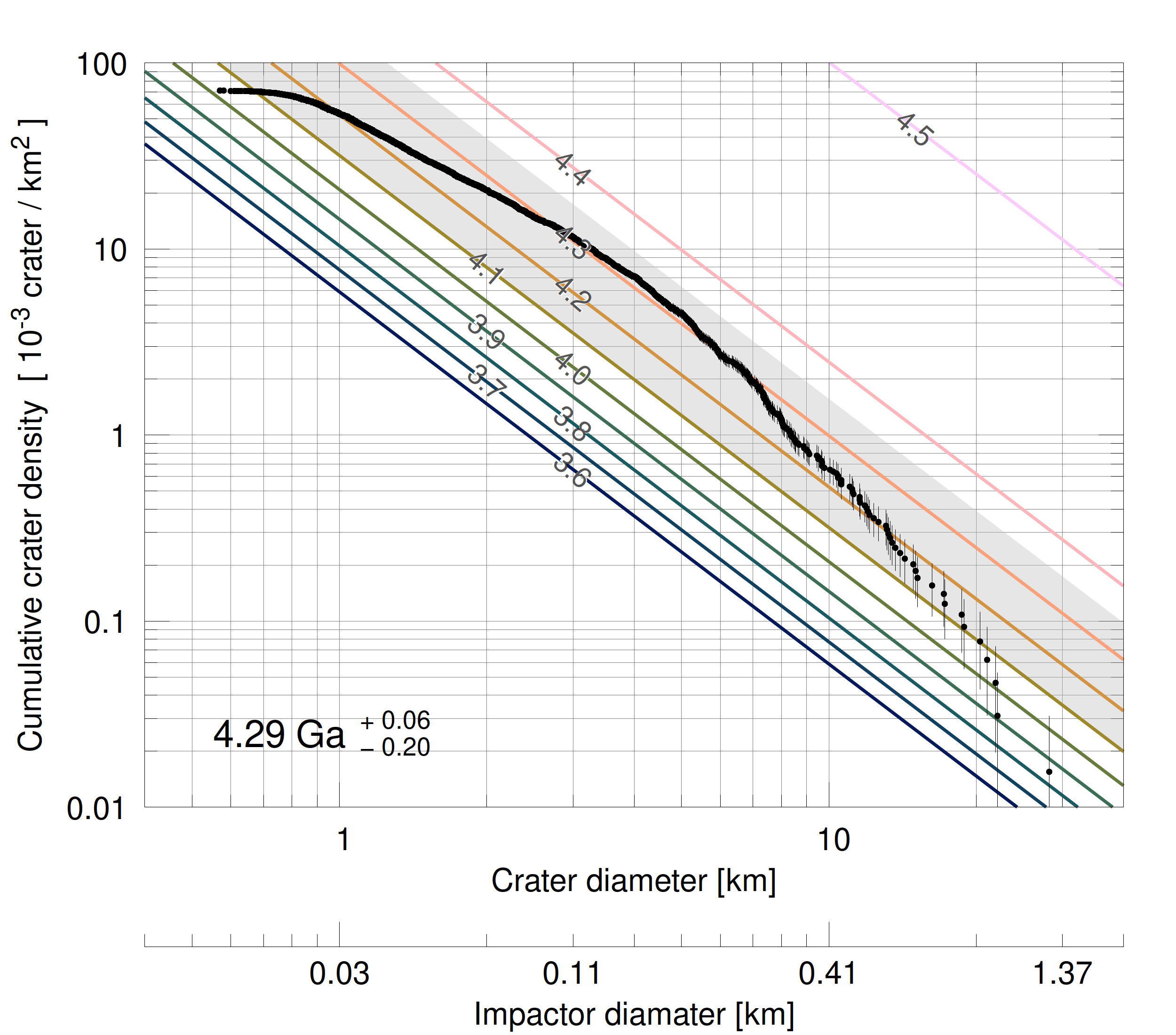}  
	\end{subfigure}
	\newline
	\begin{subfigure}{.45\textwidth}
		\caption{\label{fig:tetS19S19}Tethys cratered plain}
		\centering
		\includegraphics[width=\linewidth]{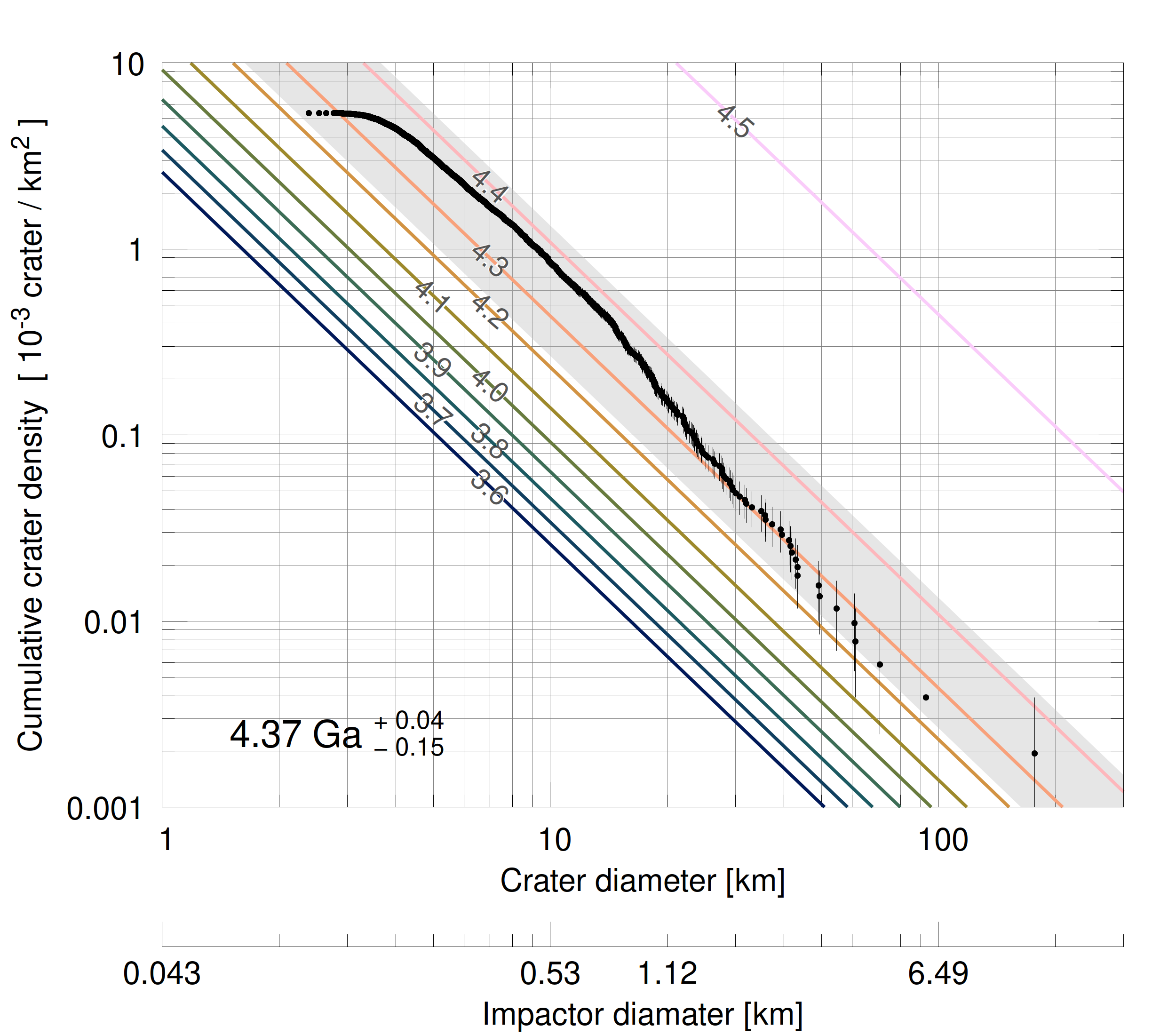}  
	\end{subfigure}
	\begin{subfigure}{.45\textwidth}
		\caption{\label{fig:dioS19S19}Dione cratered plain}
		\centering
		\includegraphics[width=\linewidth]{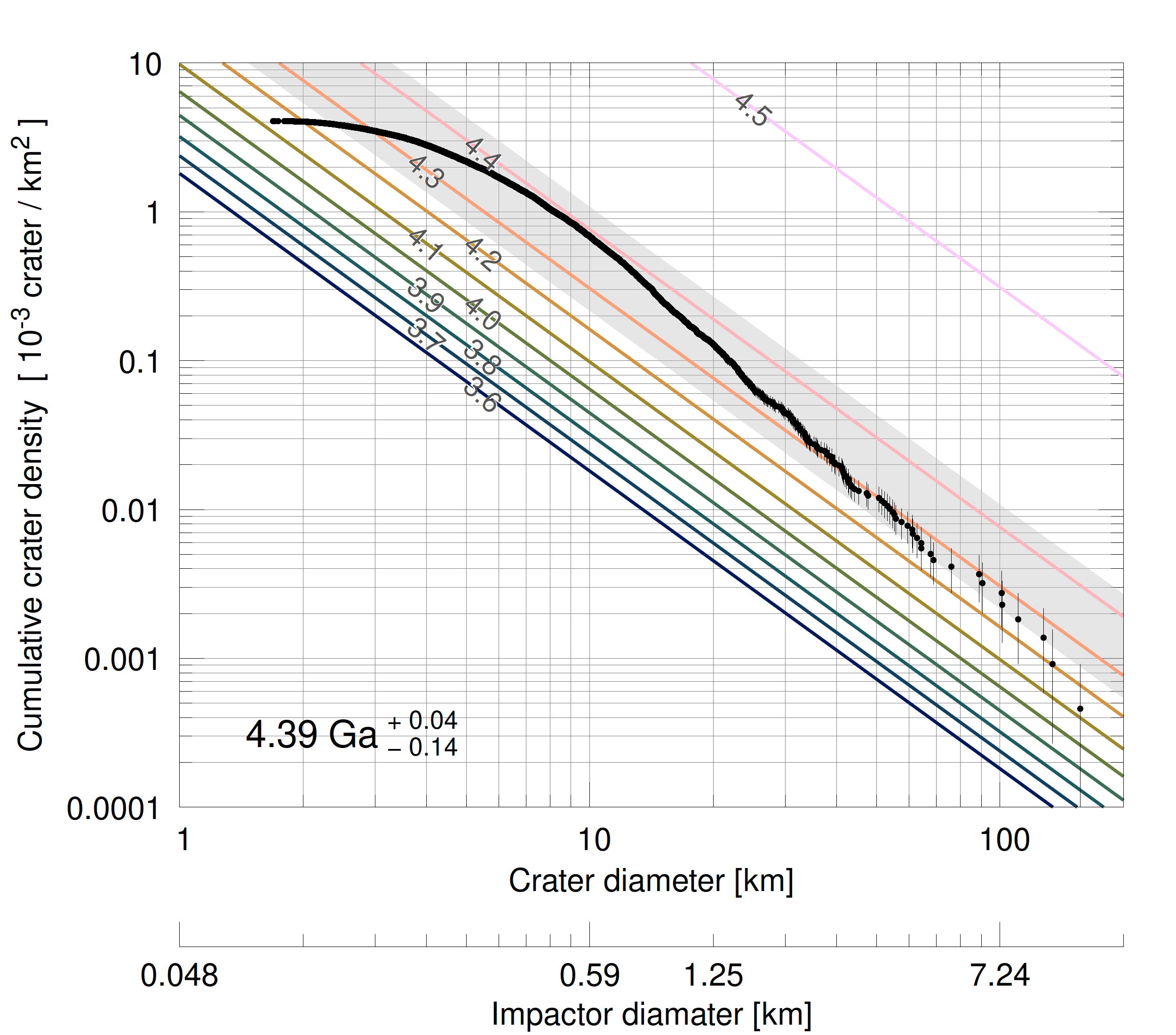} 
	\end{subfigure}
	\newline
	\begin{subfigure}{.45\textwidth}
		\caption{\label{fig:rheS19S19}Rhea cratered plain}
		\centering
		\includegraphics[width=\linewidth]{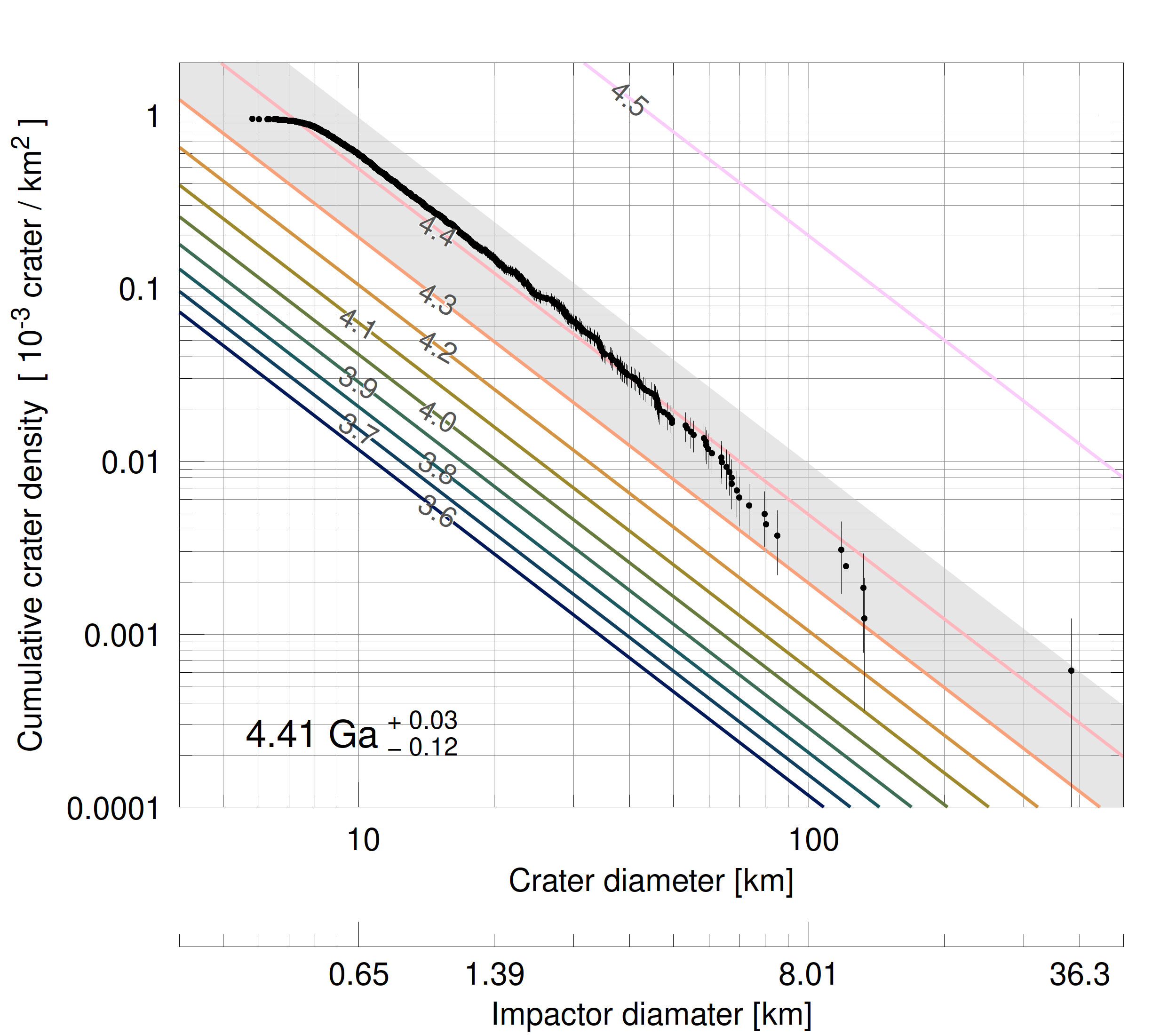}
	\end{subfigure}
	\begin{subfigure}{.45\textwidth}
		\captionsetup{width=5.5cm}
		\caption*{\scriptsize{Figure 5: 
				The best-fitted ages of the cratered plains on Mimas, Enceladus, Tethys, Dione, and Rhea range from 4.21 Ga to 4.41 Ga. 
				The ages were determined by fitting size-frequency measurements (black-filled circles) to polychromatic isochrones --- crater chronologies that project the \citet{Singer2019} crater production function of each plain to different ages, ranging from 3.5 Ga to 4.5 Ga. 
				A single power law function with a cumulative slope of -2 from \citet{Singer2019} was used for the fitting.
				The error bars represent poisson errors, and the grey area indicates the uncertainty in the ages due to variations between the compact and extended scattered disc in the giant planet migration simulations. The scaling between crater diameter and impact diameter is based on the $\pi$-scaling law proposed by \citet{Zahnle2003}. 
		}}
		\centering
		\includegraphics[width=\linewidth]{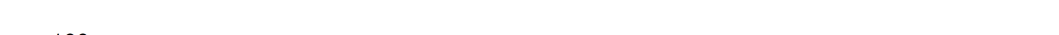}  
	\end{subfigure}
\end{figure*}

\begin{figure*}
	\begin{subfigure}{.45\textwidth}
		\caption{\label{fig:mimS19KS9}Mimas cratered plain}
		\centering
		\includegraphics[width=\linewidth]{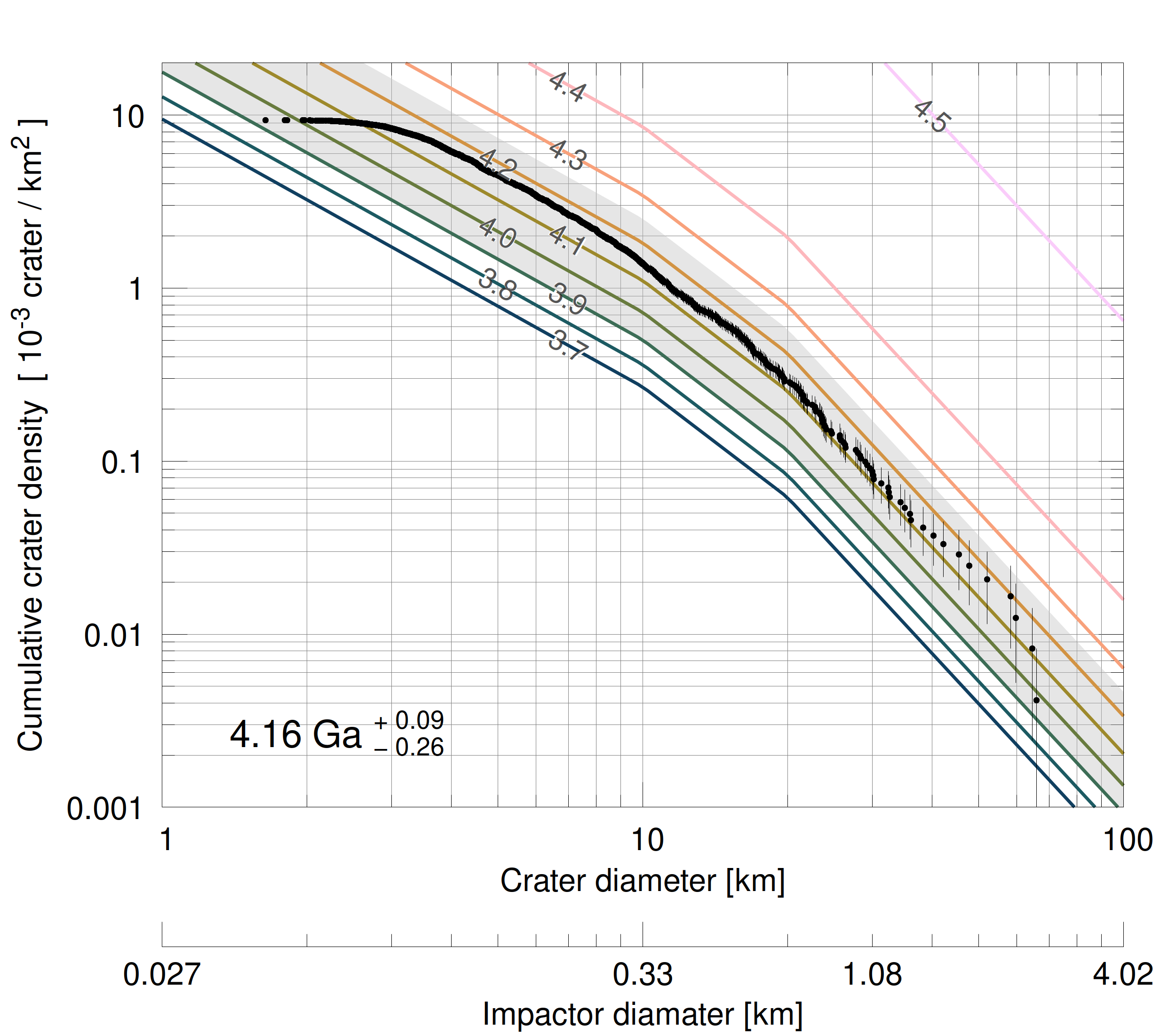}  
	\end{subfigure}
	\begin{subfigure}{.45\textwidth}
		\caption{\label{fig:encS19KS9}Enceladus mid-latitude cratered plain}
		\centering
		\includegraphics[width=\linewidth]{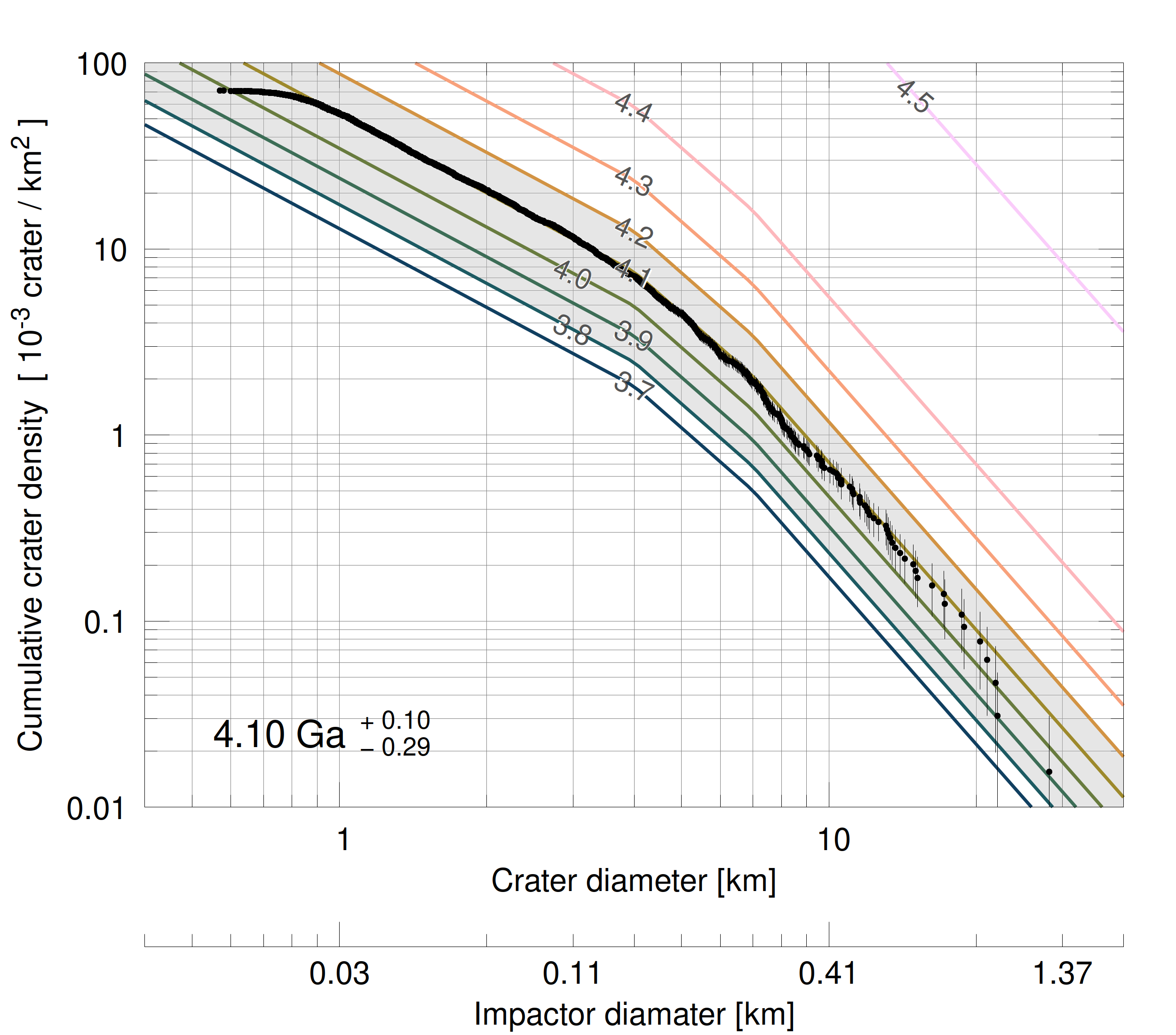}  
	\end{subfigure}
	\newline
	\begin{subfigure}{.45\textwidth}
		\caption{\label{fig:tetS19KS9}Tethys cratered plain}
		\centering
		\includegraphics[width=\linewidth]{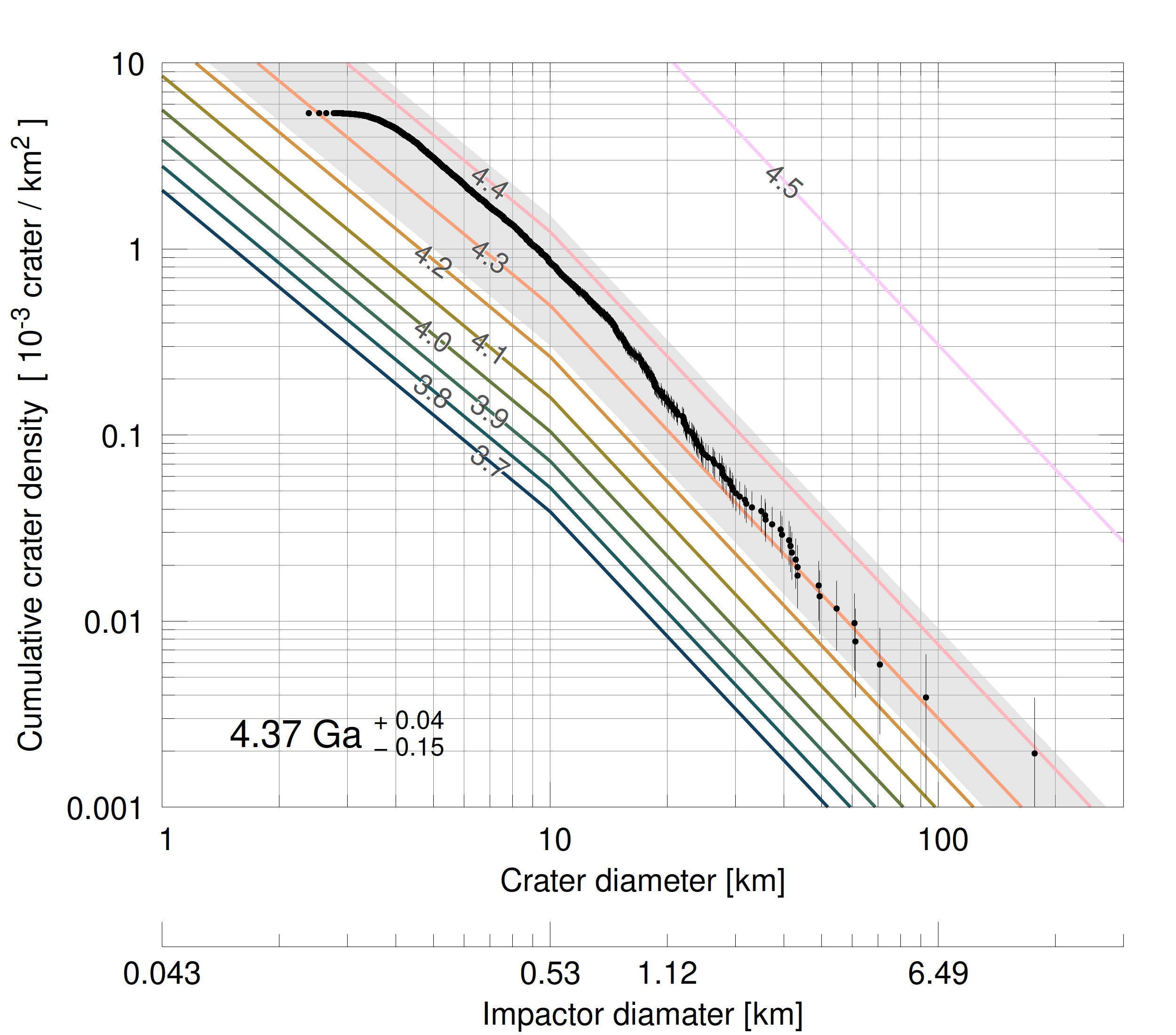}  
	\end{subfigure}
	\begin{subfigure}{.45\textwidth}
		\caption{\label{fig:dioS19KS9}Dione cratered plain}
		\centering
		\includegraphics[width=\linewidth]{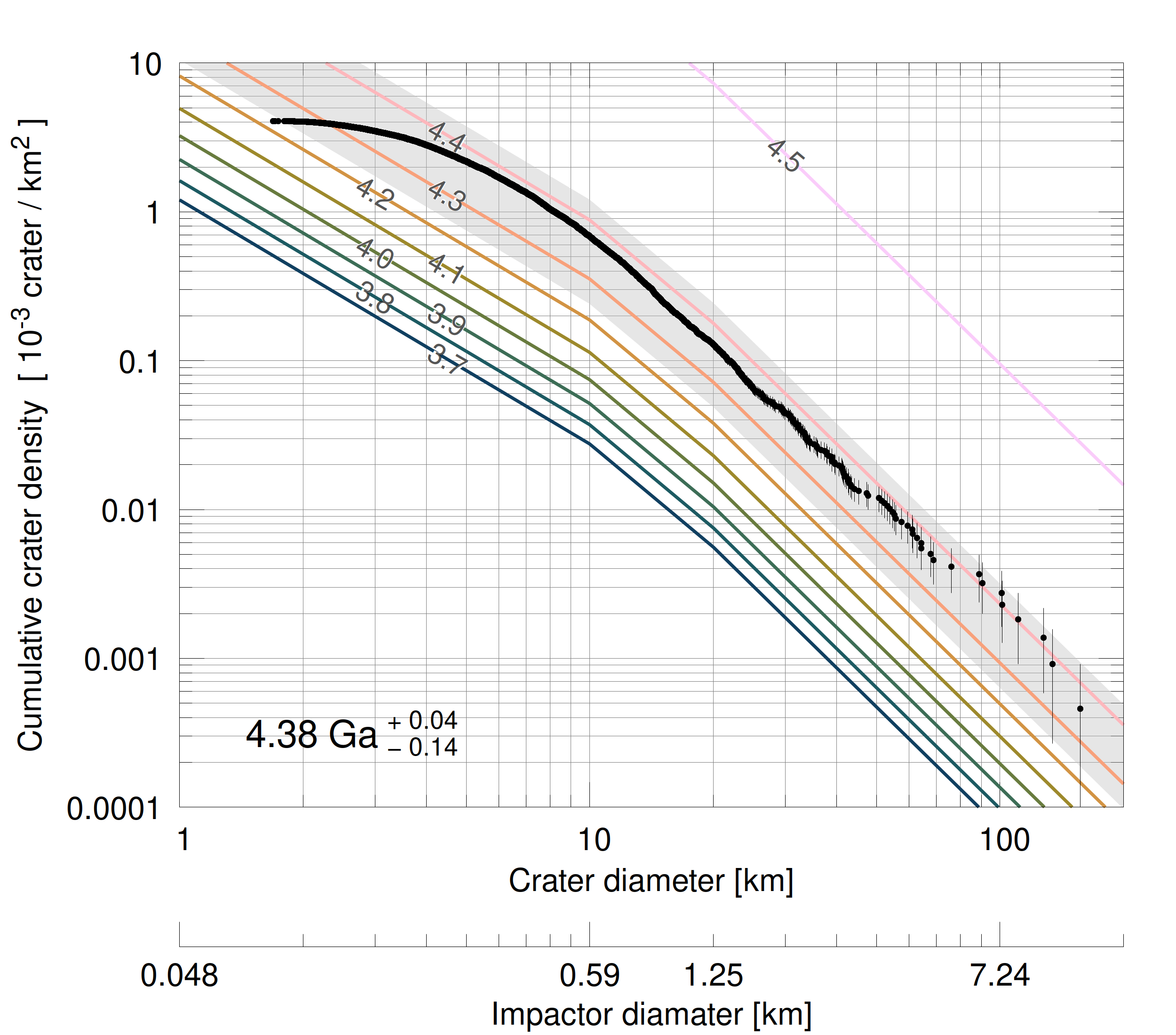} 
	\end{subfigure}
	\newline
	\begin{subfigure}{.45\textwidth}
		\caption{\label{fig:rheS19KS9}Rhea cratered plain}
		\centering
		\includegraphics[width=\linewidth]{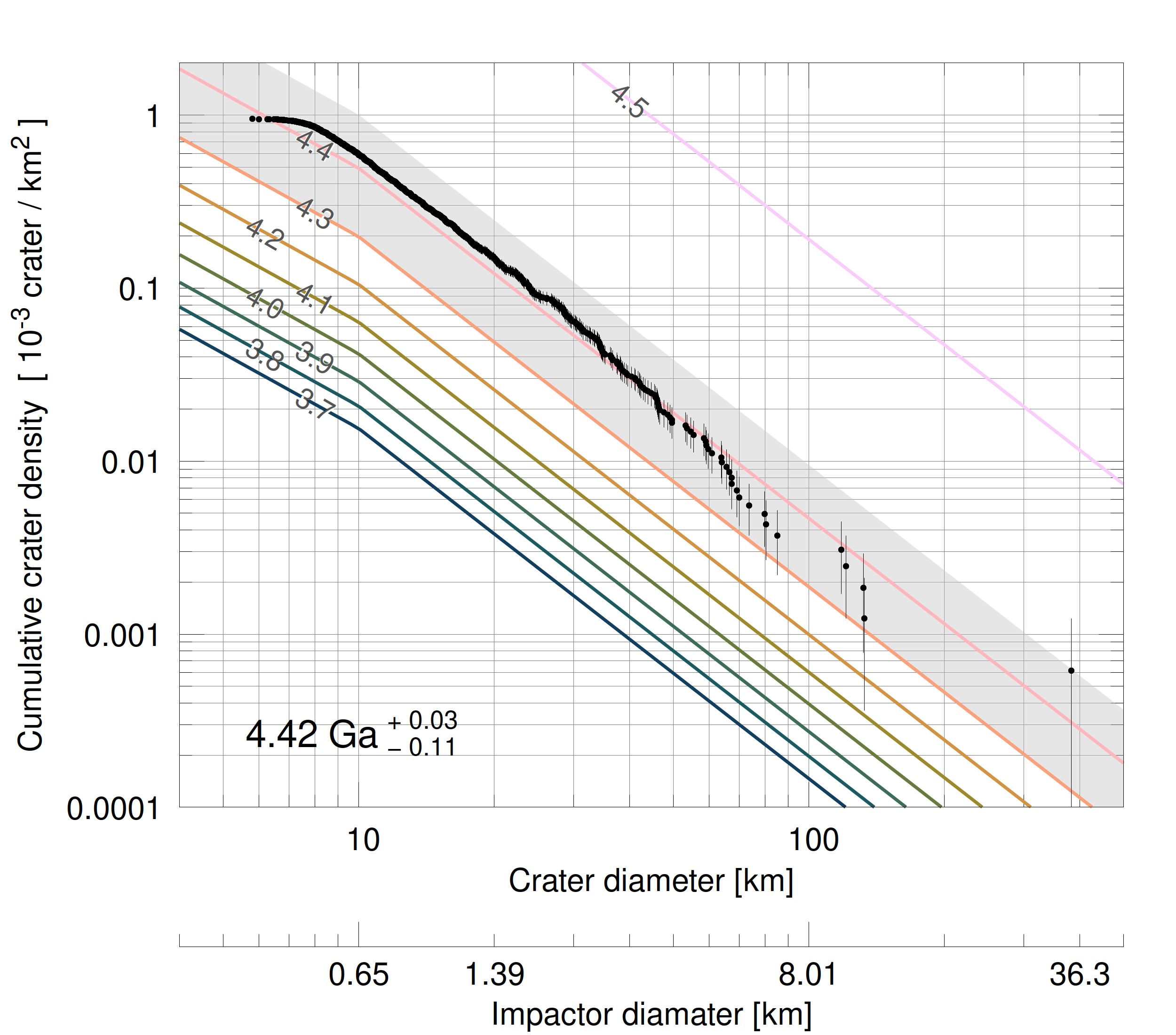}
	\end{subfigure}
	\begin{subfigure}{.45\textwidth}
		\captionsetup{width=5.5cm}
		\caption*{\scriptsize{Figure 6: The best-fitted ages for Mimas and Enceladus are 4.16 Ga and 4.10 Ga, respectively, while those for Tethys, Dione, and Rhea are around 4.4 Ga.
				We obtained these ages by fitting the corresponding size-frequency measurements (black-filled circles) to the isochrones, spanning a range of 3.7 Ga to 4.5 Ga terrain. 
				The \citet{Kirchoff2009} crater production function was used, with varying cumulative slopes and profiles for each icy satellite. The construction of isochrones involved using individual unit crater distributions. The presentation format is the same as in Fig. 5.
		}}
		\centering
		\includegraphics[width=\linewidth]{dummy.png}  
	\end{subfigure}
\end{figure*}

\begin{figure*}
	\centering
	\resizebox{\hsize}{!}{\includegraphics[angle=0]{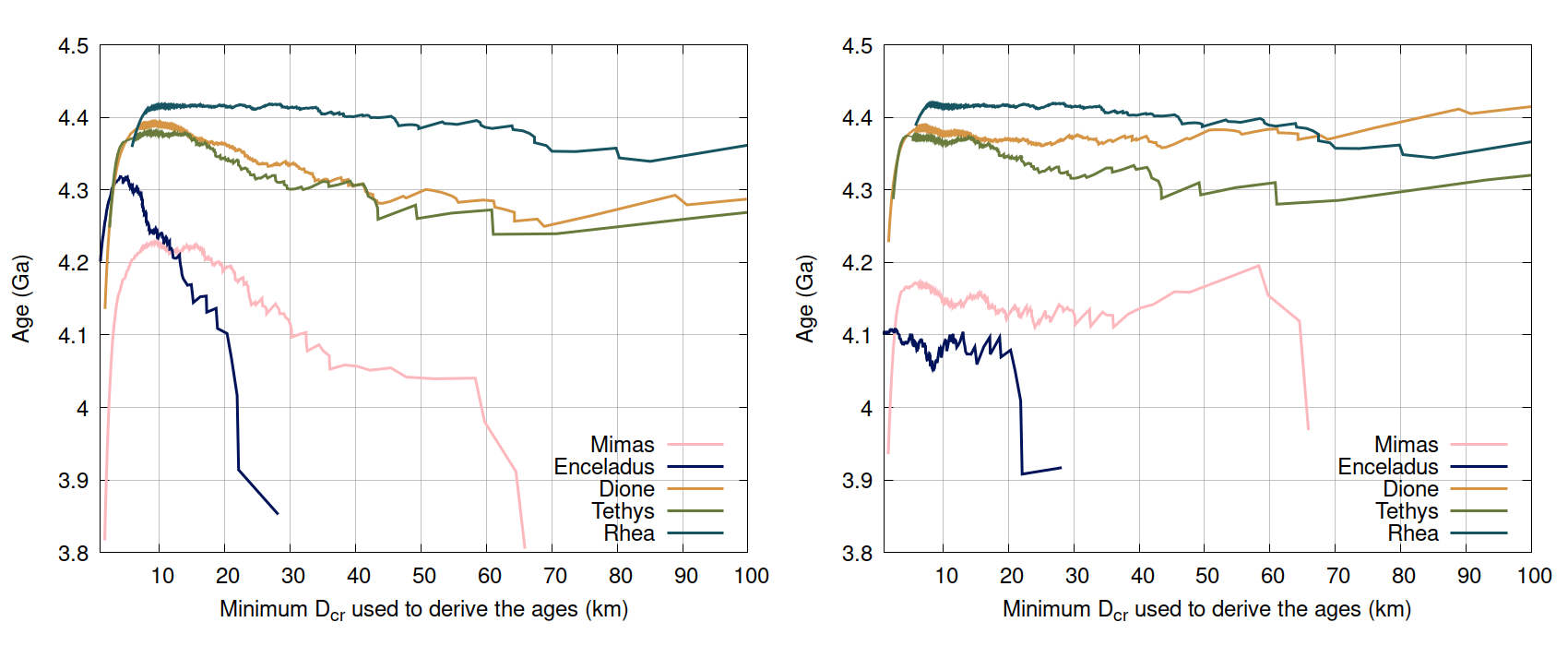}} 
	\caption{\label{fig:age}The changes in the estimated surface age derived based on a specific cumulative crater density of various crater diameters, as described in the second approach of Section~\ref{sub:2nd}
		The left panel shows the results using a single power law crater production function with a cumulative slope of -2 from \citet{Singer2019}, while the right panel uses the crater production function from \citet{Kirchoff2009}. 
		The colour lines illustrate the age variations of the cratered plains on the icy satellites. Notably, Mimas and Enceladus appear younger with ages $\lesssim$ 4.2 Ga, while Tethys, Dione, and Rhea are older ages. 
	}
\end{figure*}

\section{Discussion on the ecliptic comets and uncertainties}
\label{sub:ec_discuss}
Here we list the quantities that we adopted when calculating the number of ecliptic comet in Section~\ref{sub:nec}.
We further clarify the values we adopted for these quantities, compare with and discussed the alternated values.

\subsection{Number of scattered disc objects ($N_{\rm SD}$)}
Previous estimates of $N_{\rm SD}$ rest on the Jupiter-family comets (JFCs) as a proxy for the number of scattered disc objects (SDOs). The JFCs are observationally complete up to some absolute total magnitude (nucleus + coma), $H_{\rm T}$, and up to some perihelion distance, usually 2.5 AU. 
But, this approach is plagued by the very uncertain conversion between the comet's absolute total magnitude and nucleus diameter. \citet{LevisonDuncan1997} used the outcome of their numerical simulations to compute the total number of SDOs utilising the catalogue of JFC with $H_{\rm T}<9$.\\

Traditionally the assumption was that $H_{\rm T}=9$ corresponded to a nucleus diameter of $\sim2$~km. 
Following the approach of \citet{LevisonDuncan1997}, \citet{Brasser2013} compute that there should be $1.7^{+3.0}_{-0.9} \times 10^9$ SDOs with $D_{\rm i}>2.3$~km using dynamical modelling of the flux of JFC coming from the trans-Neptunian region during and after the episode of giant planet migration. 
Using the size-frequency adopted here at the faint end, the number of SDOs with $D_{\rm i}>1$~km is then expected to be a factor of $\sim$5.3 greater, implying that currently $N_{\rm SD,>1\,km}$ = $1.06 \times 10^{10}$. \\

However, this high number appears to violate several observational studies that estimate the number of small SDOs \citep[e.g.,][]{Lawler2018,Nesvorny2019,Shankman2013,VolkMalhotra2008}, the total mass in the disc required to migrate the planets to their current positions \citep{Wong2019}, the number of Jupiter Trojans \citep{Nesvorny2018} and our Monte Carlo simulations using modern estimates of the size-frequency distribution of the SDOs. Furthermore,based on the amount of depletion in our numerical simulations as well as those published in the literature, the estimate above would require an initial number of $>3\times10^{12}$ objects in the disc with $D_{\rm i}>1$~km, which is unreasonable. Instead, the numbers match when equating a JFC with $H_{\rm T}=9$ to a nucleus diameter of 1~km. Is there any evidence to support this claim? \\

From observations the typical difference in total absolute magnitude and the nucleus absolute magnitude: $H-H_{\rm T} = 8 \pm 1$ \citep{FernandezMorbidelli2006}. Assuming an albedo of 4\%, as is typical for comets, the absolute diameter of the nucleus is then $H=19.1$, implying that $H-H_{\rm T}=10$. \citet{BrasserWang2015} plot the total absolute magnitudes of JFC nuclei as a function of nucleus diameter for 68 objects of which both the total absolute magnitude and the nucleus diameter are reasonably well known. Even though the data is scarce for objects with $D_{\rm i} \sim 1$~km (see \citet{Snodgrass2011} for a possible explanation), the observations support the notion that comets with $D_{\rm i}>1$~km {\it could} have $H_{\rm T}<9$, but these objects must be young and highly active. Depending on the activity law used \citep{Fernandez1999} these objects have an active fraction between 1\% and 10\%, which is acceptable given that the typical active fraction found by \citet{BrasserWang2015} is $1.86\%^{+10.7\%}_{-1.59\%}$. \\

In light of the latest observational results estimating the number of SDOs with diameter $D_{\rm i}>10$~km \citep{Nesvorny2019} and their size-frequency distribution at the faint end, we must reject the high estimated primordial number of SDOs that arises from the traditional active JFC $H_{\rm T}$-to-$D$ conversion (i.e., $H_{\rm T}>9$ having $D_{\rm i}>2$ km). 
Our dynamical results, coupled with the observed number of SDOs from \citet{Nesvorny2019} as well as the size-frequency distribution slope at the faint end as observed from the Jupiter Trojans \citep{Nesvorny2018} seems to favour our size conversion suggested above. A more in-depth study between the potential excess in the number SDOs and its uncertainty, and the conversion between total and nucleus absolute magnitude for JFCs is beyond the scope of this paper.\\


\subsection{The rate of decline of the scattered disc objects ($r_{\rm SD}$)}
The value of $r_{\rm SD}$ published in the literature shows a large disparity over the last Gyr of numerical models. It is also time-dependent because, in the early phase, the rate of decline of the planetesimals is much higher than at later times as the planetesimals on highly unstable orbits are removed first. With time the rate of decline decreases \citep[e.g.][]{HolmanWisdom1993}. \\

We compute an average rate of decline of $r_{\rm SD} = -0.092$ Gyr$^{-1}$ without the clones. \citet{DuncanLevison1997} obtain $r_{\rm SD} = -0.29$ Gyr$^{-1}$ considering the last 0.5 Gyr of their simulations, which agrees with our value from the clone simulations, $-0.35$ Gyr$^{-1}$. \citet{Levison2000a} revised this to $r_{\rm SD}=-0.068$~Gyr$^{-1}$ based on longer simulations and a different methodology. \citet{VolkMalhotra2008} find $r_{\rm SD} = -0.1$ Gyr$^{-1}$ for the last 3 Gyr when simulating the evolution of known SDOs. 
\citet{Brasser2013} calculated on average $r_{\rm SD} = -0.166 \pm 0.073$ Gyr$^{-1}$ from the last 0.5 Gyr of their simulation. 
It is a little higher than the decline averaged over the last 1 Gyr of our migration simulation: $-0.110$ Gyr$^{-1}$, but still within the uncertainty. Overall the value of $r_{\rm SD}$ ranges between $-0.1$ to $-0.2$ Gyr$^{-1}$.\\

\subsection{Fraction of ecliptic comet ($f_{\rm EC}$)}
During the last 3 Gyr of the simulations, we establish what fraction of planetesimals that begin with $q>30$~AU will end up with $q<30$~AU and for how long they spend on Neptune-crossing orbits as ecliptic comets. On average, we find that in the original simulations, 68.4\% of particles become Neptune crossers. 
If we computed this fraction over the first Gyr of the simulation, it is generally higher, but at that time, the steady-state approximation is not valid. 

\subsection{The dynamical lifetimes of the ecliptic comets ($\tau_{\rm EC}$)}
We obtain a value of 267~Myr, consistent with that of \citet{Levison2000a}. However, these current data provide no consensus on the mean planetesimal lifetime. From their simulations, \cite{LevisonDuncan1997} list a {\it median} planetesimal lifetime of 47 Myr in their simulations which lasted 1 Gyr. 
This was revised by \cite{Levison2000a}, who computed a {\it mean} lifetime of 200 Myr using the same initial condition but having extended the simulations to 4 Gyr. In comparable studies, \cite{DiSisto2007} find the {\it mean} Centaur lifetime of 72 Myr; \cite{Tiscareno2003} calculate 9 Myr from a smaller data set; \citet{Fernandez2004} obtain 1.8~Gyr for the typical lifetime of planetesimals {\it beyond} Neptune. 

\subsection{Other uncertainties}
The uncertainties in the quantities used to calculate the number and injection rate of ecliptic comets ($N_{\rm EC}$ \& $F_{\rm EC}$) and the initial number of scattered disc objects ($N_{\rm SD}$) are attributed to the differences in results between the compact and extended disc models used in the giant planet migration simulations.
The upper and lower bounds of the quantities are established by the two models. For example, the final leftover percentage, 0.31\%, has an uncertainty of $\pm$0.2\%: the average of the number of leftover particles ($N_{\rm left}$) after 4.5 Gyr of evolution is 50, which is the average of 17 leftovers from the compact disc model and 83 leftovers from the extended disc model. The uncertainty corresponds to $\pm$33 leftovers. We used the final leftover percentages to derive the initial $N_{\rm SD}$, which is 1.9$\times$10$^{12}$ for compact discs and 3.8$\times$10$^{11}$ for extended discs. Hence, the complete expression of $N_{\rm SD}$ is  $6.4^{+12.9}_{-2.57} \times 10^{11}$. We computed the maximum and minimum expected crater densities from the initial $N_{\rm SD}$ at the two extremes and propagate the uncertainty to estimate the surface ages. \\

It should be noted that since the average is taken as the average between the two outcomes of the compact and extended disc models, most quantities have symmetric uncertainties, except for $N_{\rm SD}$ and surface ages. The uncertainty of $N_{\rm SD}$ is asymmetric due to the $\propto 1/\rm N_{\rm left}$ relation.
The derived ages are asymmetric due to the involvement of the Weibull distribution. The stretched exponential decline function slows down over time and the relation between ages and crater densities is not linear, and thus the sensitivity to ages reduces with (higher) crater densities.\\

The same uncertainty consideration applies to the calculation of ecliptic comets, where the uncertainties in the declining rate of the scattered disc population ($r_{\rm SD}$), fraction of scattered disc objects that become ecliptic comets ($f_{\rm EC}$), and the mean lifetime of ecliptic comets ($\tau_{\rm EC}$) reflect the different outcomes between the two disc models used. 
While reporting the results regarding ecliptic comets in Section~\ref{sub:result_nec}, we did not include all the uncertainties in the text for the sake of readability.
Here we briefly re-evaluate $N_{\rm EC}$ with appropriate uncertainties from the quantities obtained from the giant planets migration simulation results. Applying Eq.~\ref{eq:nec} with error propagation, for $\vert r_{\rm SD} \vert = (0.092 \pm 0.036)$ Gyr$^{-1}$, $f_{\rm EC} = (68.4 \pm 10)$\%, and $\tau_{\rm EC} = (267 \pm 18)$ Myr, we obtain $N_{\rm EC} = (3.7 \pm 2.2) \times 10^{7}$ ecliptic comets larger than 1~km in diameter, and the injection rate as $F_{\rm EC} = (0.13 \pm 0.07)$ yr$^{-1}$.

\section{Discussion on the giant planet migration simulation}
\label{sec:gpm}

The objective of the giant planet migration simulations was to recreate the dynamics of planetesimals during and after the dynamical instability; specifically, recording their scattering and depletion in the disc. It is well known that the migration and mutual scattering amongst the giant planets leads to a chaotic evolution \citep[e.g.][]{Nesvorny2012}. To mitigate against this and to ensure robust results we made simplifications and adjustments to the orbital evolution of the giant planets. Our intentions were twofold: ensuring a deterministic evolutionary pathway of the giant planets that aligns with the current Solar System configuration, and maximising computational efficiency to run more simulations with more test particles for increased statistical certainty. However, we acknowledge the limitations and discrepancies in our approximations, which we will discuss below.\\

To avoid chaotic orbital evolution in the giant planet migration simulations, we adopted the method proposed by \citet{Levison2008}, where modifications to the velocities were applied to the planets at each time step, evolving their semi-major axes, eccentricities and inclinations along a deterministic path; this was combined with eccentricity damping. Thus, we no longer required the crossing of mean motion resonances of the Jupiter--Saturn pair to drive up the eccentricities of the ice giants. Jupiter and Saturn remained close to 5 AU and 9 AU, respectively, while migrating Uranus and Neptune were guided by a fictitious force parameterised by an e-folding timescale of 50 Myr. The fictitious force mimicked Neptune's planetesimal-driven migration, allowing us to treat the planetesimals as massless. This approach facilitated faster simulations with more planetesimals, increasing the impact data collected within our computational constraints. We note that in several previous dynamical studies, the evolution of the giant planets is often guided by fictitious forces to investigate the dynamics of small bodies under different initial conditions or planetary evolutionary pathways \citep{Levison2008,Nesvorny2015b,Nesvorny2016b}.\\

The stable gas giants and massless planetesimal approximation might need more accuracy. Regarding the migration of the gas giants, Jupiter migrated inwards by less than 0.5 AU and Saturn moving outwards by approximately 2 AU, which is much less than the migration of the ice giants \citep{Brasser2015,Wong2019}. The intense scattering of planetesimals is primarily driven by the intrusion of the ice giants into the disc \citep{Nesvorny2012,Brasser2013}. \\

In addition to the above approximations, we now realise that the smooth and slow migration of Neptune, its low eccentricity, and the timing of instability may not be consistent with some versions of migration models. For instance, alternative models propose grainy migration with Neptune scattering massive Pluto-sized planetesimals \citep{Nesvorny2016b}, or ``jumping Neptune" which pushes the initially 2:1 resonant planetesimals outwards to create the 44 AU Kuiper Belt kernel \citep{Nesvorny2015b}. Some models also suggest a ``violent" evolution involving the ejection of one or two extra giant planets \citep{Nesvorny2012}. In our cases with smooth migration, Neptune's low eccentricity results in high-inclination planetesimals $($Fig.~\ref{fig:tp1g}$)$ and an increased number of Hot Classical detected scattered objects from Kozai resonances \citep{Nesvorny2016b}. We will investigate alternative migration scenarios in the future.\\

In our giant planet migration simulation, Neptune migrated slowly over 78 Myr to reach 28 AU and 600 Myr to reach 30 AU. This migration speed was chosen to ensure the capture of Saturn with a low obliquity by maintaining the secular spin-orbit resonance, which tilted Saturn’s spin axis to 27$^{\circ}$ \citep{Brasser2015}, and the observed inclination distribution of Kuiper Belt objects \citep{Nesvorny2015b}. Our previous work in \citet{Wong2019} also demonstrated a similar slow migration ($\sim$200 Myr to reach 30 AU) using GENGA simulations. It is worth noting that different instability models show variations in the radial evolution rate of Neptune. For instance, \citet{Nesvorny2012} recorded faster migration (20 to 100 Myr) for systems starting with four or five planets, while slower migration ($>$1 Gyr) was observed in models with initially six planets, or a five-planet simulation that resembles the GENGA simulation of \citet{Wong2019}. In other words, there appears to be a great variation in migration rate, as well as in the trigger of the instability \citep{Levison2011,Quarles2019}. This is still a topic of active research.\\

To postulate our result with an alternate scenario for Neptune's migration would be as follows. A faster migration where Neptune intrudes into the disc earlier would result in a steeper decline in the planetesimals' population, resulting in a shorter timescale ($\tau$) and potentially older estimated ages for the icy satellites' terrains. As for Neptune's eccentricity, we suggest that its high eccentricity phase is relatively short-lived, lasting around 10 Myr \citep[e.g.,][]{Nesvorny2012}. Depending on when this phase occurs, the number of remaining planetesimals may not significantly alter the overall impact rate. We believe that Neptune's short-lived but excited state would only slightly decrease the e-folding timescale of the impact chronology, thereby marginally increasing our estimated surface ages. The effect would be more pronounced if the high eccentricity phase occurred earlier. In our migration simulations we initiated the instability at the beginning. The timing of instability might vary slightly --- delayed by 5 to 20 Myr -- but this range of variation is smaller than the uncertainties associated with the planetesimal disc and the fitting of impact chronologies.\\

To summarize, a wide range of variations in the initial conditions and events could have occurred during the giant planet migration phase. Different migration scenarios may capture specific characteristics of the observed Solar System while overlooking others. In this study, we conducted simulations in good faith based on the prevailing notion of a slow and smooth migration of Neptune with an early instability. These approximations allowed us to study the dynamics of planetesimals for building crater chronologies and age calculations. As the dynamical community refines its understanding of the early conditions of the giant planets and the planetesimal disc, we can provide surface ages with greater accuracy. Once the crater counting methodology is well-established, it will be essential to explore and rerun simulations with altered migration scenarios.


\section{Absolute surface ages comparison across the studies}
Compared with the surface ages calculated from 20~km cumulative crater density values and published in \citet{Wong2020}, the new ages determined using a best fit are older by approximately 110 to 300 Myr (3\% to 8\% difference). The discrepancy in the estimated ages is mainly attributed to the updated initial number of scattered disc objects ($N_{\rm SD}$) in the disc, which changed the expected number of craters and consequently affected the estimation of the surface ages. \citet{Wong2020} adopted two size-frequency distributions by \citet{Nesvorny2016a} and \citet{Fraser2014}, which yielded $N_{\rm SD, >1\,km}$ as 2.235 $\times$ 10$^{12}$ and 2.555 $\times$ 10$^{11}$, respectively. The updated $N_{\rm SD, >1\,km}$ value of 6.4$\times$ 10$^{11}$, calculated using planetesimal size-frequency distribution from \citet{Singer2019} and with constraint from the current observed $N_{\rm SD}$, falls between the previous two $N_{\rm SD, >1 km}$ values and thus results in surface ages that also fall between the previous estimates. 
We consider the updated $N_{\rm SD, >1\,km}$ in this work tto be more reliable.\\

Additional factors that could contribute to the difference in estimated ages are the stretched exponential function parameters: $\beta$ and $\tau$. The new $\beta$=0.317 and $\tau$=1.596 Myr were obtained by fitting the impacting frequency onto Saturn from the latest simulation, while \citet{Wong2020} obtained $\beta$=0.349 and $\tau$=2.58 Myr from the combined impact frequencies of all four giant planets, which were considered as the overall values for the outer Solar System.  
The new pair of parameters are both smaller than previous estimates; they counteract each other: increasing $\beta$ generally decreases the surface ages, while increasing $\tau$ increases the surface ages. We consider the new $\beta$ and $\tau$ to better represent the crater chronology of the Saturnian satellites system since (i) they were specifically fitted for the impact on Saturn, (ii) the latest scattered disc simulations combined the migration simulations and clone simulations provide higher impact resolution for better fits, and (iii) the new fitting algorithm reduces the biases at early times when fitting the impact data through the size-frequency distribution of impacts.\\

The observed cumulative crater density could also affect the estimated surface ages, but we maintain the same source of observational values in this paper, referencing \citet{Kirchoff2009}, and expanding the analysis to include a wider range of crater sizes. Using our previous approach in \citet{Wong2020} using only the cumulative crater density of craters with diameter $D_{\rm cr}\geq 20$~km, the ages estimated here on average are $\sim$250 million years older. The difference can be explained by using a reduced total number of scattered disc objects $N_{\rm SD, >1\, km}$ compared to the size-frequency distribution from \citet{Nesvorny2016a} used by \citet{Wong2020}.\\

A reason that we reported the ages calculated with the second approach (in Section~\ref{sub:2nd}) is to compare them with the published ages in case A from \citet{Kirchoff2009}, as shown in the last two columns of Table~\ref{tab:ks9}. 
They adopted a cratering rate that was extrapolated back in time as $t^{-1}$ \citep{Zahnle1998} from the current cratering rate given by \citet{Zahnle2003}, while our approach is to go forward in time starting at 4.5 Ga. However, heavily cratered records on the Galilean satellites suggested the cratering rate, especially in the early Solar System, is higher than the value extrapolated with a decline going as $t^{-1}$. If such an extrapolation underestimates the expected number of craters formed since the formation of the crust, the calculated surface ages will be overestimated. Most ages in case A of \citet{Kirchoff2009} are reported as 4.6 Ga, the same as 4.56 Ga of the Solar System. They also suggest the large discrepancy between calculated ages using different minimum crater diameters could be due to inaccuracies in the cratering rate at different crater diameters.\\

We argue that our proposed cratering rate, which declines as a Weibull function, is a more optimal description for the outer Solar System, as our estimated surface ages provide improved values and did not approach 4.56 Gyr. Our method for dating the surface ages is generally more accurate than that of \citet{Zahnle2003} because of our better fit to the impact chronology function and much higher impact resolution. We have been running more simulations to constrain the evolving cratering rate and possibility the impact probability in more recent time. We seek to validate our obtained surface ages by extrapolating from the current cratering rate in the manner of \citet{Zahnle2003}, and thereby to provide more accurate ages for young surfaces. We reserve this for future work.\\

Throughout this work we have assumed that giant planet migration began at 4.5~Ga, because radiogenic reset ages of asteroids suggest that this violent epoch started before 4.48~Ga \citep{Mojzsis2019}. The onset of giant planet migration could have been earlier \citep[e.g.][]{DeSousa2020}, so that the absolute surface ages deduced here could be up to 80~Myr older. Still, the difference in the rate of decline in the impact flux is the primary factor of disparity in the ages, and the uncertainties in the ages derived here are generally larger than 80~Myr. 

\subsection{Contribution by planetocentric projectiles }
Due to modelling limitations and our incomplete understanding of the source of the planetocentric population, we are still exploring ways to incorporate them into the dynamical simulation and our studies.\\

However, \citet{Wong2020} found that irregular satellites did not significantly contribute to the impact count because their population declined rapidly (dynamical e-folding time of $\sim$10,000 years) and the lack of a replenishing source. 
Although numerous elongated/elliptical craters suggest impacts from a planetocentric population \citep{Ferguson2020,Ferguson2022a,Ferguson2022b}, it is plausible that these craters originated from heliocentric populations captured by giant planets into highly inclined and eccentric orbits \citep{Carruba2003,Nesvorny2003} rather than having to be solely caused by primordial planetocentric objects.
If both populations have the same origin, we can still rely on the exact impact chronology. 
Once a planetesimal is captured in a circumplanetary orbit, it is expected to be removed within 1 to 10 thousand years.\\

The formation of sesquinaries requires primary impactors to eject debris from the regular satellites into circumplanetary orbit, and the likelihood of generating such debris was higher in the distant past, when the impact flux was greater, rather than today. Thus, while elongated/elliptical craters could be from planetocentric debris, it is unclear whether they must necessarily be young features and whether they are from ejecta or other planetocentric sources.

\section{Conclusions}
In this work we have created crater isochrones for the regular icy satellites of Saturn combining state of the art dynamical simulations of the evolution of the outer Solar System with the most recent projectile distributions to evaluate crater densities from \citet{Kirchoff2009,Kirchoff2010}. Our updated dynamical simulations provide a high-resolution impact chronology function for all the giant planets for 4.5 Ga that is well fit by a stretched exponential decline after about 30 Myr from the onset of the simulations. We employed the latest estimates of the current number of trans-Neptunian objects and their size-frequency distribution to calculate theoretical crater densities on the regular Saturnian satellites assuming their surfaces retain all these craters. Together with the observed crater densities and their production function the chronology function provides us with crater isochrones of the five regular Saturnian satellites.\\

Our results show that that the most heavily cratered areas of Mimas, Enceladus, Tethys, Dione and Rhea are 4.1 Ga for the inner two and 4.4 Ga for the outer three satellites, with uncertainties $<300$~Myr. Our results imply that these satellites are old, which has implications for their tidal history \citep{Nakajima2019, Lainey2020} and their formation mechanism \citep{Charnoz2011,Crida2012}. Our ages are mostly consistent with those of \citet{Kirchoff2009} using the case A chronology of \citet{Zahnle2003}, while the ages of case B are inconsistent. The techniques described here are also applicable to satellites of Jupiter, Uranus and possibly Neptune.

\section*{Acknowledgements}
EW gratefully acknowledges financial support from the Grant-in-Aid for JSPS Research Fellow (Grant Numbers: 22KJ12886).
SCW thanks the Research Council of Norway through its Centres of Excellence funding scheme, project No. 223272 (CEED) and No. 332523 (PHAB).
MRK would like to acknowledge NASA Cassini Data Analysis Program (80NSSC21K0535) and Java Mission-planning and Analysis for Remote Sensing for providing the GIS software to collect data.
We thank reviewers William F. Bottke and Sebastien Charnoz for their helpful comments, which improved this manuscript, and the editor Brandon Johnson for moderating the manuscript. 

The Scientific colour map {\it batlow} \citep{colormap} is used in the figures to prevent visual distortion of the data and exclusion of readers with colourvision deficiencies \citep{Crameri2020}. 


\bibliographystyle{elsarticle-harv}
\bibliography{author}

\begin{thebibliography}{71}
\expandafter\ifx\csname natexlab\endcsname\relax\def\natexlab#1{#1}\fi
\providecommand{\url}[1]{\texttt{#1}}
\providecommand{\href}[2]{#2}
\providecommand{\path}[1]{#1}
\providecommand{\DOIprefix}{doi:}
\providecommand{\ArXivprefix}{arXiv:}
\providecommand{\URLprefix}{URL: }
\providecommand{\Pubmedprefix}{pmid:}
\providecommand{\doi}[1]{\href{http://dx.doi.org/#1}{\path{#1}}}
\providecommand{\Pubmed}[1]{\href{pmid:#1}{\path{#1}}}
\providecommand{\bibinfo}[2]{#2}
\ifx\xfnm\relax \def\xfnm[#1]{\unskip,\space#1}\fi
\bibitem[{{Brasser} and {Lee}(2015)}]{Brasser2015}
\bibinfo{author}{{Brasser}, R.}, \bibinfo{author}{{Lee}, M.H.},
  \bibinfo{year}{2015}.
\newblock \bibinfo{title}{{Tilting Saturn without Tilting Jupiter: Constraints
  on Giant Planet Migration}}.
\newblock \bibinfo{journal}{\aj} \bibinfo{volume}{150}, \bibinfo{pages}{157}.
\bibitem[{{Brasser} and {Morbidelli}(2013)}]{Brasser2013}
\bibinfo{author}{{Brasser}, R.}, \bibinfo{author}{{Morbidelli}, A.},
  \bibinfo{year}{2013}.
\newblock \bibinfo{title}{{Oort cloud and Scattered Disc formation during a
  late dynamical instability in the Solar System}}.
\newblock \bibinfo{journal}{\icarus} \bibinfo{volume}{225},
  \bibinfo{pages}{40--49}.
\bibitem[{{Brasser} and {Wang}(2015)}]{BrasserWang2015}
\bibinfo{author}{{Brasser}, R.}, \bibinfo{author}{{Wang}, J.H.},
  \bibinfo{year}{2015}.
\newblock \bibinfo{title}{{An updated estimate of the number of Jupiter-family
  comets using a simple fading law}}.
\newblock \bibinfo{journal}{\aap} \bibinfo{volume}{573}, \bibinfo{pages}{A102}.
\bibitem[{{Carruba} et~al.(2003){Carruba}, {Burns}, {Bottke} and
  {Nesvorn{\'y}}}]{Carruba2003}
\bibinfo{author}{{Carruba}, V.}, \bibinfo{author}{{Burns}, J.A.},
  \bibinfo{author}{{Bottke}, W.}, \bibinfo{author}{{Nesvorn{\'y}}, D.},
  \bibinfo{year}{2003}.
\newblock \bibinfo{title}{{Orbital evolution of the Gefion and Adeona asteroid
  families: close encounters with massive asteroids and the Yarkovsky effect}}.
\newblock \bibinfo{journal}{\icarus} \bibinfo{volume}{162},
  \bibinfo{pages}{308--327}.
\bibitem[{{Charnoz} et~al.(2011){Charnoz}, {Crida}, {Castillo-Rogez}, {Lainey},
  {Dones}, {Karatekin}, {Tobie}, {Mathis}, {Le Poncin-Lafitte} and
  {Salmon}}]{Charnoz2011}
\bibinfo{author}{{Charnoz}, S.}, \bibinfo{author}{{Crida}, A.},
  \bibinfo{author}{{Castillo-Rogez}, J.C.}, \bibinfo{author}{{Lainey}, V.},
  \bibinfo{author}{{Dones}, L.}, \bibinfo{author}{{Karatekin}, {\"O}.},
  \bibinfo{author}{{Tobie}, G.}, \bibinfo{author}{{Mathis}, S.},
  \bibinfo{author}{{Le Poncin-Lafitte}, C.}, \bibinfo{author}{{Salmon}, J.},
  \bibinfo{year}{2011}.
\newblock \bibinfo{title}{{Accretion of Saturn{\textquoteright}s mid-sized
  moons during the viscous spreading of young massive rings: Solving the
  paradox of silicate-poor rings versus silicate-rich moons}}.
\newblock \bibinfo{journal}{\icarus} \bibinfo{volume}{216},
  \bibinfo{pages}{535--550}.
\bibitem[{Colwell(1994)}]{Colwell1994}
\bibinfo{author}{Colwell, J.}, \bibinfo{year}{1994}.
\newblock \bibinfo{title}{The disruption of planetary satellites and the
  creation of planetary rings}.
\newblock \bibinfo{journal}{Planetary and Space Science} \bibinfo{volume}{42},
  \bibinfo{pages}{1139--1149}.
\bibitem[{Crameri(2021)}]{colormap}
\bibinfo{author}{Crameri, F.}, \bibinfo{year}{2021}.
\newblock \bibinfo{title}{Scientific colour maps}.
\newblock \bibinfo{note}{{The development of the Scientific colour maps is not
  funded any longer, but will continue as a pro bono project for the scientific
  community. - Fabio}}.
\bibitem[{{Crameri} et~al.(2020){Crameri}, {Shephard} and
  {Heron}}]{Crameri2020}
\bibinfo{author}{{Crameri}, F.}, \bibinfo{author}{{Shephard}, G.E.},
  \bibinfo{author}{{Heron}, P.J.}, \bibinfo{year}{2020}.
\newblock \bibinfo{title}{{The misuse of colour in science communication}}.
\newblock \bibinfo{journal}{Nature Communications} \bibinfo{volume}{11},
  \bibinfo{pages}{5444}.
\bibitem[{{Crida} and {Charnoz}(2012)}]{Crida2012}
\bibinfo{author}{{Crida}, A.}, \bibinfo{author}{{Charnoz}, S.},
  \bibinfo{year}{2012}.
\newblock \bibinfo{title}{{Formation of Regular Satellites from Ancient Massive
  Rings in the Solar System}}.
\newblock \bibinfo{journal}{Science} \bibinfo{volume}{338},
  \bibinfo{pages}{1196}.
\bibitem[{{{\'C}uk} et~al.(2016){{\'C}uk}, {Dones} and
  {Nesvorn{\'y}}}]{Cuk2016}
\bibinfo{author}{{{\'C}uk}, M.}, \bibinfo{author}{{Dones}, L.},
  \bibinfo{author}{{Nesvorn{\'y}}, D.}, \bibinfo{year}{2016}.
\newblock \bibinfo{title}{{Dynamical Evidence for a Late Formation of
  Saturn{\textquoteright}s Moons}}.
\newblock \bibinfo{journal}{\apj} \bibinfo{volume}{820}, \bibinfo{pages}{97}.
\bibitem[{{de Sousa} et~al.(2020){de Sousa}, {Morbidelli}, {Raymond},
  {Izidoro}, {Gomes} and {Vieira Neto}}]{DeSousa2020}
\bibinfo{author}{{de Sousa}, R.R.}, \bibinfo{author}{{Morbidelli}, A.},
  \bibinfo{author}{{Raymond}, S.N.}, \bibinfo{author}{{Izidoro}, A.},
  \bibinfo{author}{{Gomes}, R.}, \bibinfo{author}{{Vieira Neto}, E.},
  \bibinfo{year}{2020}.
\newblock \bibinfo{title}{{Dynamical evidence for an early giant planet
  instability}}.
\newblock \bibinfo{journal}{\icarus} \bibinfo{volume}{339}.
\bibitem[{{Di Sisto} and {Brunini}(2007)}]{DiSisto2007}
\bibinfo{author}{{Di Sisto}, R.P.}, \bibinfo{author}{{Brunini}, A.},
  \bibinfo{year}{2007}.
\newblock \bibinfo{title}{{The origin and distribution of the Centaur
  population}}.
\newblock \bibinfo{journal}{\icarus} \bibinfo{volume}{190},
  \bibinfo{pages}{224--235}.
\bibitem[{{Dobrovolskis} et~al.(2007){Dobrovolskis}, {Alvarellos} and
  {Lissauer}}]{Dobrovolskis2007}
\bibinfo{author}{{Dobrovolskis}, A.R.}, \bibinfo{author}{{Alvarellos}, J.L.},
  \bibinfo{author}{{Lissauer}, J.J.}, \bibinfo{year}{2007}.
\newblock \bibinfo{title}{{Lifetimes of small bodies in planetocentric (or
  heliocentric) orbits}}.
\newblock \bibinfo{journal}{\icarus} \bibinfo{volume}{188},
  \bibinfo{pages}{481--505}.
\bibitem[{{Duncan} and {Levison}(1997)}]{DuncanLevison1997}
\bibinfo{author}{{Duncan}, M.J.}, \bibinfo{author}{{Levison}, H.F.},
  \bibinfo{year}{1997}.
\newblock \bibinfo{title}{{A scattered comet disk and the origin of Jupiter
  family comets}}.
\newblock \bibinfo{journal}{Science} \bibinfo{volume}{276},
  \bibinfo{pages}{1670--1672}.
\bibitem[{{Duncan} et~al.(1995){Duncan}, {Levison} and {Budd}}]{Duncan1995}
\bibinfo{author}{{Duncan}, M.J.}, \bibinfo{author}{{Levison}, H.F.},
  \bibinfo{author}{{Budd}, S.M.}, \bibinfo{year}{1995}.
\newblock \bibinfo{title}{{The Dynamical Structure of the Kuiper Belt}}.
\newblock \bibinfo{journal}{\aj} \bibinfo{volume}{110}, \bibinfo{pages}{3073}.
\bibitem[{{Ferguson} et~al.(2020){Ferguson}, {Rhoden} and
  {Kirchoff}}]{Ferguson2020}
\bibinfo{author}{{Ferguson}, S.N.}, \bibinfo{author}{{Rhoden}, A.R.},
  \bibinfo{author}{{Kirchoff}, M.R.}, \bibinfo{year}{2020}.
\newblock \bibinfo{title}{{Small Impact Crater Populations on Saturn's Moon
  Tethys and Implications for Source Impactors in the System}}.
\newblock \bibinfo{journal}{Journal of Geophysical Research (Planets)}
  \bibinfo{volume}{125}, \bibinfo{pages}{e06400}.
\bibitem[{{Ferguson} et~al.(2022a){Ferguson}, {Rhoden} and
  {Kirchoff}}]{Ferguson2022a}
\bibinfo{author}{{Ferguson}, S.N.}, \bibinfo{author}{{Rhoden}, A.R.},
  \bibinfo{author}{{Kirchoff}, M.R.}, \bibinfo{year}{2022}a.
\newblock \bibinfo{title}{{Regional Impact Crater Mapping and Analysis on
  Saturn's Moon Dione and the Relation to Source Impactors}}.
\newblock \bibinfo{journal}{Journal of Geophysical Research (Planets)}
  \bibinfo{volume}{127}, \bibinfo{pages}{e07204}.
\bibitem[{{Ferguson} et~al.(2022b){Ferguson}, {Rhoden}, {Kirchoff} and
  {Salmon}}]{Ferguson2022b}
\bibinfo{author}{{Ferguson}, S.N.}, \bibinfo{author}{{Rhoden}, A.R.},
  \bibinfo{author}{{Kirchoff}, M.R.}, \bibinfo{author}{{Salmon}, J.J.},
  \bibinfo{year}{2022}b.
\newblock \bibinfo{title}{{A unique Saturnian impactor population from
  elliptical craters}}.
\newblock \bibinfo{journal}{Earth and Planetary Science Letters}
  \bibinfo{volume}{593}, \bibinfo{pages}{117652}.
\bibitem[{{Fern{\'a}ndez} et~al.(2004){Fern{\'a}ndez}, {Gallardo} and
  {Brunini}}]{Fernandez2004}
\bibinfo{author}{{Fern{\'a}ndez}, J.A.}, \bibinfo{author}{{Gallardo}, T.},
  \bibinfo{author}{{Brunini}, A.}, \bibinfo{year}{2004}.
\newblock \bibinfo{title}{{The scattered disk population as a source of Oort
  cloud comets: evaluation of its current and past role in populating the Oort
  cloud}}.
\newblock \bibinfo{journal}{\icarus} \bibinfo{volume}{172},
  \bibinfo{pages}{372--381}.
\bibitem[{{Fern{\'a}ndez} and {Morbidelli}(2006)}]{FernandezMorbidelli2006}
\bibinfo{author}{{Fern{\'a}ndez}, J.A.}, \bibinfo{author}{{Morbidelli}, A.},
  \bibinfo{year}{2006}.
\newblock \bibinfo{title}{{The population of faint Jupiter family comets near
  the Earth}}.
\newblock \bibinfo{journal}{\icarus} \bibinfo{volume}{185},
  \bibinfo{pages}{211--222}.
\bibitem[{{Fern{\'a}ndez} et~al.(1999){Fern{\'a}ndez}, {Tancredi}, {Rickman}
  and {Licandro}}]{Fernandez1999}
\bibinfo{author}{{Fern{\'a}ndez}, J.A.}, \bibinfo{author}{{Tancredi}, G.},
  \bibinfo{author}{{Rickman}, H.}, \bibinfo{author}{{Licandro}, J.},
  \bibinfo{year}{1999}.
\newblock \bibinfo{title}{{The population, magnitudes, and sizes of Jupiter
  family comets}}.
\newblock \bibinfo{journal}{\aap} \bibinfo{volume}{352},
  \bibinfo{pages}{327--340}.
\bibitem[{{Fraser} et~al.(2014){Fraser}, {Brown}, {Morbidelli}, {Parker} and
  {Batygin}}]{Fraser2014}
\bibinfo{author}{{Fraser}, W.C.}, \bibinfo{author}{{Brown}, M.E.},
  \bibinfo{author}{{Morbidelli}, A.}, \bibinfo{author}{{Parker}, A.},
  \bibinfo{author}{{Batygin}, K.}, \bibinfo{year}{2014}.
\newblock \bibinfo{title}{{The Absolute Magnitude Distribution of Kuiper Belt
  Objects}}.
\newblock \bibinfo{journal}{\apj} \bibinfo{volume}{782}, \bibinfo{pages}{100}.
\bibitem[{{Gladman} et~al.(2001){Gladman}, {Kavelaars}, {Petit}, {Morbidelli},
  {Holman} and {Loredo}}]{Gladman2001}
\bibinfo{author}{{Gladman}, B.}, \bibinfo{author}{{Kavelaars}, J.J.},
  \bibinfo{author}{{Petit}, J.M.}, \bibinfo{author}{{Morbidelli}, A.},
  \bibinfo{author}{{Holman}, M.J.}, \bibinfo{author}{{Loredo}, T.},
  \bibinfo{year}{2001}.
\newblock \bibinfo{title}{{The Structure of the Kuiper Belt: Size Distribution
  and Radial Extent}}.
\newblock \bibinfo{journal}{\aj} \bibinfo{volume}{122},
  \bibinfo{pages}{1051--1066}.
\bibitem[{{Gomes} et~al.(2004){Gomes}, {Morbidelli} and {Levison}}]{Gomes2004}
\bibinfo{author}{{Gomes}, R.S.}, \bibinfo{author}{{Morbidelli}, A.},
  \bibinfo{author}{{Levison}, H.F.}, \bibinfo{year}{2004}.
\newblock \bibinfo{title}{{Planetary migration in a planetesimal disk: why did
  Neptune stop at 30 AU?}}
\newblock \bibinfo{journal}{\icarus} \bibinfo{volume}{170},
  \bibinfo{pages}{492--507}.
\bibitem[{{Holman} and {Wisdom}(1993a)}]{HolmanWisdom1993}
\bibinfo{author}{{Holman}, M.J.}, \bibinfo{author}{{Wisdom}, J.},
  \bibinfo{year}{1993}a.
\newblock \bibinfo{title}{{Dynamical Stability in the Outer Solar System and
  the Delivery of Short Period Comets}}.
\newblock \bibinfo{journal}{\aj} \bibinfo{volume}{105}, \bibinfo{pages}{1987}.
\bibitem[{{Holman} and {Wisdom}(1993b)}]{Holman1993}
\bibinfo{author}{{Holman}, M.J.}, \bibinfo{author}{{Wisdom}, J.},
  \bibinfo{year}{1993}b.
\newblock \bibinfo{title}{{Dynamics of the Kuiper Belt}}, in:
  \bibinfo{booktitle}{AAS/Division for Planetary Sciences Meeting Abstracts
  \#25}, p. \bibinfo{pages}{10.03}.
\bibitem[{Housen et~al.(1983)Housen, Schmidt and Holsapple}]{Housen1983}
\bibinfo{author}{Housen, K.R.}, \bibinfo{author}{Schmidt, R.M.},
  \bibinfo{author}{Holsapple, K.A.}, \bibinfo{year}{1983}.
\newblock \bibinfo{title}{Crater ejecta scaling laws: Fundamental forms based
  on dimensional analysis}.
\newblock \bibinfo{journal}{Journal of Geophysical Research: Solid Earth}
  \bibinfo{volume}{88}, \bibinfo{pages}{2485--2499}.
\bibitem[{{Kenyon} and {Bromley}(2004)}]{KenyonBromley2004}
\bibinfo{author}{{Kenyon}, S.J.}, \bibinfo{author}{{Bromley}, B.C.},
  \bibinfo{year}{2004}.
\newblock \bibinfo{title}{{The Size Distribution of Kuiper Belt Objects}}.
\newblock \bibinfo{journal}{\aj} \bibinfo{volume}{128},
  \bibinfo{pages}{1916--1926}.
\bibitem[{{Kirchoff} et~al.(2018){Kirchoff}, {Bierhaus}, {Dones}, {Robbins},
  {Singer}, {Wagner} and {Zahnle}}]{Kirchoff2018}
\bibinfo{author}{{Kirchoff}, M.R.}, \bibinfo{author}{{Bierhaus}, E.B.},
  \bibinfo{author}{{Dones}, L.}, \bibinfo{author}{{Robbins}, S.J.},
  \bibinfo{author}{{Singer}, K.N.}, \bibinfo{author}{{Wagner}, R.J.},
  \bibinfo{author}{{Zahnle}, K.J.}, \bibinfo{year}{2018}.
\newblock \bibinfo{title}{{Cratering Histories in the Saturnian System}}, in:
  \bibinfo{editor}{{Schenk}, P.M.}, \bibinfo{editor}{{Clark}, R.N.},
  \bibinfo{editor}{{Howett}, C.J.A.}, \bibinfo{editor}{{Verbiscer}, A.J.},
  \bibinfo{editor}{{Waite}, J.H.} (Eds.), \bibinfo{booktitle}{Enceladus and the
  Icy Moons of Saturn}, p. \bibinfo{pages}{267}.
\bibitem[{{Kirchoff} and {Schenk}(2009)}]{Kirchoff2009}
\bibinfo{author}{{Kirchoff}, M.R.}, \bibinfo{author}{{Schenk}, P.},
  \bibinfo{year}{2009}.
\newblock \bibinfo{title}{{Crater modification and geologic activity in
  Enceladus' heavily cratered plains: Evidence from the impact crater
  distribution}}.
\newblock \bibinfo{journal}{\icarus} \bibinfo{volume}{202},
  \bibinfo{pages}{656--668}.
\bibitem[{{Kirchoff} and {Schenk}(2010)}]{Kirchoff2010}
\bibinfo{author}{{Kirchoff}, M.R.}, \bibinfo{author}{{Schenk}, P.},
  \bibinfo{year}{2010}.
\newblock \bibinfo{title}{{Impact cratering records of the mid-sized, icy
  saturnian satellites}}.
\newblock \bibinfo{journal}{\icarus} \bibinfo{volume}{206},
  \bibinfo{pages}{485--497}.
\bibitem[{{Lainey} et~al.(2020){Lainey}, {Casajus}, {Fuller}, {Zannoni},
  {Tortora}, {Cooper}, {Murray}, {Modenini}, {Park}, {Robert} and
  {Zhang}}]{Lainey2020}
\bibinfo{author}{{Lainey}, V.}, \bibinfo{author}{{Casajus}, L.G.},
  \bibinfo{author}{{Fuller}, J.}, \bibinfo{author}{{Zannoni}, M.},
  \bibinfo{author}{{Tortora}, P.}, \bibinfo{author}{{Cooper}, N.},
  \bibinfo{author}{{Murray}, C.}, \bibinfo{author}{{Modenini}, D.},
  \bibinfo{author}{{Park}, R.S.}, \bibinfo{author}{{Robert}, V.},
  \bibinfo{author}{{Zhang}, Q.}, \bibinfo{year}{2020}.
\newblock \bibinfo{title}{{Resonance locking in giant planets indicated by the
  rapid orbital expansion of Titan}}.
\newblock \bibinfo{journal}{Nature Astronomy} .
\bibitem[{{Lamy} et~al.(2004){Lamy}, {Toth}, {Fernandez} and
  {Weaver}}]{Lamy2004}
\bibinfo{author}{{Lamy}, P.L.}, \bibinfo{author}{{Toth}, I.},
  \bibinfo{author}{{Fernandez}, Y.R.}, \bibinfo{author}{{Weaver}, H.A.},
  \bibinfo{year}{2004}.
\newblock \bibinfo{title}{{The sizes, shapes, albedos, and colors of cometary
  nuclei}}, in: \bibinfo{editor}{{Festou}, M.C.}, \bibinfo{editor}{{Keller},
  H.U.}, \bibinfo{editor}{{Weaver}, H.A.} (Eds.), \bibinfo{booktitle}{Comets
  II}, p. \bibinfo{pages}{223}.
\bibitem[{{Lawler} et~al.(2018){Lawler}, {Shankman}, {Kavelaars},
  {Alexandersen}, {Bannister}, {Chen}, {Gladman}, {Fraser}, {Gwyn}, {Kaib},
  {Petit} and {Volk}}]{Lawler2018}
\bibinfo{author}{{Lawler}, S.M.}, \bibinfo{author}{{Shankman}, C.},
  \bibinfo{author}{{Kavelaars}, J.J.}, \bibinfo{author}{{Alexandersen}, M.},
  \bibinfo{author}{{Bannister}, M.T.}, \bibinfo{author}{{Chen}, Y.T.},
  \bibinfo{author}{{Gladman}, B.}, \bibinfo{author}{{Fraser}, W.C.},
  \bibinfo{author}{{Gwyn}, S.}, \bibinfo{author}{{Kaib}, N.},
  \bibinfo{author}{{Petit}, J.M.}, \bibinfo{author}{{Volk}, K.},
  \bibinfo{year}{2018}.
\newblock \bibinfo{title}{{OSSOS. VIII. The Transition between Two Size
  Distribution Slopes in the Scattering Disk}}.
\newblock \bibinfo{journal}{\aj} \bibinfo{volume}{155}, \bibinfo{pages}{197}.
\bibitem[{{Levison} and {Duncan}(1994)}]{LevisonDuncan1994}
\bibinfo{author}{{Levison}, H.F.}, \bibinfo{author}{{Duncan}, M.J.},
  \bibinfo{year}{1994}.
\newblock \bibinfo{title}{{The long-term dynamical behavior of short-period
  comets}}.
\newblock \bibinfo{journal}{\icarus} \bibinfo{volume}{108},
  \bibinfo{pages}{18--36}.
\bibitem[{{Levison} and {Duncan}(1997)}]{LevisonDuncan1997}
\bibinfo{author}{{Levison}, H.F.}, \bibinfo{author}{{Duncan}, M.J.},
  \bibinfo{year}{1997}.
\newblock \bibinfo{title}{{From the Kuiper Belt to Jupiter-Family Comets: The
  Spatial Distribution of Ecliptic Comets}}.
\newblock \bibinfo{journal}{\icarus} \bibinfo{volume}{127},
  \bibinfo{pages}{13--32}.
\bibitem[{{Levison} et~al.(2006){Levison}, {Duncan}, {Dones} and
  {Gladman}}]{Levison2006}
\bibinfo{author}{{Levison}, H.F.}, \bibinfo{author}{{Duncan}, M.J.},
  \bibinfo{author}{{Dones}, L.}, \bibinfo{author}{{Gladman}, B.J.},
  \bibinfo{year}{2006}.
\newblock \bibinfo{title}{{The scattered disk as a source of Halley-type
  comets}}.
\newblock \bibinfo{journal}{\icarus} \bibinfo{volume}{184},
  \bibinfo{pages}{619--633}.
\bibitem[{{Levison} et~al.(2000){Levison}, {Duncan}, {Zahnle}, {Holman} and
  {Dones}}]{Levison2000a}
\bibinfo{author}{{Levison}, H.F.}, \bibinfo{author}{{Duncan}, M.J.},
  \bibinfo{author}{{Zahnle}, K.}, \bibinfo{author}{{Holman}, M.},
  \bibinfo{author}{{Dones}, L.}, \bibinfo{year}{2000}.
\newblock \bibinfo{title}{{NOTE: Planetary Impact Rates from Ecliptic Comets}}.
\newblock \bibinfo{journal}{\icarus} \bibinfo{volume}{143},
  \bibinfo{pages}{415--420}.
\bibitem[{{Levison} et~al.(2011){Levison}, {Morbidelli}, {Tsiganis},
  {Nesvorn{\'y}} and {Gomes}}]{Levison2011}
\bibinfo{author}{{Levison}, H.F.}, \bibinfo{author}{{Morbidelli}, A.},
  \bibinfo{author}{{Tsiganis}, K.}, \bibinfo{author}{{Nesvorn{\'y}}, D.},
  \bibinfo{author}{{Gomes}, R.}, \bibinfo{year}{2011}.
\newblock \bibinfo{title}{{Late Orbital Instabilities in the Outer Planets
  Induced by Interaction with a Self-gravitating Planetesimal Disk}}.
\newblock \bibinfo{journal}{\aj} \bibinfo{volume}{142}, \bibinfo{pages}{152}.
\bibitem[{{Levison} et~al.(2008){Levison}, {Morbidelli}, {Van Laerhoven},
  {Gomes} and {Tsiganis}}]{Levison2008}
\bibinfo{author}{{Levison}, H.F.}, \bibinfo{author}{{Morbidelli}, A.},
  \bibinfo{author}{{Van Laerhoven}, C.}, \bibinfo{author}{{Gomes}, R.},
  \bibinfo{author}{{Tsiganis}, K.}, \bibinfo{year}{2008}.
\newblock \bibinfo{title}{{Origin of the structure of the Kuiper belt during a
  dynamical instability in the orbits of Uranus and Neptune}}.
\newblock \bibinfo{journal}{\icarus} \bibinfo{volume}{196},
  \bibinfo{pages}{258--273}.
\bibitem[{{Mah} and {Brasser}(2019)}]{MahBrasser2019}
\bibinfo{author}{{Mah}, J.}, \bibinfo{author}{{Brasser}, R.},
  \bibinfo{year}{2019}.
\newblock \bibinfo{title}{{The origin of the cratering asymmetry on Triton}}.
\newblock \bibinfo{journal}{\mnras} \bibinfo{volume}{486},
  \bibinfo{pages}{836--842}.
\bibitem[{{Malyshkin} and {Tremaine}(1999)}]{Malyshkin1999}
\bibinfo{author}{{Malyshkin}, L.}, \bibinfo{author}{{Tremaine}, S.},
  \bibinfo{year}{1999}.
\newblock \bibinfo{title}{{The Keplerian Map for the Planar Restricted
  Three-Body Problem as a Model of Comet Evolution}}.
\newblock \bibinfo{journal}{\icarus} \bibinfo{volume}{141},
  \bibinfo{pages}{341--353}.
\bibitem[{{Mojzsis} et~al.(2019){Mojzsis}, {Brasser}, {Kelly}, {Abramov} and
  {Werner}}]{Mojzsis2019}
\bibinfo{author}{{Mojzsis}, S.J.}, \bibinfo{author}{{Brasser}, R.},
  \bibinfo{author}{{Kelly}, N.M.}, \bibinfo{author}{{Abramov}, O.},
  \bibinfo{author}{{Werner}, S.C.}, \bibinfo{year}{2019}.
\newblock \bibinfo{title}{{Onset of Giant Planet Migration before 4480 Million
  Years Ago}}.
\newblock \bibinfo{journal}{\apj} \bibinfo{volume}{881}, \bibinfo{pages}{44}.
\bibitem[{{Morfill} et~al.(1983){Morfill}, {Fechtig}, {Gruen} and
  {Goertz}}]{Morfill1983}
\bibinfo{author}{{Morfill}, G.E.}, \bibinfo{author}{{Fechtig}, H.},
  \bibinfo{author}{{Gruen}, E.}, \bibinfo{author}{{Goertz}, C.K.},
  \bibinfo{year}{1983}.
\newblock \bibinfo{title}{{Some consequences of meteoroid impacts on Saturn's
  rings}}.
\newblock \bibinfo{journal}{\icarus} \bibinfo{volume}{55},
  \bibinfo{pages}{439--447}.
\bibitem[{{Nakajima} et~al.(2019){Nakajima}, {Ida}, {Kimura} and
  {Brasser}}]{Nakajima2019}
\bibinfo{author}{{Nakajima}, A.}, \bibinfo{author}{{Ida}, S.},
  \bibinfo{author}{{Kimura}, J.}, \bibinfo{author}{{Brasser}, R.},
  \bibinfo{year}{2019}.
\newblock \bibinfo{title}{{Orbital evolution of Saturn's mid-sized moons and
  the tidal heating of Enceladus}}.
\newblock \bibinfo{journal}{\icarus} \bibinfo{volume}{317},
  \bibinfo{pages}{570--582}.
\bibitem[{{Nesvorn{\'y}}(2015)}]{Nesvorny2015b}
\bibinfo{author}{{Nesvorn{\'y}}, D.}, \bibinfo{year}{2015}.
\newblock \bibinfo{title}{{Jumping Neptune Can Explain the Kuiper Belt
  Kernel}}.
\newblock \bibinfo{journal}{\aj} \bibinfo{volume}{150}, \bibinfo{pages}{68}.
\bibitem[{Nesvorn\'{y}(2018)}]{Nesvorny2018}
\bibinfo{author}{Nesvorn\'{y}, D.}, \bibinfo{year}{2018}.
\newblock \bibinfo{title}{Dynamical evolution of the early solar system}.
\newblock \bibinfo{journal}{Annual Review of Astronomy and Astrophysics}
  \bibinfo{volume}{56}, \bibinfo{pages}{137--174}.
\bibitem[{{Nesvorn{\'y}} et~al.(2003){Nesvorn{\'y}}, {Alvarellos}, {Dones} and
  {Levison}}]{Nesvorny2003}
\bibinfo{author}{{Nesvorn{\'y}}, D.}, \bibinfo{author}{{Alvarellos}, J.L.A.},
  \bibinfo{author}{{Dones}, L.}, \bibinfo{author}{{Levison}, H.F.},
  \bibinfo{year}{2003}.
\newblock \bibinfo{title}{{Orbital and Collisional Evolution of the Irregular
  Satellites}}.
\newblock \bibinfo{journal}{\aj} \bibinfo{volume}{126},
  \bibinfo{pages}{398--429}.
\bibitem[{{Nesvorn{\'y}} and {Morbidelli}(2012)}]{Nesvorny2012}
\bibinfo{author}{{Nesvorn{\'y}}, D.}, \bibinfo{author}{{Morbidelli}, A.},
  \bibinfo{year}{2012}.
\newblock \bibinfo{title}{{Statistical Study of the Early Solar System's
  Instability with Four, Five, and Six Giant Planets}}.
\newblock \bibinfo{journal}{\aj} \bibinfo{volume}{144}, \bibinfo{pages}{117}.
\bibitem[{{Nesvorn{\'y}} and {Vokrouhlick{\'y}}(2016)}]{Nesvorny2016b}
\bibinfo{author}{{Nesvorn{\'y}}, D.}, \bibinfo{author}{{Vokrouhlick{\'y}}, D.},
  \bibinfo{year}{2016}.
\newblock \bibinfo{title}{{Neptune's Orbital Migration Was Grainy, Not
  Smooth}}.
\newblock \bibinfo{journal}{\apj} \bibinfo{volume}{825}, \bibinfo{pages}{94}.
\bibitem[{{Nesvorn{\'y}} et~al.(2013){Nesvorn{\'y}}, {Vokrouhlick{\'y}} and
  {Morbidelli}}]{Nesvorny2013}
\bibinfo{author}{{Nesvorn{\'y}}, D.}, \bibinfo{author}{{Vokrouhlick{\'y}}, D.},
  \bibinfo{author}{{Morbidelli}, A.}, \bibinfo{year}{2013}.
\newblock \bibinfo{title}{{Capture of Trojans by Jumping Jupiter}}.
\newblock \bibinfo{journal}{\apj} \bibinfo{volume}{768}, \bibinfo{pages}{45}.
\bibitem[{{Nesvorn{\'y}} et~al.(2016){Nesvorn{\'y}}, {Vokrouhlick{\'y}} and
  {Roig}}]{Nesvorny2016a}
\bibinfo{author}{{Nesvorn{\'y}}, D.}, \bibinfo{author}{{Vokrouhlick{\'y}}, D.},
  \bibinfo{author}{{Roig}, F.}, \bibinfo{year}{2016}.
\newblock \bibinfo{title}{{The Orbital Distribution of Trans-Neptunian Objects
  Beyond 50 au}}.
\newblock \bibinfo{journal}{\apjl} \bibinfo{volume}{827}, \bibinfo{pages}{L35}.
\bibitem[{{Nesvorn{\'y}} et~al.(2019){Nesvorn{\'y}}, {Vokrouhlick{\'y}},
  {Stern}, {Davidsson}, {Bannister}, {Volk}, {Chen}, {Gladman}, {Kavelaars},
  {Petit}, {Gwyn} and {Alexandersen}}]{Nesvorny2019}
\bibinfo{author}{{Nesvorn{\'y}}, D.}, \bibinfo{author}{{Vokrouhlick{\'y}}, D.},
  \bibinfo{author}{{Stern}, A.S.}, \bibinfo{author}{{Davidsson}, B.},
  \bibinfo{author}{{Bannister}, M.T.}, \bibinfo{author}{{Volk}, K.},
  \bibinfo{author}{{Chen}, Y.T.}, \bibinfo{author}{{Gladman}, B.J.},
  \bibinfo{author}{{Kavelaars}, J.J.}, \bibinfo{author}{{Petit}, J.M.},
  \bibinfo{author}{{Gwyn}, S.D.J.}, \bibinfo{author}{{Alexandersen}, M.},
  \bibinfo{year}{2019}.
\newblock \bibinfo{title}{{OSSOS. XIX. Testing Early Solar System Dynamical
  Models Using OSSOS Centaur Detections}}.
\newblock \bibinfo{journal}{\aj} \bibinfo{volume}{158}, \bibinfo{pages}{132}.
\bibitem[{{Quarles} and {Kaib}(2019)}]{Quarles2019}
\bibinfo{author}{{Quarles}, B.}, \bibinfo{author}{{Kaib}, N.},
  \bibinfo{year}{2019}.
\newblock \bibinfo{title}{{Instabilities in the Early Solar System Due to a
  Self-gravitating Disk}}.
\newblock \bibinfo{journal}{\aj} \bibinfo{volume}{157}, \bibinfo{pages}{67}.
\bibitem[{{Schenk}(1991)}]{Schenk1991}
\bibinfo{author}{{Schenk}, P.M.}, \bibinfo{year}{1991}.
\newblock \bibinfo{title}{{Fluid volcanism on Miranda and Ariel: flow
  morphology and composition.}}
\newblock \bibinfo{journal}{\jgr} \bibinfo{volume}{96},
  \bibinfo{pages}{1887--1906}.
\bibitem[{{Schenk}(2003)}]{Schenk2003}
\bibinfo{author}{{Schenk}, P.M.}, \bibinfo{year}{2003}.
\newblock \bibinfo{title}{{Importance of Target Properties on Planetary Impact
  Craters, Both Simple and Complex}}, in: \bibinfo{editor}{{Herrick}, R.},
  \bibinfo{editor}{{Pierazzo}, E.} (Eds.), \bibinfo{booktitle}{Impact
  Cratering: Bridging the Gap Between Modeling and Observations},
  p.~\bibinfo{pages}{61}.
\bibitem[{{Schenk} and {Zahnle}(2007)}]{Schenk07}
\bibinfo{author}{{Schenk}, P.M.}, \bibinfo{author}{{Zahnle}, K.},
  \bibinfo{year}{2007}.
\newblock \bibinfo{title}{{On the negligible surface age of Triton}}.
\newblock \bibinfo{journal}{\icarus} \bibinfo{volume}{192},
  \bibinfo{pages}{135--149}.
\bibitem[{{Shankman} et~al.(2013){Shankman}, {Gladman}, {Kaib}, {Kavelaars} and
  {Petit}}]{Shankman2013}
\bibinfo{author}{{Shankman}, C.}, \bibinfo{author}{{Gladman}, B.J.},
  \bibinfo{author}{{Kaib}, N.}, \bibinfo{author}{{Kavelaars}, J.J.},
  \bibinfo{author}{{Petit}, J.M.}, \bibinfo{year}{2013}.
\newblock \bibinfo{title}{{A Possible Divot in the Size Distribution of the
  Kuiper Belt's Scattering Objects}}.
\newblock \bibinfo{journal}{\apjl} \bibinfo{volume}{764}, \bibinfo{pages}{L2}.
\bibitem[{{Sheppard} et~al.(2000){Sheppard}, {Jewitt}, {Trujillo}, {Brown} and
  {Ashley}}]{Sheppard00}
\bibinfo{author}{{Sheppard}, S.S.}, \bibinfo{author}{{Jewitt}, D.C.},
  \bibinfo{author}{{Trujillo}, C.A.}, \bibinfo{author}{{Brown}, M.J.I.},
  \bibinfo{author}{{Ashley}, M.C.B.}, \bibinfo{year}{2000}.
\newblock \bibinfo{title}{{A Wide-Field CCD Survey for Centaurs and Kuiper Belt
  Objects}}.
\newblock \bibinfo{journal}{\aj} \bibinfo{volume}{120},
  \bibinfo{pages}{2687--2694}.
\bibitem[{{Singer} et~al.(2019){Singer}, {McKinnon}, {Gladman}, {Greenstreet},
  {Bierhaus}, {Stern}, {Parker}, {Robbins}, {Schenk}, {Grundy}, {Bray},
  {Beyer}, {Binzel}, {Weaver}, {Young}, {Spencer}, {Kavelaars}, {Moore},
  {Zangari}, {Olkin}, {Lauer}, {Lisse}, {Ennico}, {New Horizons Geology}, Team,
  {New Horizons Surface Composition Science Theme Team} and {New Horizons Ralph
  and LORRI Teams}}]{Singer2019}
\bibinfo{author}{{Singer}, K.N.}, \bibinfo{author}{{McKinnon}, W.B.},
  \bibinfo{author}{{Gladman}, B.}, \bibinfo{author}{{Greenstreet}, S.},
  \bibinfo{author}{{Bierhaus}, E.B.}, \bibinfo{author}{{Stern}, S.A.},
  \bibinfo{author}{{Parker}, A.H.}, \bibinfo{author}{{Robbins}, S.J.},
  \bibinfo{author}{{Schenk}, P.M.}, \bibinfo{author}{{Grundy}, W.M.},
  \bibinfo{author}{{Bray}, V.J.}, \bibinfo{author}{{Beyer}, R.A.},
  \bibinfo{author}{{Binzel}, R.P.}, \bibinfo{author}{{Weaver}, H.A.},
  \bibinfo{author}{{Young}, L.A.}, \bibinfo{author}{{Spencer}, J.R.},
  \bibinfo{author}{{Kavelaars}, J.J.}, \bibinfo{author}{{Moore}, J.M.},
  \bibinfo{author}{{Zangari}, A.M.}, \bibinfo{author}{{Olkin}, C.B.},
  \bibinfo{author}{{Lauer}, T.R.}, \bibinfo{author}{{Lisse}, C.M.},
  \bibinfo{author}{{Ennico}, K.}, \bibinfo{author}{{New Horizons Geology}, G.},
  \bibinfo{author}{Team, I.S.T.}, \bibinfo{author}{{New Horizons Surface
  Composition Science Theme Team}}, \bibinfo{author}{{New Horizons Ralph and
  LORRI Teams}}, \bibinfo{year}{2019}.
\newblock \bibinfo{title}{{Impact craters on Pluto and Charon indicate a
  deficit of small Kuiper belt objects}}.
\newblock \bibinfo{journal}{Science} \bibinfo{volume}{363},
  \bibinfo{pages}{955--959}.
\bibitem[{{Snodgrass} et~al.(2011){Snodgrass}, {Fitzsimmons}, {Lowry} and
  {Weissman}}]{Snodgrass2011}
\bibinfo{author}{{Snodgrass}, C.}, \bibinfo{author}{{Fitzsimmons}, A.},
  \bibinfo{author}{{Lowry}, S.C.}, \bibinfo{author}{{Weissman}, P.},
  \bibinfo{year}{2011}.
\newblock \bibinfo{title}{{The size distribution of Jupiter Family comet
  nuclei}}.
\newblock \bibinfo{journal}{\mnras} \bibinfo{volume}{414},
  \bibinfo{pages}{458--469}.
\bibitem[{{Tiscareno} and {Malhotra}(2003)}]{Tiscareno2003}
\bibinfo{author}{{Tiscareno}, M.S.}, \bibinfo{author}{{Malhotra}, R.},
  \bibinfo{year}{2003}.
\newblock \bibinfo{title}{{The Dynamics of Known Centaurs}}.
\newblock \bibinfo{journal}{\aj} \bibinfo{volume}{126},
  \bibinfo{pages}{3122--3131}.
\bibitem[{{Trujillo} et~al.(2001){Trujillo}, {Luu}, {Bosh} and
  {Elliot}}]{Trujillo01}
\bibinfo{author}{{Trujillo}, C.A.}, \bibinfo{author}{{Luu}, J.X.},
  \bibinfo{author}{{Bosh}, A.S.}, \bibinfo{author}{{Elliot}, J.L.},
  \bibinfo{year}{2001}.
\newblock \bibinfo{title}{{Large Bodies in the Kuiper Belt}}.
\newblock \bibinfo{journal}{\aj} \bibinfo{volume}{122},
  \bibinfo{pages}{2740--2748}.
\bibitem[{{Vokrouhlick{\'y}} et~al.(2008){Vokrouhlick{\'y}}, {Nesvorn{\'y}} and
  {Levison}}]{Vokrouhlicky2008}
\bibinfo{author}{{Vokrouhlick{\'y}}, D.}, \bibinfo{author}{{Nesvorn{\'y}}, D.},
  \bibinfo{author}{{Levison}, H.F.}, \bibinfo{year}{2008}.
\newblock \bibinfo{title}{{Irregular Satellite Capture by Exchange Reactions}}.
\newblock \bibinfo{journal}{\aj} \bibinfo{volume}{136},
  \bibinfo{pages}{1463--1476}.
\bibitem[{{Volk} and {Malhotra}(2008)}]{VolkMalhotra2008}
\bibinfo{author}{{Volk}, K.}, \bibinfo{author}{{Malhotra}, R.},
  \bibinfo{year}{2008}.
\newblock \bibinfo{title}{{The Scattered Disk as the Source of the Jupiter
  Family Comets}}.
\newblock \bibinfo{journal}{\apj} \bibinfo{volume}{687},
  \bibinfo{pages}{714--725}.
\bibitem[{{Wong} et~al.(2019){Wong}, {Brasser} and {Werner}}]{Wong2019}
\bibinfo{author}{{Wong}, E.W.}, \bibinfo{author}{{Brasser}, R.},
  \bibinfo{author}{{Werner}, S.C.}, \bibinfo{year}{2019}.
\newblock \bibinfo{title}{{Impact bombardment on the regular satellites of
  Jupiter and Uranus during an episode of giant planet migration}}.
\newblock \bibinfo{journal}{Earth and Planetary Science Letters}
  \bibinfo{volume}{506}, \bibinfo{pages}{407--416}.
\bibitem[{{Wong} et~al.(2021){Wong}, {Brasser} and {Werner}}]{Wong2020}
\bibinfo{author}{{Wong}, E.W.}, \bibinfo{author}{{Brasser}, R.},
  \bibinfo{author}{{Werner}, S.C.}, \bibinfo{year}{2021}.
\newblock \bibinfo{title}{{Early impact chronology of the icy regular
  satellites of the outer solar system}}.
\newblock \bibinfo{journal}{\icarus} \bibinfo{volume}{358},
  \bibinfo{pages}{114184}.
\bibitem[{{Wong} and {Brown}(2015)}]{WongBrown2015}
\bibinfo{author}{{Wong}, I.}, \bibinfo{author}{{Brown}, M.E.},
  \bibinfo{year}{2015}.
\newblock \bibinfo{title}{{The Color-Magnitude Distribution of Small Jupiter
  Trojans}}.
\newblock \bibinfo{journal}{\aj} \bibinfo{volume}{150}, \bibinfo{pages}{174}.
\bibitem[{{Yoshida} and {Terai}(2017)}]{YoshidaTerai2017}
\bibinfo{author}{{Yoshida}, F.}, \bibinfo{author}{{Terai}, T.},
  \bibinfo{year}{2017}.
\newblock \bibinfo{title}{{Small Jupiter Trojans Survey with the Subaru/Hyper
  Suprime-Cam}}.
\newblock \bibinfo{journal}{\aj} \bibinfo{volume}{154}, \bibinfo{pages}{71}.
\bibitem[{{Zahnle} et~al.(1998){Zahnle}, {Dones} and {Levison}}]{Zahnle1998}
\bibinfo{author}{{Zahnle}, K.}, \bibinfo{author}{{Dones}, L.},
  \bibinfo{author}{{Levison}, H.F.}, \bibinfo{year}{1998}.
\newblock \bibinfo{title}{{Cratering Rates on the Galilean Satellites}}.
\newblock \bibinfo{journal}{\icarus} \bibinfo{volume}{136},
  \bibinfo{pages}{202--222}.
\bibitem[{{Zahnle} et~al.(2003){Zahnle}, {Schenk}, {Levison} and
  {Dones}}]{Zahnle2003}
\bibinfo{author}{{Zahnle}, K.}, \bibinfo{author}{{Schenk}, P.},
  \bibinfo{author}{{Levison}, H.}, \bibinfo{author}{{Dones}, L.},
  \bibinfo{year}{2003}.
\newblock \bibinfo{title}{{Cratering rates in the outer Solar System}}.
\newblock \bibinfo{journal}{\icarus} \bibinfo{volume}{163},
  \bibinfo{pages}{263--289}.

\end{thebibliography}
\clearpage

\appendix
\renewcommand{\thefigure}{A\arabic{figure}}
\setcounter{figure}{0}

\section{Alternative crater production function or size-frequency distribution of the planetesimals}
We investigated the difference and uncertainties in the estimated surface ages of Saturn's regular icy satellites with different isochrons. To achieve this, we utilised an alternate crater production function or size-frequency distribution of the scattered disc objects. Compare to the corresponding figure in Fig.~5, each figure only varies one parameter.\\

\begin{figure*}
	\centering
	\begin{subfigure}{.45\textwidth}
		\caption{Mimas cratered plain}
		\centering
		\includegraphics[width=\linewidth]{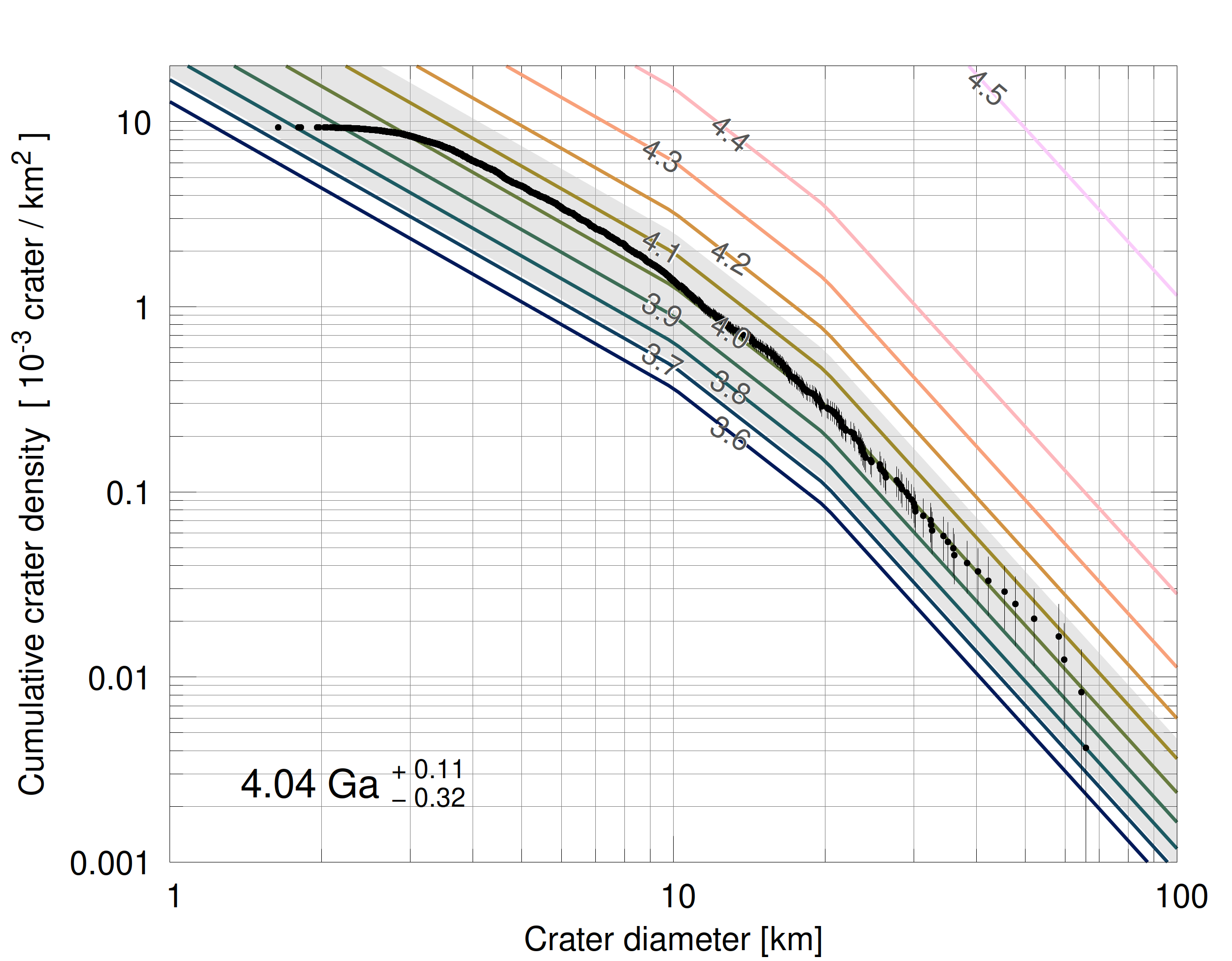}  
	\end{subfigure}
	\begin{subfigure}{.45\textwidth}
		\caption{Enceladus mid-latitude cratered plain}
		\centering
		\includegraphics[width=\linewidth]{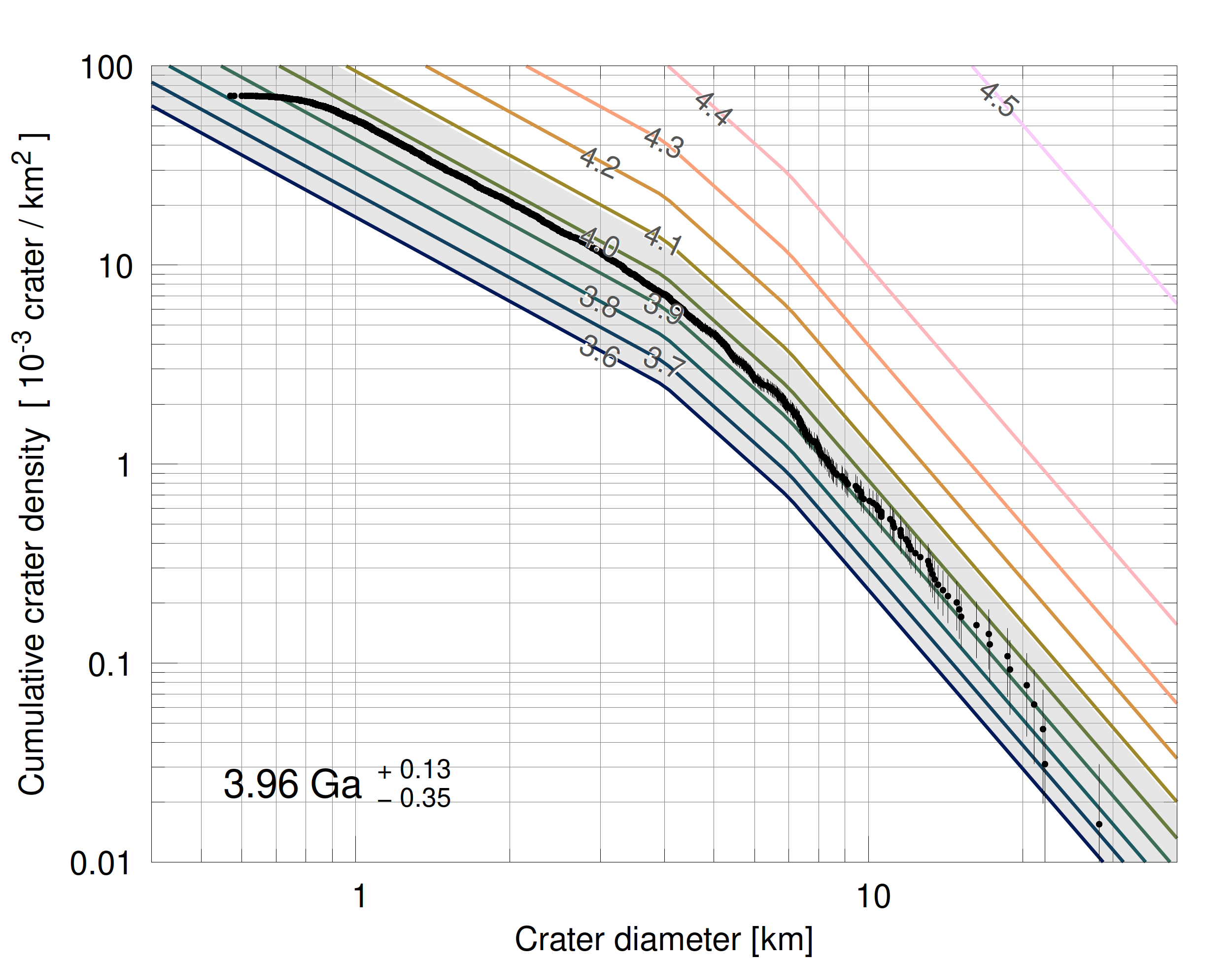}  
	\end{subfigure}
	\newline
	\par\bigskip
	\begin{subfigure}{.45\textwidth}
		\caption{Tethys cratered plain}
		\centering
		\includegraphics[width=\linewidth]{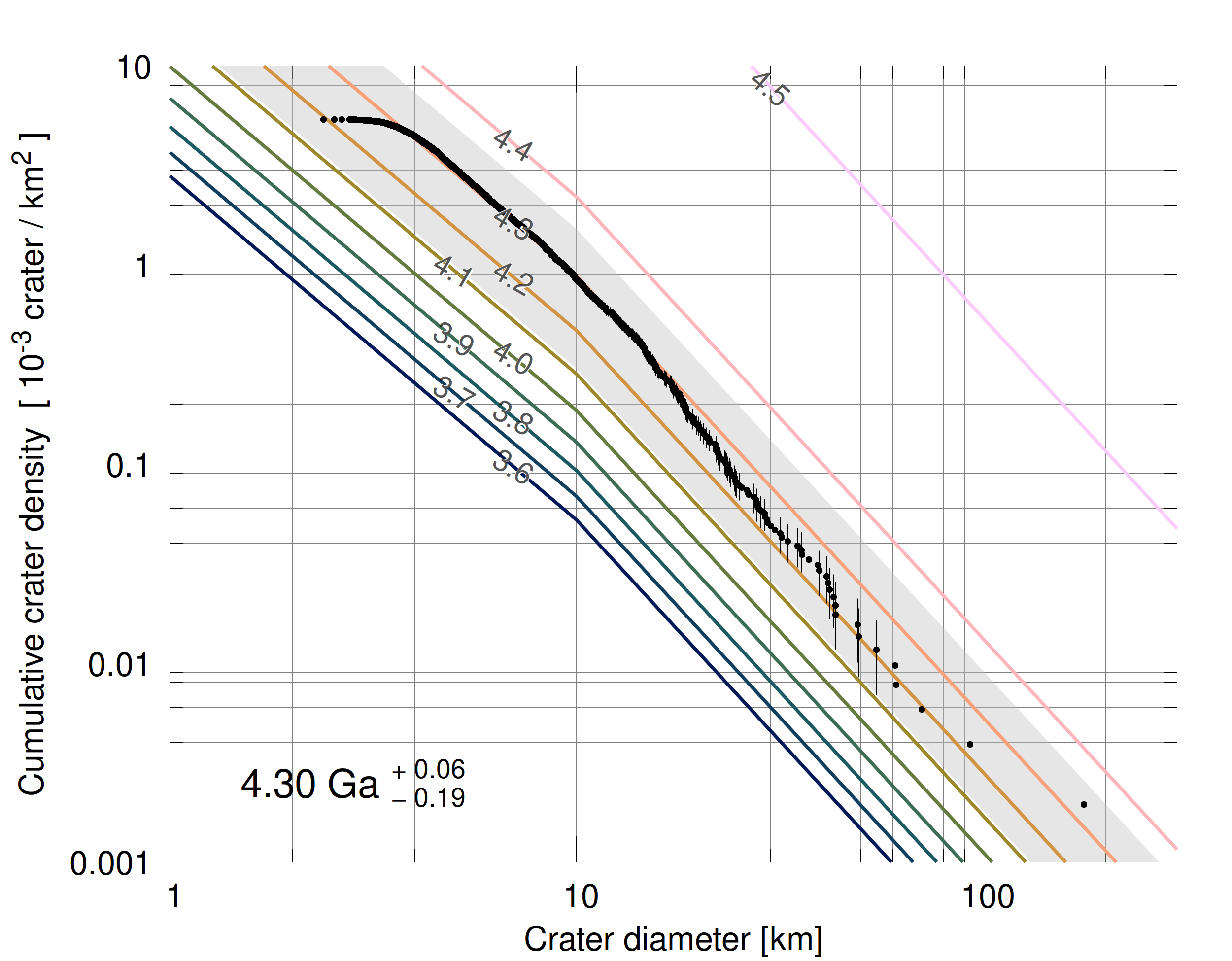}  
	\end{subfigure}
	\begin{subfigure}{.45\textwidth}
		\caption{Dione cratered plain}
		\centering
		\includegraphics[width=\linewidth]{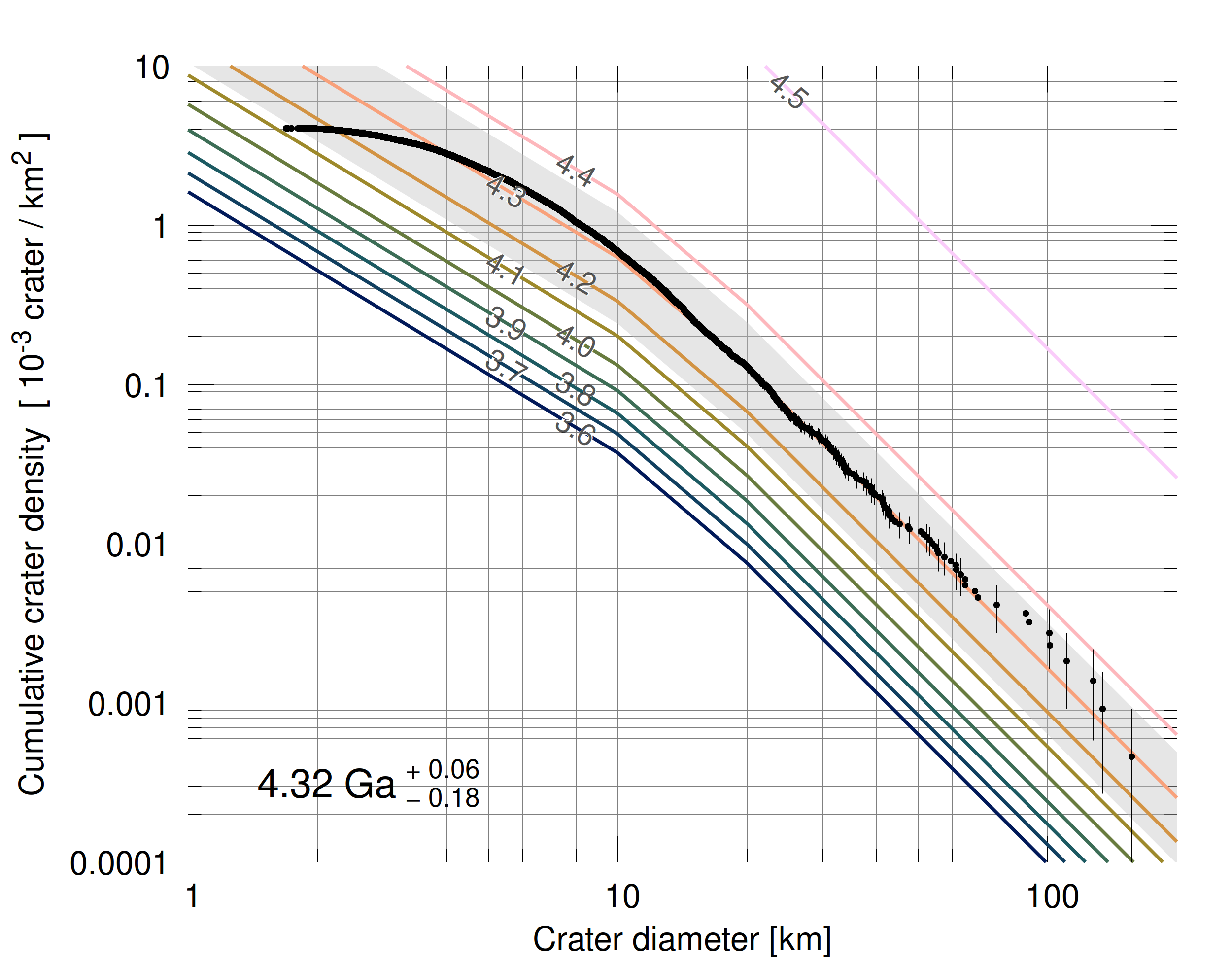}  
	\end{subfigure}
	\newline
	\par\bigskip
	\begin{subfigure}{.45\textwidth}
		\caption{Rhea cratered plain}
		\centering
		\includegraphics[width=\linewidth]{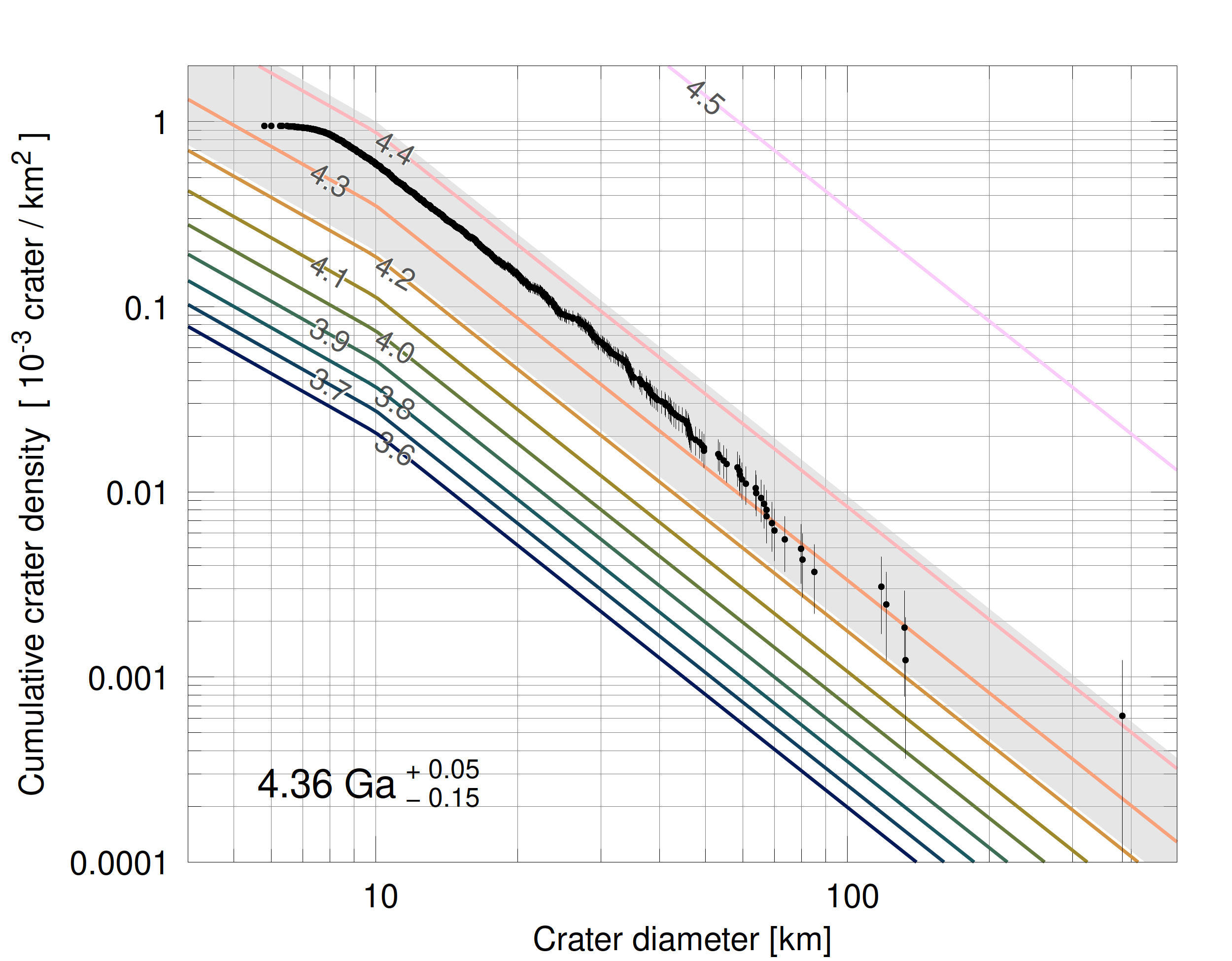}  
	\end{subfigure}
	\begin{subfigure}{.45\textwidth}
		\captionsetup{width=5.5cm}
		\caption*{\scriptsize{Figure A1: Same presentation as in Fig.~5. It shows the best-fitted ages estimated using an impactor cumulative slope of -2.25 from \citet{Schenk07}. The crater production function are referenced from \citet{Kirchoff2009}. Other parameters were kept the same as in the main paper. A steeper cumulative slope led to a higher initial number of scattered disc objects, resulting in higher expected crater densities and younger estimated surface ages.}}
		\label{fig:isochron_sz7}
		\centering
		\includegraphics[width=\linewidth]{dummy.png}  
	\end{subfigure}
	\caption*{}
\end{figure*}

\begin{figure*}
	\begin{subfigure}{.45\textwidth}
		\caption{Mimas cratered plain}
		\centering
		\includegraphics[width=\linewidth]{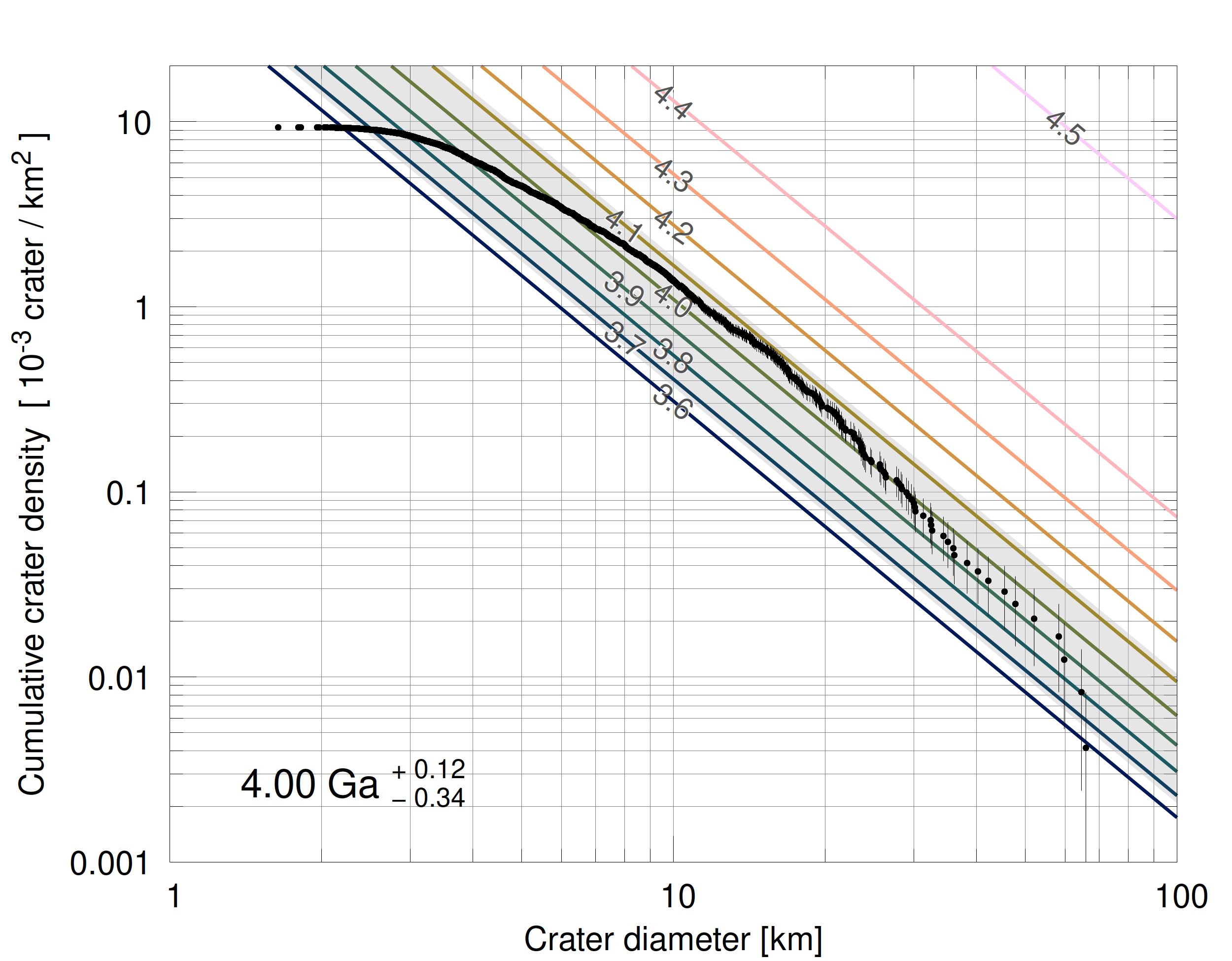}  
	\end{subfigure}
	\begin{subfigure}{.45\textwidth}
		\caption{Enceladus mid-latitude cratered plain}
		\centering
		\includegraphics[width=\linewidth]{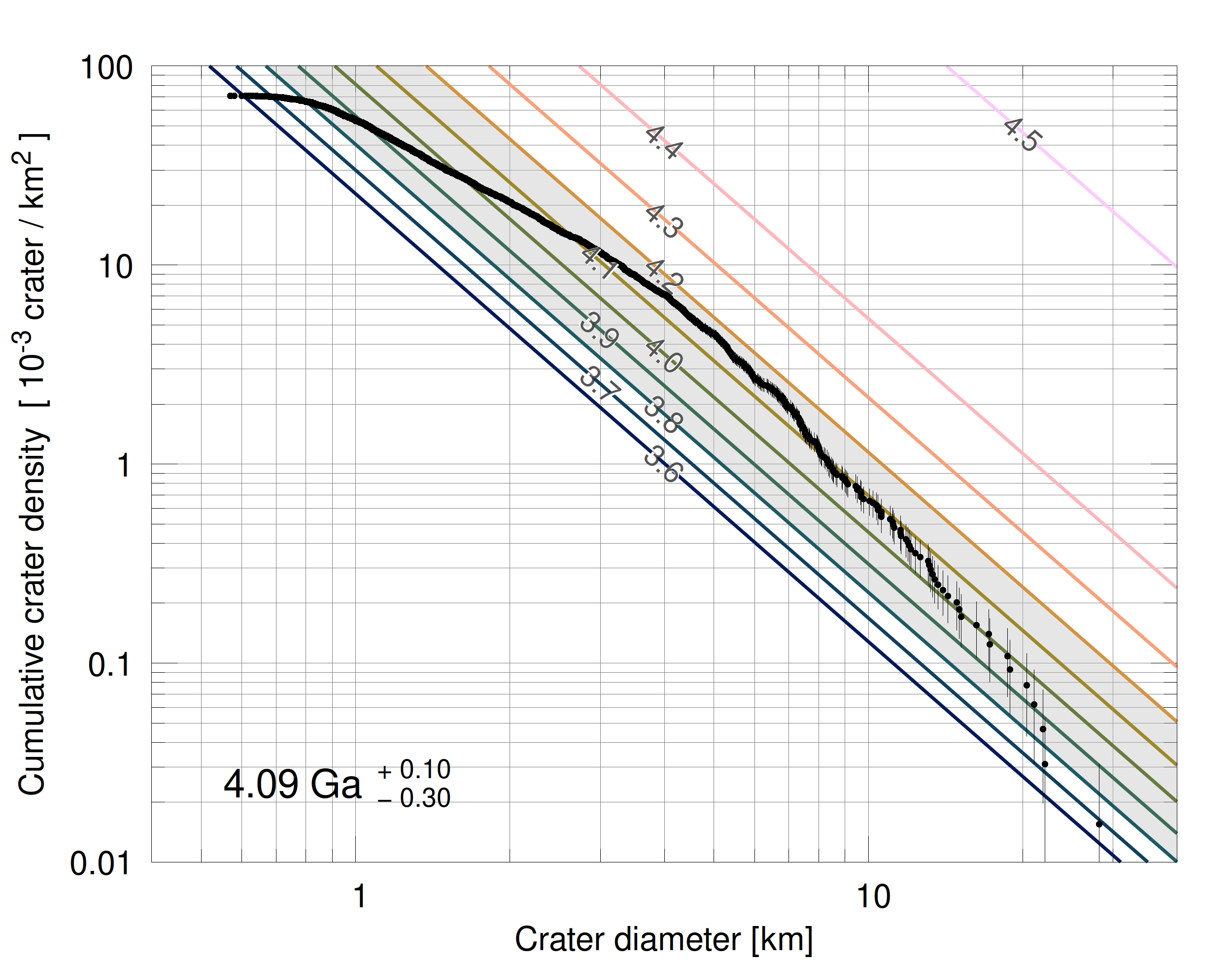}  
	\end{subfigure}
	\newline
	\par\bigskip
	\begin{subfigure}{.45\textwidth}
		\caption{Tethys cratered plain}
		\centering
		\includegraphics[width=\linewidth]{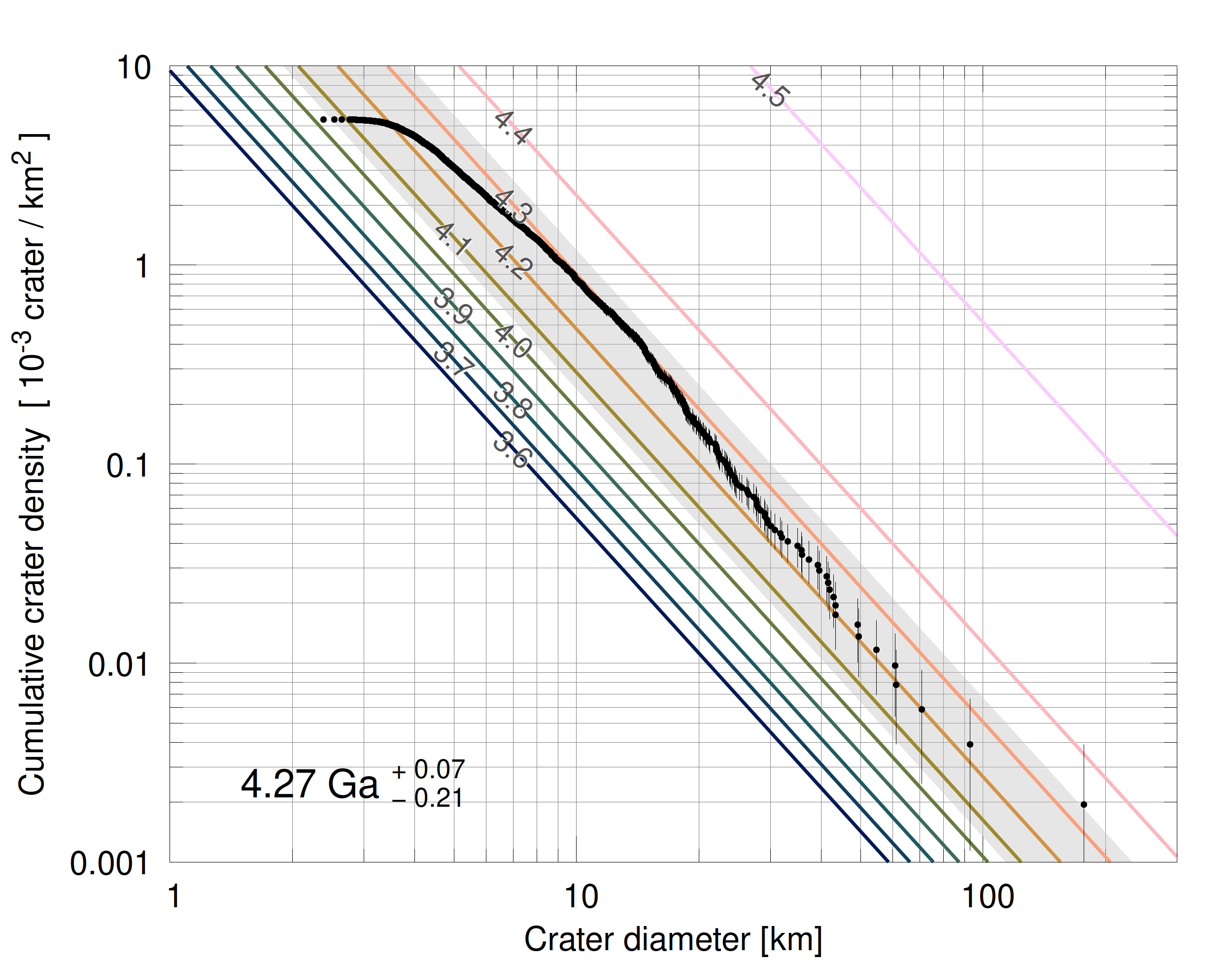}  
	\end{subfigure}
	\begin{subfigure}{.45\textwidth}
		\caption{Dione cratered plain}
		\centering
		\includegraphics[width=\linewidth]{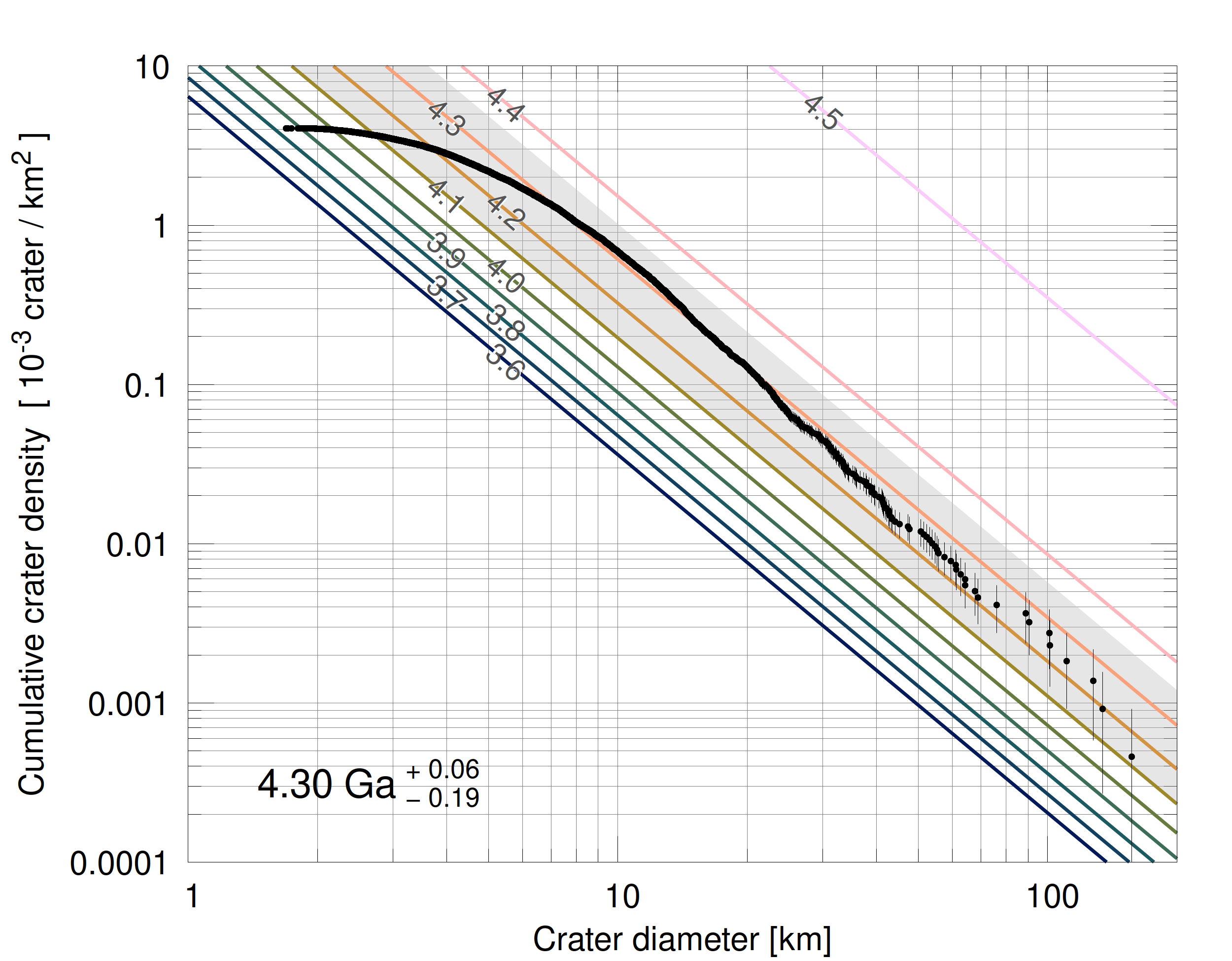}  
	\end{subfigure}
	\newline
	\par\bigskip
	\begin{subfigure}{.45\textwidth}
		\caption{Rhea cratered plain}
		\centering
		\includegraphics[width=\linewidth]{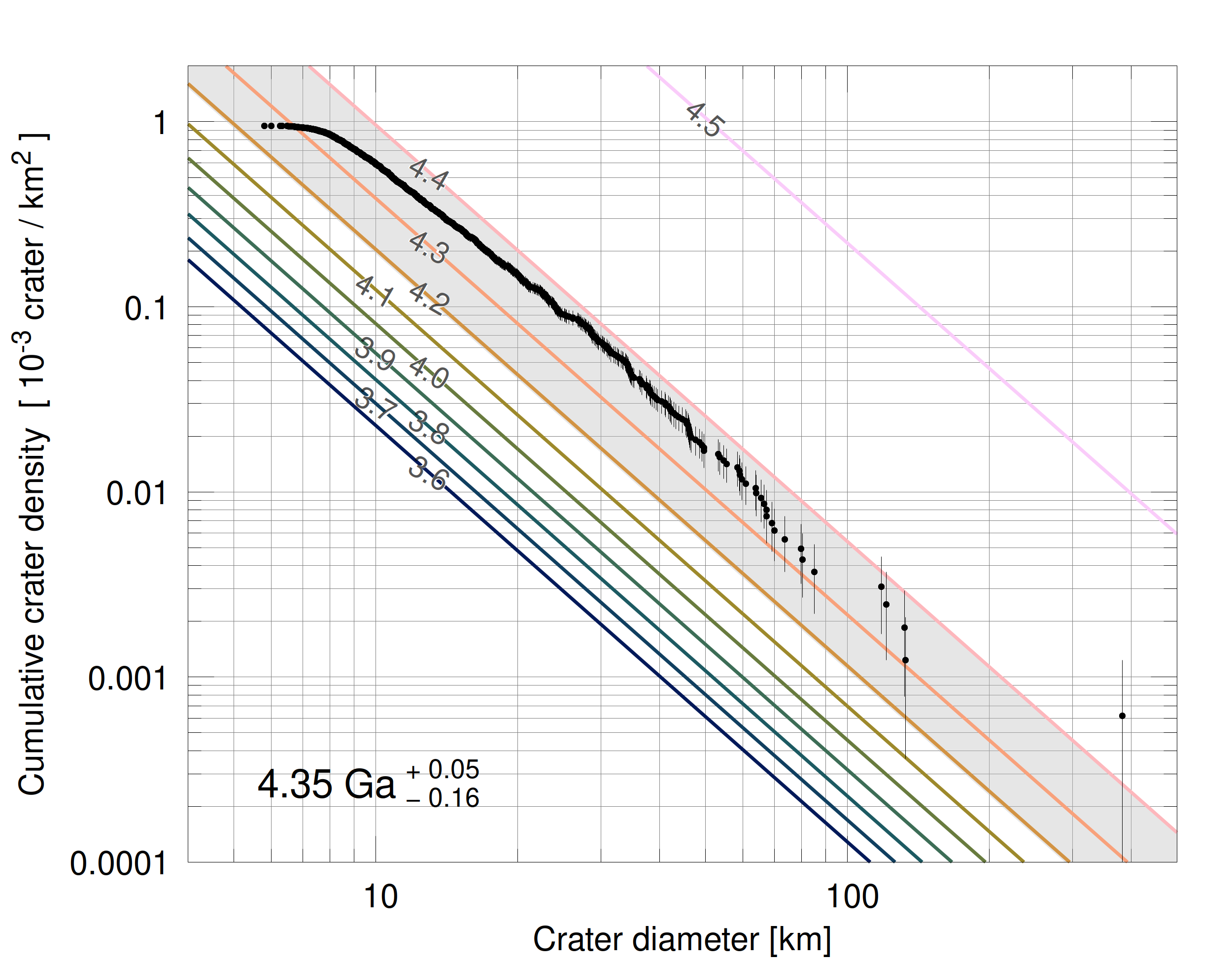}  
	\end{subfigure}
	\begin{subfigure}{.45\textwidth}
		\captionsetup{width=5.5cm}
		\caption*{\scriptsize{Figure A2: Same presentation as in Fig.~5. 
				It demonstrates the effect of using single power law function of -2.25 \citep{Schenk07} for the size-frequency distribution of both the scattered disc objects and the crater production function. Other parameters were kept the same.}} 
	\label{fig:isochron_225}
	\centering
	\includegraphics[width=\linewidth]{dummy.png}  
\end{subfigure}
\caption*{}
\end{figure*}

\begin{figure*}
\centering
\resizebox{\hsize}{!}{\includegraphics{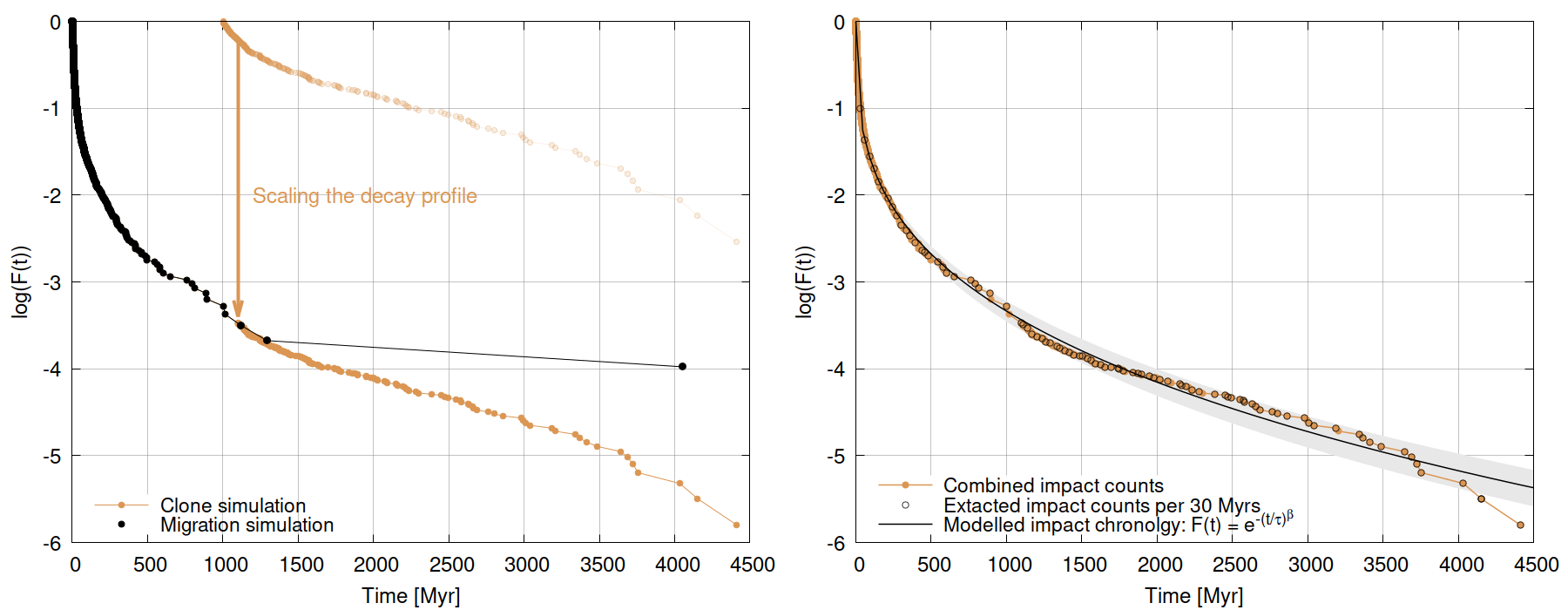}} 
\caption{\label{fig:fit_step} {\it Left}: A higher impact resolution and more extended impact decay profile is created by scaling the clone simulation's impact decay profile, which is further linked to the first 1100 Myr impact decay profile from the migration simulation. {\it Right}: the combined decay profile with the selection of data points per 30 Myr. Overplotted with the fitting of the impact chronology as a curved line in Weibull distribution. }
\end{figure*}

\clearpage
\section{Role of the clone simulations}
\begin{figure*}
\centering
\resizebox{\hsize}{!}{\includegraphics{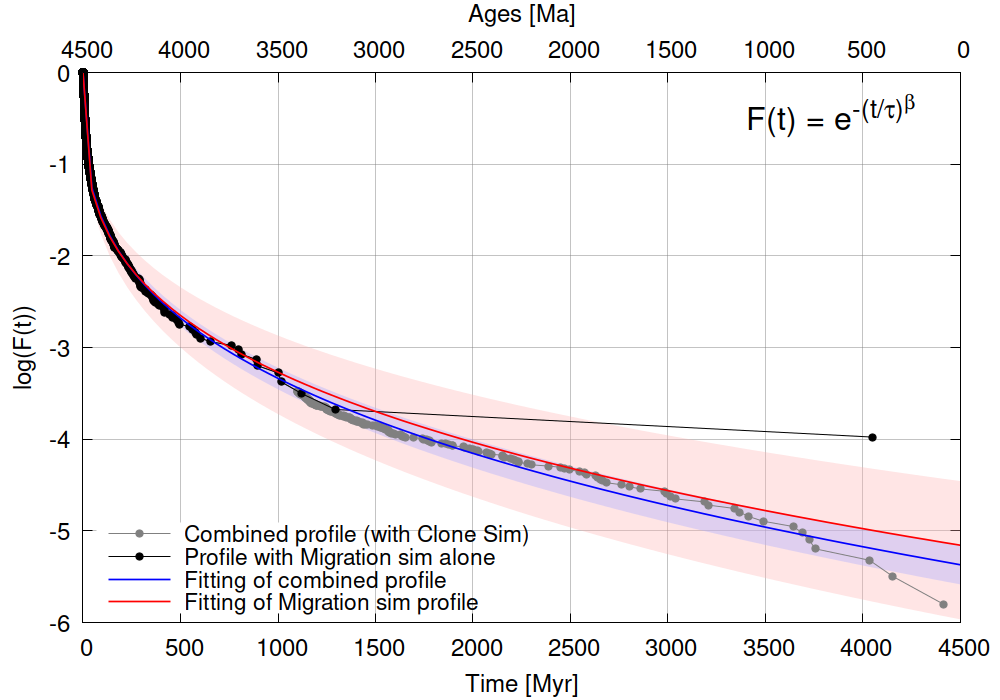}} 
\caption{\label{fig:clone} The difference in the modelled crater chronologies of Saturn was constructed from the combined decaying impact profile with both migration simulation and clone simulation and from the profile with only the migration simulation. The changes in the modelled curves and uncertainties become more evident after 1000 Myr into the simulation. }
\end{figure*}
Our study employed the impact recorded in the clone simulations to extend the impact chronologies' length and accuracy, especially in recent times. We like to demonstrate the improvement in the surface ages estimation by using Enceladus' crater plains as an example. We found that the estimated age with the clone simulation reported as the main result of this study was 4.10 Ga $\pm$ 0.04 ($\sim$1\%), while the ages derived without the clone simulation (solely the migration simulation) was 4.08 Ga +/- 0.16 ($\sim$4\%). Please note: these uncertainties account for the fitting of $\beta$ and $\tau$ but do not consider the uncertainty introduced by different planetesimal disc models. Despite the two ages, with and without the impact data from clone simulations, agreeing with the uncertainties, we argue here for the importance of the clone simulations, using Fig.~\ref{fig:clone} as a guide. \\

The best fits for $\tau$ and $\beta$ differ between simulations with and without clones, leading to changes in the modelled chronology. Relying solely on the impact records by the migration simulations results in far fewer impacts, compresses the declining impact profile and alters the e-folding time ($\tau$) and stretching parameter ($\beta$) of the Weibull distribution fitting. For sparsely cratered and younger surfaces, even slight differences in $\tau$ and $\beta$ would significantly impact the estimated surface ages. Fortunately, the studied cratered plains all turned out to be ancient, but before performing the study, we could not anticipate that the estimated ages would exceed 3.5 Ga. In Fig.~\ref{fig:clone} the difference between the red (without clone simulations) and the blue lines (with clone simulation) is evident, particularly at younger ages, after 1 Gyr into the simulation. Therefore, if we want to use these numerical results for younger surfaces, a highly accurate impact chronology is required, which requires the clones.\\

Furthermore, including the clones reduces the uncertainties in fitting $\tau$ and $\beta$, particularly for $\tau$, and translates to greater accuracy in the resulting ages, in particular for younger surfaces. Note the size difference between pale red (without clone simulations) and pale blue shaded areas (with clone simulations) in Fig.~\ref{fig:clone}, which cover the uppermost and lowermost range for the two modelled crater chronologies.\\

Last but not least, as stated in the manuscript, impacts after 1 Gyr of simulation time are extremely rare: note the number of solid black dots in Fig.~\ref{fig:clone} after 1 Gyr of evolution. Each dot represents an impact recorded from the migration simulation. Estimating surface ages that turned out to be younger than 3.5 Ga using the modelled chronologies constructed without the clone simulation would not be justified.\\

Overall, we like to highlight the usefulness of clone simulations in improving the accuracy and reducing uncertainties in estimating surface ages based on the modelled impact chronology.

\cleardoublepage 
\end{document}